%% file: MSSM-Inflation.tex
\documentclass[aps,groupedaddress,nofootinbib,preprintnumbers,floatfix,showkeys,preprintnumbers]{revtex4}

\usepackage[normalem]{ulem}

\usepackage{hhline}
\usepackage{graphicx}
\usepackage{amsfonts}
\usepackage{amsmath}
\usepackage[cal=boondoxo]{mathalpha}
\usepackage{tabularx,url,color}
\usepackage{xcolor}
\usepackage{multirow}
\usepackage{slashed}
\usepackage{cancel}
\usepackage{soul}
\usepackage{ulem}
\usepackage{subfig}
\usepackage{yfonts}
\usepackage{xspace}
\usepackage{setspace}
\usepackage{hyperref}
\usepackage[capitalise]{cleveref}
\usepackage{kpfonts}

\makeatletter
\DeclareRobustCommand{\element}[1]{\@element#1\@nil}
\def\@element#1#2\@nil{%
  \MakeLowercase{#1}%
  \if\relax#2\relax\else\MakeLowercase{#2}\fi}
\makeatother

\makeatletter
\newcommand{\doublewidetilde}[1]{{%
  \mathpalette\double@widetilde{#1}%
}}
\newcommand{\double@widetilde}[2]{%
  \sbox\z@{$\m@th#1\widetilde{#2}$}%
  \ht\z@=.9\ht\z@
  \widetilde{\box\z@}%
}
\makeatother

\usepackage{amssymb}
\usepackage{pifont}
\newcommand{\xmark}{\ding{55}}%

\newcommand{\omegacdm}{\ensuremath{\Omega_\text{cdm}h^2}}
\newcommand{\Mtop}{\ensuremath{m_{\mathrm{t}}}}
\newcommand{\Mone}{\ensuremath{M_1}}

\newcommand{\Mtwo}{\ensuremath{M_2}}
\newcommand{\Mthree}{\ensuremath{M_3}}
\newcommand{\muorq}{\ensuremath{Q}}

\definecolor{amethyst}{rgb}{0.6, 0.4, 0.8}

\newcommand{\p}{\ensuremath{\mathrm{p}}}

\newcommand{\JuneCorrections}[1]{{\color{black}{#1}}}

\definecolor{darkbrown}{rgb}{0.7, 0.36, 0.23}

\newcommand{\eqVtree}{V_\mathrm{tree}}
\newcommand{\eqVrge}{V_{\scriptscriptstyle{\mathrm{RGE}}}}
\newcommand{\Vtree}{$\eqVtree$}
\newcommand{\Vrge}{$\eqVrge$}

\newcommand{\hep}{HEP}

\newcommand{\phistar}{\phi_*}
\newcommand{\DeltaNStar}{\Delta N_*}

\newcommand{\Mp}{M_{\scriptscriptstyle{\mathrm{Pl}}}}
\newcommand{\As}{A_{\scriptscriptstyle{\mathrm{S}}}}
\newcommand{\ns}{n_{\scriptscriptstyle{\mathrm{S}}}}

\newcommand{\nsmean}{\overline{\ns}}
\newcommand{\alphas}{n_{{\scriptscriptstyle{\mathrm{S}}},\mathrm{run}}}
\newcommand{\nt}{n_{\scriptscriptstyle{\mathrm{T}}}}
\newcommand{\alphat}{n_{{\scriptscriptstyle{\mathrm{T}}},\mathrm{run}}}
\newcommand{\dd}{\mathrm{d}}

\newcommand{\Planck}{\ensuremath{\mathrm{Planck}}}
\newcommand{\lnrrad}{\ensuremath{\ln R_{\mathrm{rad}}}}
\newcommand\srlo{\mathrel{\overset{\makebox[0pt]{\mbox{\normalfont\tiny\sffamily SRLO}}}{\simeq}}}

\input{macros}

\def\eg{{\sl e.g.}\ }
\def\ie{{\sl i.e.}\ }

\DeclareMathOperator{\sgn}{sgn}

\begin{document}
\title{
MSSM-inflation revisited:\\ Towards a coherent description of high-energy physics and cosmology}
\author{Gilles Weymann-Despres$^1$, Sophie Henrot-Versill\'e$^1$,  Gilbert Moultaka$^{2,3}$, Vincent Vennin$^4$, Laurent Duflot$^1$, Richard von Eckardstein$^5$ }
\affiliation{$^1$ Universit\'e Paris-Saclay, CNRS/IN2P3, IJCLab, Orsay, France,}
\affiliation{$^2$ Laboratoire Charles Coulomb (L2C), Universit\'e de Montpellier, CNRS, Montpellier, France.}
\affiliation{$^3$ Laboratoire Univers \& Particules de Montpellier (LUPM), Universit\'e de Montpellier, CNRS, Montpellier, France}
\affiliation{$^4$ Laboratoire de Physique de l'Ecole Normale Sup\'erieure, ENS, CNRS, Universit\'e PSL, Sorbonne Universit\'e, Universit\'e Paris Cit\'e, F-75005 Paris, France}
\affiliation{$^5$ Institute for 
Theoretical Physics, University of M\"unster, 48149 M\"unster, Germany}

\begin{abstract}
The aim of this paper is to highlight the challenges and potential gains surrounding a coherent description of physics from the high-energy scales of inflation down to the lower energy scales probed in particle-physics experiments. As an example, we revisit the way inflation can be realised within an effective Minimal Supersymmetric Standard Model (eMSSM), in which the \lletexte \ and \uddtexte \ flat directions are lifted by the combined effect of soft-supersymmetric-breaking masses already present in the MSSM, together with the addition of effective non-renormalizable operators. We clarify some features of the model and address the question of the one-loop Renormalization Group improvement of the inflationary potential, discussing its impact on the fine-tuning of the model. We also compare the parameter space that is compatible with current observations (in particular the amplitude, $\As$, and the spectral index, $\ns$, of the primordial cosmological fluctuations) at tree level and at one loop, and discuss the role of reheating. Finally we perform combined fits of particle and cosmological observables (mainly $\As$, $\ns$, the Higgs mass, and the cold-dark-matter energy density) with the one-loop inflationary potential applied to some examples of dark-matter annihilation channels (Higgs-funnel, Higgsinos and A-funnel), and discuss the status of the ensuing MSSM spectra with respect to the LHC searches.
\end{abstract}

\keywords{Inflation -- Cosmic Microwave Background -- High-Energy Physics -- Supersymmetry -- MSSM -- Physics of the early universe}

\date{\today}
\maketitle

\tableofcontents
\newpage

\section{Introduction}
Tackling the understanding of the physics of inflation is the next challenge of Cosmic Microwave Background (CMB) experiments.
The observation of large-scale B-mode polarization in the CMB will offer a unique probe of fundamental physics at energies far
beyond the reach of CERN's Large Hadron Collider (LHC), opening a new window towards primordial cosmology.

Today, the results from the Planck satellite support the inflation hypothesis~\cite{Starobinsky:1980te, Sato:1980yn, Guth:1980zm, Linde:1981mu, Linde:1983gd, Albrecht:1982wi, Linde:1990flp, Lyth:1998xn}. Measurements of the amplitude and
spectral index of the primordial scalar power spectrum, and constraints on its running, on the amount of non-Gaussianities, on the amplitude of isocurvature modes and
on the tensor-to-scalar ratio, 
favor single-field slow-roll inflationary models ~\cite{Martin:2013nzq, Planck:2018jri}. Such phenomenological models are often inspired by high-energy constructions~\cite{Ratra:1987rm, Liddle:1994dx, Martin:2013tda} in which one or more scalar fields acquire a flat-enough potential. Though, very few of them come with a complete embedding within or beyond the Standard Model of particle physics that would allow a coherent description of physics from the high-energy scales of inflation down to the lower energy scales probed in particle-physics experiments. The aim of this paper is to use a well-defined theoretical framework as a test case and study all the aspects of a fully specified embedding. 

In this perspective, the  extensions of the Standard Model of particle physics based on supersymmetry (SUSY) (see \eg \cite{Nilles:1983ge,Martin:1997ns} for reviews), such
as the Minimal or Next-to-Minimal Supersymmetric Standard Model (MSSM, NMSSM) provide an appealing 
theoretical framework that allows one to describe both the inflation era (as they naturally include flat directions 
\cite{Buccella:1982nx,Affleck:1984xz,Luty:1995sd,Gherghetta:1995dv} that could support inflation \cite{Enqvist:2003gh, Allahverdi:2006iq, Allahverdi:2006we, BuenoSanchez:2006rze,Allahverdi:2007vy,Allahverdi:2010zp,Boehm:2012rh, Choudhury:2011jt, Choudhury:2013jya, Choudhury:2014sxa, Dubinin:2017irq, Dubinin:2017irg}) and the physical processes that can be measured at the LHC \cite{Allahverdi:2011su}. Such a theory can predict multiple observables, both for cosmology and \JuneCorrections{particle-}physics (\hep), which can then be compared to measurements to more accurately assess the favored/disfavored area in parameter space within a coherent description of our universe.

In this paper, the test case that we consider is an inflationary scenario associated with two sets of flat directions, dubbed \lletexte\ and \uddtexte, naturally encompassed in the MSSM. The potential of the inflaton candidates can then be generated only when these flat directions are lifted by the combined effects of soft-SUSY-breaking masses already present in the MSSM, and effective non-renormalizable operators that should be added to the model \cite{Allahverdi:2006iq}. We will thus refer to this scenario as the effective MSSM (eMSSM) to avoid confusion when discussing the particle-physics features of the strict MSSM. 
The analysis in \cite{Allahverdi:2006we, BuenoSanchez:2006rze, Boehm:2012rh, Choudhury:2014sxa} or the Generalized MSSM Inflation Model (GMSSM) analysis of \cite{Martin:2013tda} have been, for example, motivated by such configurations. Since the inflaton candidates are gauge invariant, the parameters of the potential depend on the energy scale at which they are evaluated through the renormalization group equations (RGEs). This well-known aspect has already been studied in this specific eMSSM case to relate the inflaton mass at the scale of inflation to the one probed at the LHC~\cite{Allahverdi:2007vy, Allahverdi:2010zp, Boehm:2012rh}. In this paper, we go one step further and study how the RGEs at one-loop level affect all parameters in the inflationary potential, and impact the way we connect cosmology and particle-physics constraints.\footnote{\JuneCorrections{The inclusion of RGEs corrections in other inflationary potentials has already been performed for instance in~\cite{Barrow:1995xb, Senoguz:2008nok, Barvinsky:2008ia, DeSimone:2008ei, Barvinsky:2009ii, Bezrukov:2010jz, Steinwachs:2013tr, Elizalde:2014xva, George:2015nza, Enckell:2018kkc}.}}

In a first part, we review the basics of slow-roll inflation and we revisit the eMSSM inflationary potential. In a second part, we set the stage of the analysis with the description of the phenomenological MSSM we are using, together with the observational constraints, the tools and the methodology. In a third part, we address the conditions required on the inflationary potential parameters to fulfill slow-roll inflation. In particular, we discuss the initial conditions for inflation and the required level of fine-tuning of the parameters, comparing the tree-level and one-loop inflationary potentials. In a fourth part, we identify the region of the 
parameter space  where inflation takes place and yields to predicted values of the amplitude, $\As$, and tilt, $\ns$, of the primordial spectrum that are compatible with CMB experiments. We discuss the LHC phenomenology in this region, and also study the impact of neglecting the RGEs
(as done in previous works \cite{Boehm:2012rh, Martin:2013tda}) in this analysis. Finally, we illustrate the impact of the use of the one-loop potential on several eMSSM points compatible with \hep\ observations, the cold-dark-matter energy density and the inflationary observables. 

\section{Inflation and \texorpdfstring{\lowercase{e}}\ MSSM}

Inflation is a phase of accelerated expansion that takes place in the early universe, and during which quantum vacuum fluctuations are amplified by gravitational instability~\cite{Starobinsky:1979ty,Mukhanov:1981xt, Starobinsky:1982ee,Guth:1982ec,Bardeen:1983qw} giving rise to classical density perturbations. Those seed the CMB anisotropies as well as the large-scale structures of the universe. For inflation to occur within General Relativity, the universe must be dominated by a fluid with negative pressure. This cannot be achieved by common low-energy fluids, but rather requires working with fields. 

\subsection{Single-field slow-roll inflation}

\JuneCorrections{To keep the paper as self-contained as possible, to set the notations but also to stress the role played by reheating, we start with a reminder of the main ingredients of single-field slow-roll inflation.}

\subsubsection{Scalar-field inflation}

The simplest field compatible with space-time symmetries is a homogeneous scalar field $\phi$, referred to as the ``inflaton''. When minimally coupled to gravity, it evolves according to the Klein-Gordon equation:
\begin{equation}
    \label{eq:klein-gordon}
    \ddot\phi+3H\dot\phi+V_\phi=0,
\end{equation}
where dots denote derivatives with respect to cosmic time, $H=\dot{a}/a$ is the Hubble parameter, with $a$ the scale factor of the universe, and $V(\phi)$ is the potential energy stored in $\phi$ (and $V_\phi\equiv \dd V/\dd\phi)$. According to Friedmann's equation, the Hubble parameter is related to the energy density of the universe,
\begin{equation}
    \label{eq:friedmann}
    3\Mp^2H^2=V+\frac{\dot\phi^2}{2},
\end{equation}
where $\Mp$ is  the reduced Planck mass. Inflation occurs when $\ddot{a}>0$, which from the above is equivalent to

\begin{equation}
    \label{eq:inflate}
    V>\dot\phi^2.
\end{equation}
This implies that the potential function $V(\phi)$ must be sufficiently flat for the field to roll it down sufficiently slowly, such that its kinetic energy does not exceed half its potential energy.

\subsubsection{Slow-roll approximation}

\label{sec:IA1}

The Klein-Gordon equation (\ref{eq:klein-gordon}) being second order, the dynamical phase-space $(\phi,\dot{\phi})$ has dimension two. When inflation takes place, there exists a dynamical attractor along which the acceleration term $\ddot{\phi}$ becomes negligible in \cref{eq:klein-gordon}, and the friction term $3H\dot{\phi}$ compensates the potential gradient $V_\phi$. Such an attractor is called ``slow-roll''~\cite{Liddle:1994dx} since for flat-enough potentials it is also such that the condition of \cref{eq:inflate} is saturated, \ie $\dot\phi^2\ll V$.

Technically, slow roll corresponds to the regime where the slow-roll parameters $\epsilon_n$ are small. Starting from $\epsilon_0=1/H$, those are iteratively defined via

\begin{equation}
\label{eq:sr_params}
\epsilon_{n+1}=\dd\ln\vert\epsilon_n\vert/\dd N\, ,
\end{equation}
where
\begin{equation}
    \label{eq:e-folds}
    N-N_0\equiv \ln(a/a_0) = \int^t_{t_0}{H \dd t}
\end{equation}
is the so-called number of $e$-folds, $N_{\textit{e}\textrm{-folds}}$, generated between the times $t_0$ and $t$. For instance, the first slow-roll parameter is given by $\epsilon_1=-\dot{H}/H^2=1-a\ddot{a}/\dot{a}^2$, so inflation ($\ddot{a}>0$) corresponds to $\epsilon_1<1$. Using ~\cref{eq:klein-gordon} and \cref{eq:friedmann}, one also has $\epsilon_1=3\dot{\phi}^2/(2V+\dot{\phi}^2)$ and one recovers the condition of \cref{eq:inflate}, with $\epsilon_1\ll 1$ corresponding to $\dot\phi^2\ll V$. Similarly, $\epsilon_2 = 2\epsilon_1-2V'/(H\dot{\phi})-6$, so $\vert \epsilon_2\vert \ll 1$ corresponds to when the acceleration term in the Klein-Gordon equation is subdominant. In this regime, $\dot{\phi}$ becomes a function of $\phi$ only,
\begin{equation}
\label{eq:SR:traj}
\dot\phi\srlo - \Mp\frac{V_\phi}{\sqrt{3 V}}\, ,
\end{equation}
and approximate expressions for the slow-roll parameters can be derived that only involve the potential function:
\begin{align}
    \label{eq:epsilon:1:pot}
    \epsilon_1\  & \srlo \ \  \frac{\Mp^2}{2}\left(\frac{V_{\phi}}{V}\right)^2,\\
    \label{eq:epsilon:2:pot}
    \epsilon_2 \ & \srlo  \ \  2 \Mp^2\left[\left(\frac{V_{\phi}}{V}\right)^2-\frac{V_{\phi\phi}}{V}\right],\\
    \label{eq:epsilon:3:pot}
    \epsilon_3\  & \srlo \ \  \frac{2}{\epsilon_2}\Mp^4 \left[\frac{V_{\phi\phi\phi} V_\phi}{V^2}-3\frac{V_{\phi\phi}}{V}\left(\frac{V_\phi}{V}\right)^2+2\left(\frac{V_\phi}{V}\right)^4\right],
\end{align}
where $\srlo$ indicates that we work at leading order in the slow-roll approximation and where only the first three slow-roll parameters are given.

Cosmological perturbations can be introduced on top of this homogeneous and isotropic expanding background. Only the scalar and tensor sectors propagate with equations of motion that are set by the dynamics of the background, which can itself be described by means of the slow-roll parameters. This is why, when the initial conditions for inflation are set in the quantum vacuum state (the so-called ``Bunch-Davies vacuum''), the Fourier modes of these perturbations only depend on the value of the slow-roll parameters at the time when they cross out the Hubble radius during inflation. At linear order in perturbation theory, their statistics is Gaussian, hence fully specified by the power spectra $\mathcal{P}_\zeta$ and $\mathcal{P}_h$, where $\zeta$ denotes the scalar curvature perturbation and $h$ stands for the tensor perturbation. At leading order in slow roll, their amplitudes are parameterised by~\cite{Stewart:1993bc, Gong:2001he}
\begin{align}
    \As\equiv \left.\Pscalar\right|_{k_*} & \srlo  \ \ \ \frac{V_*}{24\pi^2\Mp^4\epsilon_{1*}}    \label{eq:AsSR}\, ,\\
    r\equiv \left.\frac{\Ptensor}{\Pscalar}
    \right|_{k_*}& \srlo \ \ \   16\epsilon_{1*}\label{eq:rSR}\, ,
\end{align} 
where $k_*$ denotes the CMB pivot scale that we chose to be $k_* = 0.05\ \Mpc^{-1}$. Stars indicate that the quantities are evaluated at the time of $k_*$ Hubble crossing. The slight scale-dependence of the scalar and tensor power spectra is parameterised by the spectral tilts:
\begin{align}
\label{eq:nsSR}
    \ns\equiv 1+\left.\frac{\dd \ln \Pscalar}{\dd \ln k}\right|_{k_*}& \srlo  1-2\epsilon_{1*}-\epsilon_{2*}
    \quad\quad\quad\text{and}\quad\quad\quad
    \nt\equiv \left.\frac{\dd \ln \Ptensor}{\dd \ln k}\right|_{k_*}  \srlo -2\epsilon_{1*}\ ,
\end{align}
and by the running of these tilts:
\begin{align}
    \alphas\equiv \left.\frac{\dd^2 \ln \Pscalar}{(\dd \ln k)^2}\right|_{k_*}& \srlo  -2\epsilon_{1*}\epsilon_{2*}-\epsilon_{2*}\epsilon_{3*}
    \quad\quad\quad\text{and}\quad\quad\quad
    \alphat\equiv \left.\frac{\dd^2 \ln \Ptensor}{(\dd \ln k)^2}\right|_{k_*}  \srlo -2\epsilon_{1*}\epsilon_{2*}\, .
\label{eq:alphat:SR}
\end{align}
In order to derive observational predictions from a given inflationary potential $V(\phi)$, one must therefore solve the background dynamical equations (\ref{eq:klein-gordon}) and (\ref{eq:friedmann}), compute the \vevchange\  of the inflaton at the time of Hubble crossing of the pivot scale, evaluate \cref{eq:epsilon:1:pot,eq:epsilon:3:pot} at that location, and insert the result into \cref{eq:AsSR,eq:alphat:SR}.

\subsubsection{\Vevchange at Hubble crossing}

In this process, the non-trivial remaining step is the calculation of the \vevchange at Hubble crossing, $\phistar$. As mentioned in \cref{sec:IA1}, slow roll is a dynamical attractor, so in single-field models it singles out one phase-space trajectory. Therefore, if one knows the end point of inflation $\phi_{\mathrm{end}}$ as well as the number of \textit{e}-folds $\DeltaNStar$ elapsed between the time the pivot scale $k_*$ crosses out the Hubble radius and the end of inflation, the value of $\phistar$ can be derived from
\begin{equation}
\label{eq:srDeltaN*}
\DeltaNStar  = \int_{\phi_{\mathrm{end}}}^{\phistar}\frac{\dd\phi}{\sqrt{2\epsilon_1}}\simeq \int_{\phi_{\mathrm{end}}}^{\phistar} \frac{V(\phi)}{V_\phi(\phi)}\dd\phi \, .
\end{equation}
Here, the first part of the formula was obtained from \cref{eq:e-folds} jointly with the expression of $\epsilon_1$ in terms of $\phi$ and $\dot{\phi}$ given by \cref{eq:sr_params}, and the second part of the formula follows from \cref{eq:epsilon:1:pot}. The end point $\phi_{\mathrm{end}}$ can be obtained from solving $\epsilon_1=1$, using slow-roll formula (\ref{eq:epsilon:1:pot}) for $\epsilon_1$. The number of \textit{e}-folds $\DeltaNStar$ depends on the Hubble scale during inflation and on the expansion history since the end of inflation until the present time. More precisely, it is given by~\cite{Martin:2006rs, Martin:2010kz, Easther:2011yq}
\begin{equation}
    \label{eq:implicit_eq}
    \DeltaNStar =  \lnrrad-\ln\left(\frac{k_*}{a_0\Tilde\rho_{\gamma}^{1/4}}\right)-\frac{1}{4}\ln\left[\frac{9V_{\mathrm{end}}}{\epsilon_{1*}(3-\epsilon_{1\mathrm{end}}) V_*}\right]+\frac{1}{4}\ln(8\pi^2\As) \, ,  
\end{equation}
where ``end'' denotes quantities computed at the end of inflation. This equation is implicit for $\DeltaNStar$. In this expression, $\Tilde\rho_{\gamma}$ is the energy density of radiation today rescaled by the change in the number of relativistic degrees of freedom between the completion of reheating  and today, and $\lnrrad$ is the so-called ``reheating parameter''. It incorporates the effect of reheating, \ie the stage between inflation and the radiation era, and is given by
\begin{equation}
\label{eq:lnrrad}
\lnrrad =
\frac{1-3\wbarreh}{12(1+\wbarreh)}\ln\left(\frac{\rho_{\mathrm{reh}}}{\rho_{\mathrm{end}}}\right),
\end{equation}
where $\wbarreh$ is the mean equation-of-state parameter during reheating and $\rho_{\mathrm{reh}}$ is the energy density of the universe at the onset of the radiation era. If reheating is instantaneous ($\rho_{\mathrm{reh}}=\rho_{\mathrm{end}}$), or if its equation of state is the one of radiation ($\wbarreh=1/3$), then $\lnrrad=0$, so $\lnrrad$ measures how much reheating departs from a radiation-dominated phase. Note that it is common practice to approximate the third term in \cref{eq:implicit_eq} by its slow-roll limit, the impact of this approximation is further discussed in \cref{sec:tree_results_reheating}.

\subsection{eMSSM}

In this work, inflation is realised within an effective MSSM (eMSSM), in which the \lletexte\ and \uddtexte\ flat directions are lifted by the combined effect of soft-SUSY-breaking masses already present in the MSSM, together with the addition of effective non-renormalizable operators. The goal of this section is to introduce the relevant parameters, and to derive the inflationary potential.  
We will treat the renormalization-group-improved inflationary potential differently from previous studies, reaching different conclusions concerning the amount of fine-tuning in the model (cf. \cref{sec:fine-tune}).

\subsubsection{eMSSM flat directions and the inflationary potential } 
\label{sec::flatMSSM}

The scalar potential of the R-parity conserving MSSM has a large number of flat directions 
in the supersymmetric limit 
that can be lifted by soft-SUSY-breaking terms and/or by higher-dimensional supersymmetry-preserving operators 
\cite{Gherghetta:1995dv}. This offers in principle various possibilities for implementing inflationary
scenarios with the inflaton being a combination of scalar fields within the MSSM.
The model we study in the present paper extends the MSSM by specific 
non-renormalizable superpotential terms that lift the so-called 
\lletexte\ or \uddtexte\ flat directions in order to trigger an inflationary phase \cite{Allahverdi:2006iq}.

We will first go through the main ingredients of the model. This will allow us to recall the underlying physical assumptions and possible uncertainties, introduce the notations, but also clarify discrepancies between the different parameter normalizations performed in the literature.

The notations \lletexte\ and \uddtexte\ stand for the 
$\rm{SU}(3)_c\times \rm{SU}(2)_L \times \rm{U}(1)_Y$ gauge-invariant operators, $L_i\cdot L_j e_k \equiv \epsilon_{\alpha \beta} \, L_i^\alpha L_j^\beta  \, e_k$ and $(u_i \times d_j)\cdot d_k \equiv\epsilon_{a b c} \, u_i^a d_j^b  \, d_k^c$, that characterize the corresponding flat directions \cite{Buccella:1982nx,Affleck:1984xz,Luty:1995sd,Gherghetta:1995dv}
for specific lepton or quark generation indices $i,j,k$.
$L$ denotes an $\rm{SU}(2)_L$ doublet scalar field and $e,u,d$, $\rm{SU}(2)_L$ singlet scalar fields; $a,b,c$ are color indices and $\alpha, \beta$, $\rm{SU}(2)_L$ flavor indices. Because the $\epsilon$ symbols are antisymmetric with respect to all their indices, the relevant cases are obviously 
$i\neq j$ for \lletexte\ and $j\neq k$ for \uddtexte. The associated flat directions of the potential correspond to scalar field components satisfying the  following configurations for a fixed choice of generation, flavour and color indices:
\begin{align}
L_i^1 &=L_j^2  = e_k \equiv \ell(x), \ (i\neq j)\label{eq:lledir} \\ 
u_i^a   &=d_j^b=d_k^c \equiv \mathcal{q}(x), \ (j\neq k, 
a\neq b\neq c)\label{eq:udddir}
\end{align}
where $\ell(x)$ and $\mathcal{q}(x)$ denote arbitrary complex-valued scalar fields,  and all other scalar field components are set to zero.

In the absence
of renormalizable R-parity violating terms in the superpotential, these flat directions are lifted by 
dimension-6 operators of the form,
\begin{equation}
\begin{aligned}
  W_{6}^{(\lleind)}=&\lambda_{\lle}\frac{(L_i\cdot L_j e_k)^2}{\Mp^{3}}, \\ 
  W_{6}^{(\uddind)}=&\lambda_{\udd}\frac{(u_i \times d_j\cdot d_k)^2}{\Mp^{3}}.
  \label{eq:superpotential1}
\end{aligned}
\end{equation} 
These contributions to the superpotential (where we denote the superfields by the same letters as their scalar components) can be viewed as effective operators originating from an UltraViolet (UV) completion of the MSSM after integrating out the corresponding heavy fields. The dimensionless couplings (that we take to be real-valued) $\lambda_{\lle}$ and
$\lambda_{\udd}$ are thus expected to be typically of order one. For definiteness, we choose the corresponding mass scale to be the Planck scale. However, depending on the content of the UV completion, similar operators with lower mass scales such as $\Mgut$ (where GUT stands for Grand Unified Theory) can also arise. We will come back to the consequences of such a change in \cref{eq:superpotential1} in \cref{sec:examples_and_big_table}.
Since we are interested in single-field inflation, we will be considering one flat direction at a time, \ie one given choice of $(i,j,k)$ for the slepton or squark generation content. (For the latter case,
we also avoid extra color factor enhancement of $\lambda_{\udd}$ by assuming for simplicity a fixed choice
for the color indices $a\neq b\neq c$)\footnote{This does not prevent other similar $W_6$ operators from being generated by the UV physics (note that, beyond \cref{eq:superpotential1}, operators of the form $(L_i\cdot L_j e_k) (L_l\cdot L_m e_n)$ or $(u_i\times d_j \cdot d_k) (u_l\times d_m \cdot d_n)$ or $(L_i\cdot L_j e_k) (u_l\times d_m \cdot d_n)$ can also arise). Rather, it corresponds to setting initial field conditions near the considered flat directions at the onset of inflation.}. 

We will use the common notation $\varphi(x)$ to denote
either $\ell(x)$ or $\mathcal{q}(x)$. Since in \cref{eq:lledir,eq:udddir} there are always three distinct canonically normalized complex-valued fields
involved, the three corresponding kinetic terms in the MSSM immediately lead to the normalization
\begin{equation}
\ell(x), \mathcal{q}(x) = \frac{1}{\sqrt{3}} \varphi(x)
\label{eq:norma1}
\end{equation}
in order for the complex-valued $\varphi(x)$ to have a canonical kinetic term. 
Taking into account the (soft) mass terms of the three fields leads also to the canonically
normalized mass for $\varphi(x)$:
\begin{align}
\label{eq:mphi_udd}
m_{\varphi}^2 =& \frac{m_{\tilde{u}^{i}_R}^2+m_{\tilde{d}^{j}_R}^2+m_{\tilde{d}^{k}_R}^2}{{3}}  \ \ \  (\udd),\\
\label{eq:mphi_lle}
m_{\varphi}^2 =& \frac{m_{\tilde{l}^{i}_L}^2+m_{\tilde{l}^{j}_L}^2+m_{\tilde{e}^{k}_R}^2}{{3}} \ \  \  (\lle).
\end{align}
The normalizations in \cref{eq:norma1,eq:mphi_udd,eq:mphi_lle} are in agreement with the literature \cite{Allahverdi:2006we,Allahverdi:2007vy,Allahverdi:2010zp,Boehm:2012rh}. However, we found differences with respect to the same literature in other parts related to the normalization of the couplings in $W_6$ as well as in the
soft-SUSY-breaking parameter that modify the inflationary potential. To ease the discussion, we will adopt three notations to distinguish among the various occurrences of $W_6$.  We will generically denote by $\widetilde{\lambda_6}$ 
the couplings appearing in \cref{eq:superpotential1}, 
\begin{equation}
\label{eq:def-tildelambda6}
  \widetilde{\lambda_6}=  \lambda_{\lle} \ \textrm{or} \ \lambda_{\udd},
\end{equation}
and 
by $\doublewidetilde{\lambda_6}$ the coupling appearing in $W_6$ along a given flat direction, defined as:
\begin{equation}
  W_6=\frac{\doublewidetilde{\lambda_6}}{6}\frac{\varphi^6}{\Mp^{3}}. 
  \label{eq:superpotential}
\end{equation}
Note that the $1/6$ 
normalization is the one adopted in the literature \cite{Allahverdi:2006iq, Allahverdi:2006we,Allahverdi:2007vy,Allahverdi:2010zp,Boehm:2012rh}.
From \cref{eq:lledir,eq:udddir,eq:superpotential1,eq:norma1}, one obviously finds the reduction $\doublewidetilde{\lambda_6} =(2/9) \widetilde{\lambda_6}$, which is inconsistent with the identification $\widetilde{\lambda_6} = \doublewidetilde{\lambda_6} \equiv \lambda_6$ performed in the literature, where  $\lambda_6$ denotes the 
actual coupling entering the inflaton potential. We will see that $\lambda_6$ comes with a prefactor with respect to $\widetilde{\lambda_6}$ and $\doublewidetilde{\lambda_6}$. Indeed, the proper normalization that leads to a canonical
kinetic term for the real-valued inflaton field $\phi(x)$ is
\begin{equation}
\varphi(x)= \frac{1}{\sqrt{2}} \phi(x) e^{i \theta(x)}, \ \textrm{with} \ \theta(x) \in \left[0, \pi\right].\label{eq:norma2}
\end{equation}
Note that $\phi(x)$ is defined by $\phi(x)\propto \pm \vert \varphi(x) \vert$ which is why $\theta(x)$ takes values between $0$ and $\pi$ only.
From Eqs.~(\ref{eq:superpotential1}), (\ref{eq:norma1}), (\ref{eq:norma2}) and the supersymmetric contribution 
$\left|\frac{\partial W_6}{\partial \varphi}\right|^2$ to the potential, one finds that the coupling $\lambda_6$ appearing in the last term on the right-hand side of
\cref{eq:VMSSM} below, is given by
\begin{equation}
\lambda_6 =\frac{\doublewidetilde{\lambda_6} }{4 \sqrt{2}}=\frac{\widetilde{\lambda_6} }{18 \sqrt{2}} \lesssim \mathcal{O}{\left(\frac{1}{18 \sqrt{2}}\right)}.    \label{eq:norma3}
\end{equation}
To summarize: on the one hand, $\lambda_6$ differs  by a factor ${4 \sqrt{2}}$ from the normalization found in the literature
\cite{Allahverdi:2006iq, Allahverdi:2006we,Allahverdi:2007vy,Allahverdi:2010zp,Boehm:2012rh} where $\doublewidetilde{\lambda_6}$ was identified with $\lambda_6 $. On the other hand, it also 
differs by a factor $18 \sqrt{2}$ from $\widetilde{\lambda_6}$ which satisfies $\widetilde{\lambda_6} \lesssim \mathcal{O}(1)$ and which was also identified with $\lambda_6$ in the literature.  This latter difference should be taken into account when assessing the consistency of the magnitude of $\lambda_6$. The former difference could have been just an unphysical redefinition of the coupling, were it not for the presence of the extra term in the potential generated by SUSY breaking that does not scale similarly to the supersymmetric term, as we now discuss.

For simplicity, we take here, 
as in \cite{ Allahverdi:2006we,Allahverdi:2007vy,Allahverdi:2010zp,Boehm:2012rh}, the example of minimal Supergravity (mSUGRA) mediation of SUSY-breaking, and in particular the soft-SUSY-breaking terms corresponding to $W_6$.
 In mSUGRA (minimal K\"ahler potential, see \eg \cite{Nilles:1983ge,Brignole:2010sax} for reviews), the 
soft-SUSY-breaking  part of the potential at the SUSY breaking scale $\Msusy$ has the universal form 
\begin{equation}
V_{\rm soft}=  { \left\vert m_{3/2}\right\vert^2 \sum_i |\varphi_i|^2 +
\left\{ m_{3/2} \left[ \sum_i \varphi_i \frac{\partial W}{\partial \varphi_i} + (a -3) W \right] + h.c.\right\}} \label{eq:Vsoft}
\end{equation}
where the $\varphi_i$ denote all the complex-valued scalar components of the matter superfields in the visible
(MSSM) sector, $W$ is a general superpotential depending on these fields, $m_{3/2}$ the gravitino mass parameter and $a$ a parameter related to the model-dependent mechanism of SUSY breaking,  both fixed by the vacuum expectation values of scalar fields and superpotential in the hidden sector and are in general complex-valued. The first term in \cref{eq:Vsoft} corresponds to the soft-SUSY-breaking masses
of all the scalar fields in the observable sector taking the universal value $\vert m_{3/2}\vert$ at the scale $\Msusy$. As for the second term, 
it is easy to see that any monomial $W_n$ contributing to $W$ and containing a product of $n$ $\varphi_i$'s, satisfies the identity
$\sum_i \varphi_i \frac{\partial W_n}{\partial \varphi_i}=n W_n$. The overall contribution of a given $W_n$
is thus $\widetilde{A}_n W_n + h.c.$ where $\widetilde{A}_n=m_{3/2} (a + n -3)$ defines the corresponding soft-SUSY-breaking scalar coupling, the A-term, which is in general complex-valued.
In the eMSSM, the relevant terms in $W$ are $W_3$ (comprising the full MSSM renormalizable superpotential)
and $W_6$ given by \cref{eq:lledir,eq:udddir}. The expression for $\widetilde{A}_n$ noted above then immediately leads to a relation
between the trilinear soft-SUSY-breaking coupling $\widetilde{A}_3$ and the bi-trilinear coupling $\widetilde{A}_6$ at the SUSY breaking scale:
\begin{equation}
\label{eq:msugra}
\widetilde{A}_3 = a m_{3/2}, \ \widetilde{A}_6 = (3 + a) m_{3/2}, \ \text{whence} \ \widetilde{A}_6 = \frac{3+a}{a} \widetilde{A}_3.
\end{equation}
$\widetilde{A}_3$ stands for the universal value of all the soft-SUSY-breaking trilinear scalar couplings of the MSSM at the $\Msusy$ scale. \Cref{eq:msugra} is a key relation. It links, at the SUSY-breaking scale, $\widetilde{A}_6$ to a universal value  of the soft trilinear couplings in the squark/slepton sectors of the MSSM. It will therefore allow us, together with \cref{eq:mphi_udd,eq:mphi_lle}, to relate the particle-physics features of the MSSM to the inflation features
of the eMSSM. To obtain the full potential, $V_{\rm soft}$ should be added to the supersymmetric F-term, 
$V_{\rm F}= \left\vert \frac{\partial W_6}{\partial \varphi}\right\vert^2$, and D-term ($V_{\rm D}$) contributions.\footnote{$V_{\rm D}=\frac12 \sum_i g_i^2 \vec{D}^i \cdot \vec{D}^i$ with $\vec{D}^i=\sum_j \Phi_j^\dag \vec{T}^i \Phi_j$. The $i$ index runs over the three gauge groups, with the corresponding gauge coupling $g_i$, and generators $\vec{T}^i$; $\Phi_j$ is the scalar multiplet for the $j^{th}$ squark or slepton  generation. }
Along the \lletexte\ or \uddtexte\ flat directions, $V_{\rm D}$ vanishes, as well as the $W_3$ contribution to $V_{\rm F}$ and to $V_{\rm soft}$. Adding the F-term contribution of \cref{eq:superpotential}  
 to its contribution to $V_{\rm soft}$, one finally finds the potential,
\begin{equation}
\label{eq:VMSSM}
    V(\phi)=\frac{1}{2}\mphi^2\phi(x)^2+\sqrt{2}\left\vert\widetilde{A}_6\right\vert\cos\left[6\theta(x)+\theta_6\right]\frac{\lambda_6\phi(x)^6}{6\Mp^3}+\lambda_6^2\frac{\phi(x)^{10}}{\Mp^6},
\end{equation}
where we wrote $\widetilde{A}_6=\vert \widetilde{A}_6\vert e^{i \theta_6}$. To reach this form of the potential, we used \cref{eq:norma2} which leads consistently to $m_\phi^2=m_\varphi^2$, but requires reabsorbing some normalization factors in the redefinition of the coupling, $\doublewidetilde{\lambda}_6 = 4 \sqrt{2}\lambda_6$, as was already anticipated in \cref{eq:norma3}. We also took into account a factor $2$ coming from the Hermitian conjugacy in \cref{eq:Vsoft}. The outcome differs from the literature by the relative factor $\sqrt{2}$ in the A-term and a factor $18 \sqrt{2}$ between $\lambda_6$ and the effective coupling in $W_6$.

Back to \cref{eq:msugra}, we note that, in practice, the universal value $\widetilde{A}_3$ of the trilinear couplings in the various squark/slepton sectors  is lost at lower scales, as these couplings run differently with the RGEs. Since the MSSM spectrum
is mostly sensitive to the third-generation quark/squark sector, we will take $\widetilde{A}_3$ to be \Atop,
the trilinear coupling involving the top squark at $\Msusy$.
Furthermore, we will follow \cite{Allahverdi:2006we,Allahverdi:2007vy,Allahverdi:2010zp,Boehm:2012rh} by making use of the knowledge of the $a$-parameter in the simple Polonyi model for the hidden sector where $a=3-\sqrt{3}$ \cite{Polonyi:1977pj,Nilles:1983ge}, to obtain from \cref{eq:msugra}  the relation
\begin{equation}
\label{eq:polonyi}
\widetilde{A}_6 (\Msusy) = \frac{6-\sqrt{3}}{3-\sqrt{3}} \Atopm(\Msusy)
\end{equation}
between $\widetilde{A}_6$ and \Atop, where we now indicate explicitly the scale dependence.\footnote{In subsequent sections where the framework of our analysis is described, we depart from the strict assumption of high-scale universality of soft-SUSY-breaking parameters. We will however stick to \cref{eq:polonyi} as an 
illustrative example of possible correlations.} Note, however, that \cref{eq:polonyi} is found to be the reverse of the relation given in 
\cite{Allahverdi:2006we,Allahverdi:2007vy,Allahverdi:2010zp,Boehm:2012rh}.
\Cref{eq:polonyi} will be assumed in the rest of the analysis,  where we will also equate for simplicity the two scales $\Msusy$ and $\Mgut$. Since in our 
analysis all parameters, including $\Atopm$, are taken real-valued, cf. \cref{sec:pMSSM},  \Cref{eq:polonyi} implies real-valued positive or negative $\widetilde{A}_6$, and thus $\theta_6=0$ or $\pi$. The actual values of $\theta_6$ are, however, irrelevant
 due to the preferred alignment of the $\theta(x)$ field during inflation, as we now explain. 

Let us note that, since $\varphi$ is a complex field, \cref{eq:VMSSM} is {\sl a priori} a two-field model, for $\phi$ and $\theta$. The angular direction $\theta$ is however heavy and can therefore be integrated out. Indeed, in the $\theta$ direction, the potential is minimal when $\cos(6\theta+\theta_6)=-1$. Around this minimum, when $\theta$ is properly normalized so to have a canonical kinetic term, its mass is of order $H \Mp/\phi$, hence it is much larger than $H$ as long as the inflaton takes on sub-Planckian field values, which is always the case hereafter. This implies that $\theta$ decays to the configuration $\cos(6\theta+\theta_6)=-1$ in a small fraction of an e-fold, and that its excitations are sufficiently suppressed to yield small amounts of isocurvature modes and of non-Gaussianities (note that, in principle, one would have to check that all other orthogonal directions are stabilised too~\cite{Enqvist:2011pt}). We can therefore consider the one-field potential: 

\begin{equation}
    \label{eq:potential_form}
    \eqVtree (\phi) \equiv V(\phi)= \frac{1}{2}\mphi^2\phi^2-\sqrt{2}\Asix\frac{\lambda_6\phi^6}{6\Mp^3}+\lambda_6^2\frac{\phi^{10}}{\Mp^6},
\end{equation}
where, from now on, we denote $\vert\widetilde{A}_6\vert$ by $\Asix\ (>0)$ in accordance with the
common notation used in the literature.\footnote{A word of caution: 
This notation should not lead to believe
that a negative sign is not allowed for $\Atopm$ in view of \cref{eq:polonyi}; indeed the latter reads
now $\Asix (\Msusy) = \frac{6-\sqrt{3}}{3-\sqrt{3}} \left\vert \Atopm(\Msusy)\right\vert$;  $\Atopm$ can
have either sign, which is phenomenologically important when identifying parameter-space regions compatible with the observed Higgs mass, but its sign does not affect the inflationary potential.}

\Cref{eq:potential_form}, together with  the definition range of the parameters,  

\begin{align}
   \label{eq:definition_range}
    m_\phi^{2}& >0,\\
    A_6& >0,\\
    \label{eq:lambdabound}
    0 <\lambda_6 & \lesssim \mathcal{O}{\left(\frac{1}{18 \sqrt{2}}\right)},
\end{align}

will be our starting point for the analysis of inflation.

\subsubsection{Renormalization-group-improved potential}
\label{sec:RGE}

Beyond the tree-level approximation, the {\sl effective} 
potential has a more 
involved dependence on $\phi$ than that given by 
\cref{eq:potential_form}.  It is important to carefully examine this aspect as it can lead to sizeable modifications of the inflationary predictions of the model. As well known, a powerful approach to capture 
classes of loop contributions to the effective potential is the requirement of Renormalization Group Invariance of the full effective potential \cite{coleman1973radiative,Kastening:1991gv,Bando:1992np,Ford:1992mv}.
This allows a resummation to
all orders in perturbation theory of powers of logarithms appearing in the loops. Obtaining the effective potential including the leading-order logarithms boils down to replacing all parameters in the tree-level potential by their one-loop running counterparts. 
In our case, the inflationary potential becomes 
\begin{equation}
    \label{eq:potential_RGE}
    \eqVrge (\phi) \equiv V(\phi)=\frac{1}{2}\mphi^2(\phi)\phi^2-\sqrt{2}\Asix(\phi)\frac{\lambda_6(\phi)\phi^6}{6\Mp^3}+\lambda_6(\phi)^2\frac{\phi^{10}}{\Mp^6},
\end{equation}
where  the running 
quantities $\mphi(\phi)$, $\Asix(\phi)$ and $\lambda_6(\phi)$ 
are governed by the RGEs given below.
This has two potential impacts. First, although 
this introduces essentially a mild (logarithmic) dependence,
it can lead to substantial modifications of inflationary predictions, which is reminiscent of the so-called $\eta$ problem~\cite{Copeland:1994vg, Baumann:2009ds}.
Second, this allows one to relate the physics at the scale of inflation to the one observed by \hep\ experiments. We will discuss both aspects in the rest of the paper. 

The operators \lletexte\ and \uddtexte\ labeling the considered flat directions are R-parity violating. One can thus use the general results of \cite{Allanach:2003eb} to extract the one-loop Renormalization Group Equations governing $\Asix(\phi)$ and $\lambda_6(\phi)$, including multiplicative factors for the non-renormalizable operators \cite{Antusch:2002ek}, and neglecting contributions suppressed by $\Mp$. 

In the following, we use the conventions of \texttt{SuSpect3} \cite{Djouadi:2002ze,Kneur:2022vwt} and the notations summarized in \cref{sec:pMSSM}. 
In particular, our sign convention for the gaugino masses is that of \cite{Castano:1993ri}, opposite to the one adopted in \cite{Allahverdi:2006we, Allahverdi:2007vy,Allahverdi:2010zp,Boehm:2012rh}.
We denote the energy scale by $\muorq$. Since, as stated previously,  \cref{eq:polonyi} is defined at the GUT scale, we use the RGEs with the GUT scale as boundary and 
adopt the $\rm SU(5)$-GUT  normalization $g_1 = \sqrt{5/3}g_Y$ (where $g_Y$ denotes the SM hypercharge gauge coupling).
Within the above-mentioned assumptions, the RGEs of the potential parameters read:

For $L^iL^je^k$:
\begin{align}
    \label{eq:rgemphi_lle}
    \muorq\frac{\dd \mphi ^2}{\dd\muorq}&=-\frac{1}{6\pi^2}\Big(\frac{9}{10}\Mone^2g_1^2+\frac{3}{2}\Mtwo^2g_2^2+\textrm{Y}_{m_\phi}^{L^iL^je^k}\Big),\\    
    \label{eq:rgeA6_lle}
    \muorq\frac{\dd \widetilde{A}_6}{\dd\muorq}&=\frac{1}{2\pi^2}\left(\frac{9}{10}\Mone g_1^2+\frac{3}{2}\Mtwo g_2^2+\textrm{Y}_{\Asix}^{L^iL^je^k}\right),\\
    \label{eq:rgelambda6_lle}
    \muorq\frac{\dd \lambda_6}{\dd\muorq}&=-\frac{\lambda_6}{4\pi^2}\Big(\frac{9}{10}g_1^2+\frac{3}{2}g_2^2+\textrm{Y}_{\lambda_6}^{L^iL^je^k}\Big),
\end{align}

For $u^id^jd^k$:
\begin{align}
    \label{eq:rgemphi_udd}
    \muorq\frac{\dd \mphi ^2}{\dd\muorq}&=-\frac{1}{6\pi^2}\Big(\frac{2}{5}\Mone^2g_1^2+4\Mthree^2g_3^2+\textrm{Y}_{m_\phi}^{u^id^jd^k}\Big),\\
    \label{eq:rgeA6_udd}
    \muorq\frac{\dd \widetilde{A}_6}{\dd\muorq}&=\frac{1}{2\pi^2}\left(\frac{2}{5}\Mone g_1^2+4\Mthree g_3^2+\textrm{Y}_{\Asix}^{u^id^jd^k}\right),\\
    \label{eq:rgelambda6_udd}
    \muorq\frac{\dd \lambda_6}{\dd\muorq}&=-\frac{\lambda_6}{4\pi^2}\Big(\frac{2}{5}g_1^2+4g_3^2+\textrm{Y}_{\lambda_6}^{u^id^jd^k}\Big),
\end{align}

with $\Asix(Q) =\vert\widetilde{A}_6(Q)\vert$.
It is worth noting that the $U(1)$ gauge contributions that depend on a universal combination of {\sl all} soft-SUSY-breaking squared scalar masses, present in the $\beta$-function of the RGEs of each of these masses individually (see \eg \cite{Castano:1993ri}), cancel out exactly  in \cref{eq:rgemphi_lle,eq:rgemphi_udd} as a consequence of 
\cref{eq:mphi_lle,eq:mphi_udd}. It follows that the running of $\mphi ^2$ is generally given by \cref{eq:rgemphi_lle,eq:rgemphi_udd} (even without   
a universality assumption of the soft-SUSY-breaking scalar
masses at some scale, in which case the aforementioned combination would have vanished at all scales).
Moreover, the one-loop runnings of the gauge couplings $g_i$ and gaugino masses $M_i$ for $i=1,2,3$ are obtained from the following RGEs (see \eg \cite{Martin:1993zk}): 

\begin{equation}
    \label{eq:couplings}
    \muorq\frac{\dd \ln{M_i}}{\dd\muorq}=\muorq\frac{\dd \ln{g_i^2}}{\dd\muorq}=\frac{b_i}{8\pi^2}g_i^2 \quad \textrm{with $b_1 = \frac{33}{5}$, $b_2 = 1$ and $b_3 = -3$.}
\end{equation}

The Yukawa terms  contributing to 
\cref{eq:rgemphi_lle,eq:rgeA6_lle,eq:rgelambda6_lle,eq:rgemphi_udd,eq:rgeA6_udd,eq:rgelambda6_udd} depend on the inflaton type. They are mainly functions of the renormalizable Yukawa and trilinear soft-SUSY-breaking couplings, the dependence of $\textrm{Y}_{\Asix}$ and $\textrm{Y}_{\lambdasix}$ on $\lambdasix$ and $\Asix$ themselves being negligible due to Planck-mass suppression. They are given for completeness in \cref{app:analytical_Yuk0}. \Cref{eq:rgemphi_lle,eq:rgeA6_lle,eq:rgelambda6_lle,eq:rgemphi_udd,eq:rgeA6_udd,eq:rgelambda6_udd} should thus be coupled to the RGEs governing the runnings of the Yukawa and trilinear soft-SUSY-breaking couplings on top of \cref{eq:couplings} and, in general, cannot be solved analytically. Numerical solutions for the full set of RGEs can be obtained from \texttt{SuSpect3} \cite{Kneur:2022vwt}. When the Yukawa terms can be neglected, which is the case for the \lletexte \ and \uddtexte \ flat directions involving first and/or second lepton and quark generations, analytical solutions become available. For simplicity, this is the case studied in this work. One can find the RGEs' solutions within this approximation \cite{Allahverdi:2006we} in \cref{app:analytical_Yuk0}.\footnote{Note that the energy scale $\muorq$  should be replaced by $\phi$ when the runnings are used in \Vrge. Indeed the same RGEs play a double role: They allow one to improve the effective potential and to improve the energy dependence of physical scattering processes. In the latter case $\muorq$ stands for the typical energy of the process.} When the radiative corrections are included, one needs to ensure that  \cref{eq:definition_range} is verified at all scales of the theory, and that \cref{eq:lambdabound} is validated at high scale (cf. \cref{sec:consistency_theory}).

Although we rely in the sequel on \cref{eq:potential_RGE} to evaluate the effects of the loop corrections on inflation, it is useful to keep in mind some possible caveats. In principle, in the monomials appearing in \cref{eq:potential_RGE}, one should replace the $\phi$ field itself by its running counterpart taking into account the corresponding anomalous dimension. Furthermore, a constant can be in general added to the effective potential. This was considered by \cite{Enqvist:2010vd} in the context of the present model. However, when the potential is RGEs improved in the presence of such a constant, the latter does not remain constant and induces an extra field 
dependence \cite{Bando:1992np,Ford:1992mv}. Last but not least, the non-renormalizable operators in \cref{eq:potential_form}, when inserted in loops, induce in the one-loop Coleman-Weinberg potential other effective operators of the form $\phi^n$, with $n=4,8,\ ...$ that were
absent at tree-level. Some of these operators would cancel out in the large $\phi$ limit, due to supersymmetry. Others will have to be renormalized away through counterterms at some boundary scale, say $\Mgut$, but will be regenerated at lower scales.

\subsection{Slow roll in the eMSSM}
\label{sec:reheating-theo}
With a fully defined potential like \Vtree\ or \Vrge, we can further discuss slow-roll conditions and the calculation of $\phistar$ in order to determine the predicted values of the inflation observables.
To ensure that slow-roll inflation can occur, the potential needs to exhibit an inflection point at some field value $\phi_0$ defined by 
\begin{equation}
    \label{eq:flat_inflection-nu0}
    V_{\phi\phi}(\phi_0) =  0\ \ \mathrm{ and }\ \   \nu\equiv V_\phi(\phi_0) .
\end{equation}
The slow-roll parameters (see Eqs.~(\ref{eq:epsilon:1:pot}), (\ref{eq:epsilon:2:pot}) and (\ref{eq:epsilon:3:pot})) should be small enough around $\phi=\phi_0$.
In particular, the condition $\epsilon_1<1$ reads
\begin{equation}
    \label{eq:flat_inflection-nu}
     {|\nu|}  < \frac{\sqrt{2}}{\Mp}{|V(\phi_0)|}.
\end{equation}

We will refer to $\phi_0$ as ``quasi-flat inflection point'' whenever it satisfies \cref{eq:flat_inflection-nu0,eq:flat_inflection-nu}. For \Vtree, this condition delineates a region of size $|\phi-\phi_0|/\Mp \sim (\phi_0/\Mp)^{3/2}$, while a more stringent requirement comes from the condition $|\epsilon_2|<1$, which holds in the range 

\begin{equation}
\label{eq:narrow}
    |\phi-\phi_0|\simeq\frac{\phi_0^3}{60\Mp^2} \, \ . 
\end{equation}
Note that a similar condition is obtained from $|\epsilon_3|<1$, and we have numerically verified that this relation still holds for \Vrge.  
In the range defined by \cref{eq:narrow}, \cref{eq:implicit_eq} can be used and 
the inflection point is very close to the value of the field at Hubble crossing, $\phistar$. 

In order to determine $\phistar$, one needs to solve \cref{eq:implicit_eq}, of which two important ingredients are $\lnrrad$ and $\tilde\rho_\gamma$. The former describes how the reheating occurs, and the latter is a function of the variation in the number of relativistic degrees of freedom between the reheating epoch and today, which depends on the field content of the theory. An interesting feature of eMSSM inflation is that, in principle, both are entirely determined by the MSSM spectrum. 

For the reheating parameter, $\lnrrad$, the corresponding calculation is however complex and goes
beyond the scope of the present work. The analyses of \cite{Allahverdi:2011aj, Ferrantelli:2017ywq} suggest that $0<\lnrrad\ll1$ (\ie quasi-instantaneous reheating) both for \uddtexte\ and \lletexte, so, hereafter, we will assume $\lnrrad=0$ unless specified otherwise.

The effective number of relativistic degrees of freedom at reheating is $427/4$ in the Standard Model of particle physics if reheating occurs above $\sim 100\ \GeV$, while it is expected to be $915/4$ in the MSSM~\cite{Husdal:2016haj}. Instead, in the following, we do not include such details and we assume that the ratio of the relativistic degrees of freedom between the reheating epoch and today is equal to one. It only leads to differences in the value of $\DeltaNStar$ inferred from \cref{eq:implicit_eq} of the order of 0.4 \textit{e}-folds. Compared to the other sources of uncertainties discussed below, this error can be neglected. 

The impact of assuming different values of $\DeltaNStar$ (and different $\lnrrad$) is further discussed in \Cref{sec:tree_results,sec:error-budget}.

\section{Analysis framework}

In this part, we introduce the phenomenological MSSM, the benchmark points, the observational constraints and the tools and methodology used to perform the analysis.

\subsection{pMSSM and benchmark points}

\subsubsection{pMSSM}
\label{sec:pMSSM}

Even though the unconstrained MSSM may contain up to 105 parameters on top of the Standard-Model ones, we only consider  a subset of them in this analysis, as is frequently done in various studies (see \eg~\cite{MSSMWorkingGroup:1998fiq}). They are listed in \cref{tab:MSSMparam}. We will refer to this phenomenological MSSM, thus defined, as pMSSM. 

\begin{table}[htb]
\centering
\begin{tabular}{c|c } 
{\bf{Origin}} & {\bf{Parameters}}\\
 \hline
Higgs-sector  & $m_{H_u}^2$, $m_{H_d}^2$, $\tan\beta$, $\sgn(\mu)$ \\
Gaugino masses & $\Mone$, $\Mtwo$, $\Mthree$ \\
Slepton masses & $m_{\tilde l_L^{12}}$, $m_{\tilde e_R^{12}}$, $m_{\tilde \tau_L}$, $m_{\tilde \tau_R}$ \\
Squark masses & $m_{\tilde q_L^{12}}$, $m_{\tilde u_R^{12}}$, $m_{\tilde d_R^{12}}$, $m_{\tilde q_L^{3}}$, $m_{\tilde t_R}$, $m_{\tilde b_R}$ \\
A-terms & $A_{u_{12}}$, $A_{d_{12}}$, $A_{l_{12}}$, \Atop, $A_{b}$, $A_{\tau}$
\end{tabular}
\caption{Parameters of the phenomenological MSSM \label{tab:MSSMparam}}
\label{tab:Parameters}
\end{table}

We assume all the parameters to be real-valued, which implies, among other things, no extra sources of explicit \textit{CP}-violation from the SUSY extension of the Standard Model. As the $\rm K_0-\overline{\rm K}_0$ mixing limits the mass splitting between the first and second  squark generations \cite{Ciuchini:1998ix, MSSMWorkingGroup:1998fiq}, we assume that the squark masses of the first two generations are equal. The same applies to the slepton masses and the A-terms. To ensure this, and as a further simplification inspired by mSUGRA, we will use 
\begin{equation}
\label{eq:massidentity1}
    m_{\tilde{q}_L^{12}} = m_{\tilde{u}_R^{12}}  =  m_{\tilde{d}_R^{12}}\, .
\end{equation}
In the following, gauge couplings are unified at the GUT scale ($\Mgut$). Unless otherwise indicated, it is set to $3 \times 10^{16}$ \GeV\ in the following. In \cref{sec:examples_and_big_table}, it is directly calculated for a given MSSM spectrum within \texttt{SuSpect3} following \cite{Adam:2010uz}. 

The scale of ElectroWeak Symmetry Breaking (EWSB) is defined through the \texttt{SuSpect3} convention \cite{Kneur:2022vwt,Djouadi:2002ze}: 
\begin{equation}
  M_{\mathrm{EWSB}} = \sqrt{m_{\tilde{t}_1} m_{\tilde{t}_2}} \ .
\end{equation}
In what follows, we will assume that the Lightest Supersymmetric Particle (LSP) is the neutralino $\chi_1^0$ and it acts as the dark-matter (DM) candidate. 

\subsubsection{Benchmark points}
As shown in \cref{sec:RGE}, the runnings of the parameters of \Vrge\ depend on the inflaton type, but also on the values of the gauge couplings, the masses of the gauginos, the Yukawa couplings and the trilinear soft-SUSY-breaking scalar couplings (the latter two will be neglected in this work). For this reason, we define in this section two examples that will be used for illustration in the following sections. They are hereafter called 
``benchmark points'', and their characteristics at the GUT scale are given in \cref{tab:benchmark_point}. \JuneCorrections{The chosen values are representative of the MSSM points discussed in \cref{sec:examples_and_big_table}: BP1 ({\sl resp.} BP2) corresponds to $h_1$ ({\sl resp.} $A_1$), run to GUT scale. We only specify the quantities entering the potential parameters RGEs at this stage.}
 
\begin{table}[htb]
    \centering
    \begin{tabular}{c|c|c|c|c|c|c}
         & $g_{1}^{\mathrm{GUT}}$ & $g_{2}^{\mathrm{GUT}}$ & $g_{3}^{\mathrm{GUT}}$ & $M_{1}^{\mathrm{GUT}}$  & $M_{2}^{\mathrm{GUT}}$  & $M_{3}^{\mathrm{GUT}}$  \\  
         & & & & (\GeV) &(\GeV) &(\GeV) \\ 
         \hline
         BP1 & $0.70$ & $0.69$ & $0.68$ & $136$ & $1143$  & $899$  \\
         BP2 & $0.70$ & $0.69$ & $0.68$ & $898$  & $1790$  & $883$ \\ 
    \end{tabular}
    \caption{Benchmark points definition. The gauge couplings and gauginos masses are given at $\Mgut$.}
    \label{tab:benchmark_point}
\end{table}
Note that \JuneCorrections{to remain as general as possible} we clearly depart here from a universality assumption for the gaugino soft-SUSY-breaking masses.

\subsection{Observational constraints} 
\label{sec:all_observ}
\subsubsection{Cosmological observables} 
\label{sec:observables}

The main constraints from cosmological observations that are used in this paper are the cold-dark-matter energy density ($\omegacdm$), the amplitude ($\As$) and spectral index ($\ns$) of the primordial scalar perturbations. For $\omegacdm$, we are using \cite{Planck:2013pxb} 
\begin{equation}
\omegacdm = 0.1187\pm0.0017\, .
\end{equation}
On top of the experimental uncertainty quoted above, we consider a theoretical uncertainty of $0.012$ associated to the prediction of $\omegacdm$ in the MSSM \cite{Henrot-Versille:2013yma}. For $\As$ and $\ns$, we are using the measurements inferred from the combination of \Planck\ temperature, polarization, lensing, and BAO 
data assuming a $\Lambda$-CDM model \cite{Planck:2018vyg}. The corresponding measurements of the parameters inferred for $k_*=0.05\ \rm{Mpc}^{-1}$ are given in  \cref{tab:cosmo_obs}. 
In the following, when we refer to the measurements instead of the parameters, we make use of the notations $\nsmean\pm\sigma_{\ns}$, and $\overline{\As}\pm\sigma_{\As}$.

\begin{table}[htb]
    \begin{tabular}{c | c } 
     {Parameter} & {Value and error} \\
    \hline
    $\ln({10^{10}{\As}})$  & ${3.047} \pm {0.014}\ $ \\
    $\ns$ & ${0.9665} \pm {0.0038}\ $ \\
    \hline
    ${{\alphas}}$  & $-0.0042 \pm 0.0067\  $ \\
    ${r}$  & $< 0.032 $ \\
    \end{tabular}
    \caption{Measurements of the amplitude ($\As$) and spectral index (${\ns}$) of the primordial scalar spectrum \cite{Planck:2018jri} used in this analysis. For reference, we also give the running of the scalar index (${\alphas}$) \cite{Planck:2018jri} and 95\% C.L. upper limit on the tensor-to-scalar ratio $r$ \cite{Tristram:2021tvh}.
     }
    \label{tab:cosmo_obs}
\end{table}

As shown in \cite{Martin:2013tda}, the  tensor-only contribution to the tensor-to-scalar ratio, $r$, for this potential is beyond the reach of present and future experimental constraints (and below the threshold of secondary gravitational waves induced by scalar fluctuations through gravitational non-linearities (see \eg~\cite{Domenech:2021ztg}). We therefore consider a (not $r$-extended) $\Lambda$-CDM model for the $\ns$ and $\As$ constraints and we do not make use of the current experimental constraint on the tensor-to-scalar ratio.
We will discuss the consistency between the $\alphas$ predictions and the measured value in \cref{sec:examples_and_big_table}. 

Another experimental constraint is the one on $\rho_{\gamma}=\Omega_\gamma\rho_{\mathrm{cri}}$ with $\Omega_\gamma$ the density parameter of radiation today and $\rho_{\mathrm{cri}}$ the critical density which leads to $\rho_\gamma=(4.645\pm0.004)\ \times\ 10^{-34}$ \cite{ParticleDataGroup:2022pth}. This observable enters in \cref{eq:implicit_eq}. The propagation of the experimental error on this quantity leads to a negligible contribution to the error on $\DeltaNStar$ of $(1/4)\ln{(1+{0.004}/{4.645})}= 2\times10^{-4}$ (similar to $\lnrrad$ and the number of relativistic degrees of freedom discussed in the previous section, see \cref{eq:implicit_eq}). The value of $\rho_\gamma$ together with the measured values of $\As$ and $\ns$ are hereafter called ``inflationary observables''.

\subsubsection{\JuneCorrections{Particle-}physics observables} 
\label{sec:hep-obser}

On the \JuneCorrections{particle-}physics side, the values of the main measurements considered in this analysis are summarized in \cref{tab:hep_obs} with their statistical and systematic errors. Where relevant, the theoretical errors associated with the supersymmetric predictions are also indicated (with the subscript ``th'').

\begin{table}[htb]
\begin{tabular}{c | c } 
 {Measurement} & {Value and error} \\
\hline
$m_h$ [$\textrm{GeV}$] & $125.10\pm0.14\pm3.00_\mathrm{th}$ \\ 
\hline
BR($B_S\rightarrow \mu^+\mu^-$)  & $ \left(30\pm4\pm2_\mathrm{th}\right)\times 10^{-10}$  \\ 
BR($b\rightarrow s\gamma$)  & $\left(33.2\pm1.5\right)\times 10^{-5}$\\ 
$\Delta a_\mu$  & $\left(26.1\pm7.9\pm2_\mathrm{th}\right)\times 10^{-10}$\\ 
$\Mtop$ [$\textrm{GeV}$] & $172.76\pm0.30 $ \\
\end{tabular}
\caption{Main \JuneCorrections{particle-}physics measurements used in the analysis \cite{ParticleDataGroup:2022pth}. The last number is the theoretical uncertainty on the supersymmetric prediction, except for BR($b\rightarrow s\gamma)$ and $\Mtop$  for which no such error is considered.} 
\label{tab:hep_obs}
\end{table}

To complete the Higgs-sector constraints, on top of the mass quoted in \cref{tab:hep_obs}, we include
the Higgs couplings, which are taken from \cite{ParticleDataGroup:2022pth,ATLAS:2012ac,ATLAS:2012ad,ATLAS:2012rld,ATLAS:2012gfw,ATLAS:2012qaq,ATLAS:2012yve,CMS:2012uyg,CMS:2012zbs,CMS:2012bkm}. 
Also, we use the Large Electron-Positron collider limit on the mass of the first generation chargino: 
$m(\tilde{\chi}_1^+) > 103.5 \GeV$ \cite{ALEPH:2005ab}. Finally, on the DM searches' side, we also consider the limits on the direct detection rate, which are provided by the XENON1T experiment, with a rate above 0.4 keV to be less than 1 event/(tonne $\times$ day $\times$ keV${_{\textrm{ee}}}$) \cite{XENON:2019gfn}. 

The measurements of the W and Z masses, the Higgs boson width together with the forward-backward asymmetries, the left-right asymmetries, the effective weak mixing angle, and the hadronic corrections to the QED coupling are also included. 

As far as the MSSM parameter space is concerned (without considering the need to embed inflation), the modeling of the parameter space comes essentially from the intertwined constraints of the Higgs mass and the cold-dark-matter energy density \cite{Ellis:2012aa,Henrot-Versille:2013yma}. 

\subsection{Tools and methodology}
\label{sec:methodo} 

Several tools have been used to perform the analysis described in this paper: 
\begin{itemize}
\item{}{\texttt{ASPIC}}\footnote{\url{http://cp3.irmp.ucl.ac.be/~ringeval/aspic.html}}\cite{Martin:2013tda}: even though not directly interfaced  to our framework, 
we have adapted \texttt{ASPIC}
to derive reheating-consistent observable predictions. 
\item{} {\texttt{SuSpect3}} \cite{Kneur:2022vwt}: it calculates the MSSM physical masses and couplings, taking into account the dominant radiative corrections, the requirement of
EWSB, and the running of the eMSSM parameters through their RGEs, relating the low-energy physics to several high-energy model assumptions.
In particular, on top of the Z-mass and the EWSB scales, it allows one to define boundary conditions for the relevant running parameters, at up to three different physical scales that can be chosen to be the GUT, the SUSY-breaking and the inflection point scales. 
The version of \texttt{SuSpect3} used in this analysis is the \texttt{3.1.1}.
\item{} {\texttt{SFitter}} \cite{Lafaye:2004cn}: it allows one to confront the experimental data
with predictions determined from the spectrum calculated by \texttt{SuSpect3}. The statistical analysis is performed on the basis of a global $\chi^2$ calculated from the individual $\Delta\chi^2$’s of the measurements versus predictions for the observables described in the previous section. 
We then make use of \texttt{MINUIT} \cite{James:1975dr}, included in \texttt{SFitter}, to infer the values of the underlying parameters. This is done in particular through an interfacing with \texttt{Micromegas} \cite{Belanger:2013oya} for the predictions of the dark-matter energy density and the rare B-decays branching ratios. \texttt{SusyPope} \cite{Heinemeyer:2008} and \texttt{HDecay} \cite{Djouadi:2018xqq} are used to calculate the predictions of the Z-pole observables and the Higgs couplings respectively. In \texttt{SFitter}, the statistical errors on the measurements are treated as Gaussian or Poisson where appropriate.
The systematic errors are correlated if originating from the same source. Theoretical uncertainties are treated using flat errors.
\end{itemize}

The methodology we adopt to infer the parameters of an inflationary potential given some cosmological observations is inspired by \texttt{ASPIC}. Starting from an MSSM potential, \Vtree\ (\cref{eq:potential_form}) or \Vrge\ (\cref{eq:potential_RGE}), we have identified the area of the parameter space that satisfies the conditions of a quasi-flat inflection point (defined in \cref{sec:reheating-theo}), then we solve numerically \cref{eq:AsSR} and \cref{eq:nsSR}, setting $\As$ and $\ns$ to their measured values at $\phi=\phistar$ which, itself, is determined through Eqs.~(\ref{eq:srDeltaN*}) and (\ref{eq:implicit_eq}). 

In practice, since the potential has three main free parameters ($m_\phi$, $A_6$, $\lambda_6$), one needs to fix a degree of freedom, eg: $A_6$ ({\sl resp.} $\phi_0$), on top of the two constraints from $\ns$ and $\As$, in order to be able to determine the two other ones. For reasons that will be further detailed in \cref{sec:fine-tune}, the two degrees of freedom we choose to tune are $\nu$ and $\phi_0$ ({\sl resp.} $A_6$). To be able to do that, we need the expressions for the potential and its derivatives taking $\phi_0$ and $\nu$ directly as inputs: their full expressions are given in \cref{sec:fine-tune}. Choosing to fix $A_6$ or $\phi_0$ obviously leads to the same results, this choice is made according to the situation. 

Moreover, in the specific case of the one-loop \Vrge, one also needs to specify the gaugino masses and gauge couplings (cf. \cref{eq:rgemphi_lle} to \cref{eq:rgelambda6_udd}). We make use of \texttt{SuSpect3} for the calculations of the RGEs for the inflationary parameters, as well as for the calculation of the physical spectra given a point in the MSSM parameter space. The correspondence between the squark or slepton soft masses and the inflaton mass (\cref{eq:massidentity1}) as well as the relation between \Atop and $A_6$ (\cref{eq:polonyi}) are added at this stage. Finally, \texttt{SFitter} is used to perform the $\chi^2$ calculation and its minimisation by comparing the HEP predictions and actual measurements (cf. \cref{tab:hep_obs}).

\section{Slow-roll conditions, initial conditions and fine-tuning}

In this section, after an illustration of the shape of the potential, we
address two essential questions: on the one hand, the field phase-space initial conditions required for inflation to take place, and on the other hand, the conditions the inflationary potential parameters must satisfy to yield predictions in agreement with observations. This will lead us to an assessment of the level of  fine-tuning involved in this model.

\subsection{Shape of the potential}

\begin{figure}[tbp]
\begin{tabular}{cc}
\includegraphics[width=.45\textwidth]{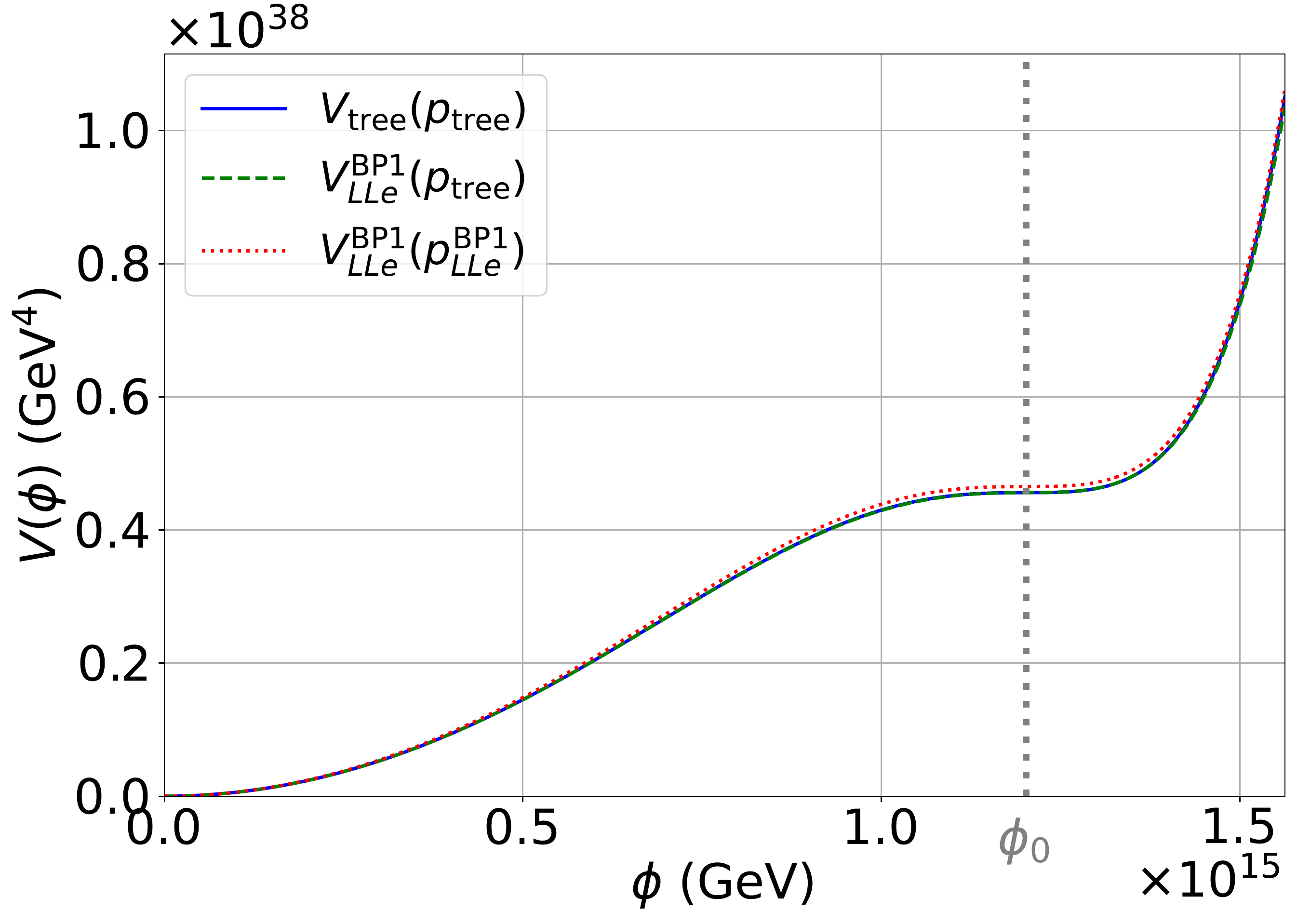}
&
\includegraphics[width=.465\textwidth]{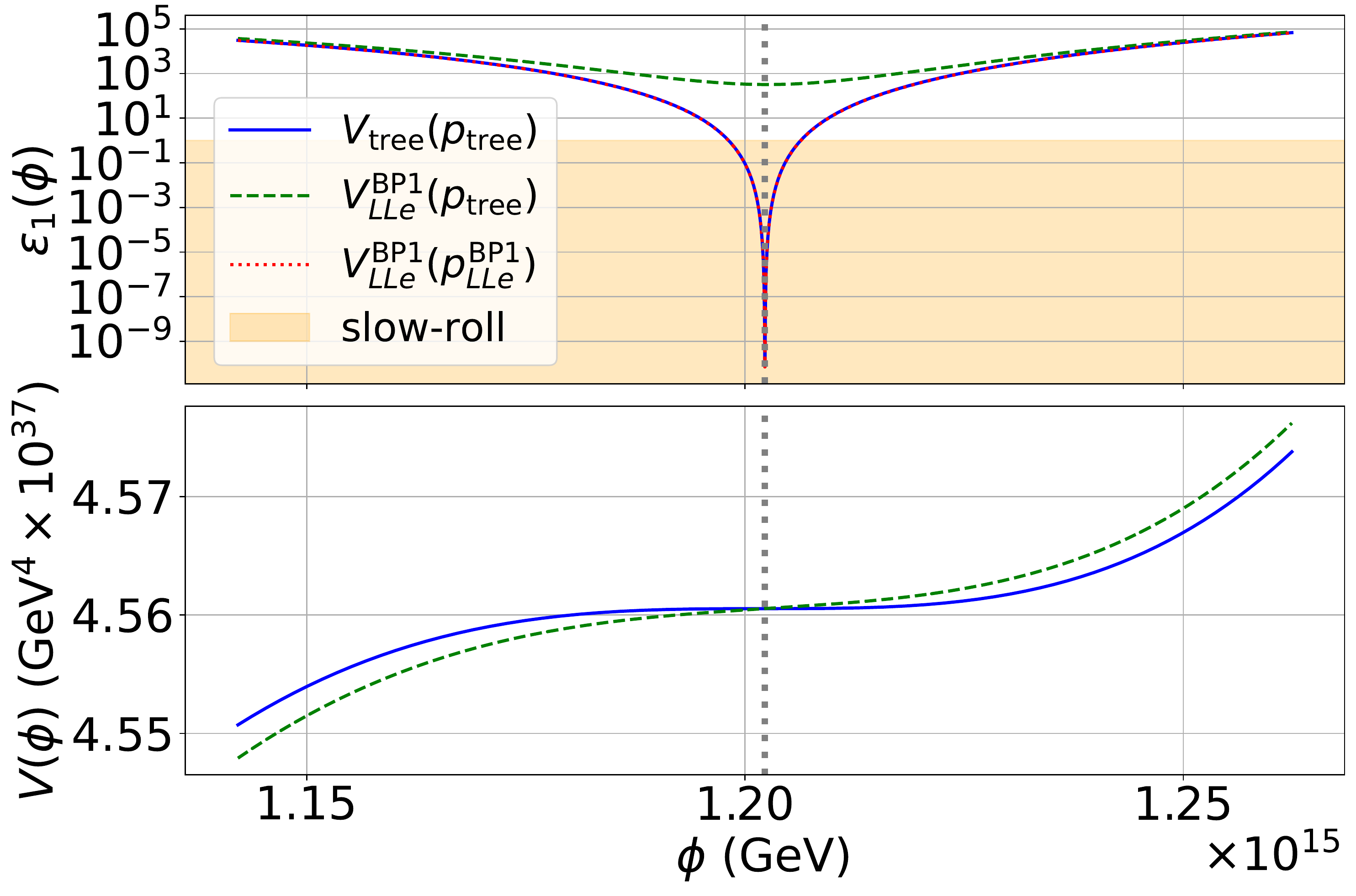}
\end{tabular}
\caption{\textsc{Left panel:}  Example of the shape of the inflationary potentials as a function of the field value for $\phi_0=1.2\times 10^{15}$ \GeV. \Vtree\ is shown in blue and \lletexte\ BP1 \Vrge\ in green (dashed line) for the parameters determined to match the inflationary observables assuming \Vtree. The red-dotted line illustrates the \lletexte\ BP1 \Vrge\ for the parameters determined by taking into account the one-loop RGEs' corrections. 
\textsc{Right panel:} 
 The associated $\epsilon_1$ function (defined by \cref{eq:epsilon:1:pot}, more details in the text) is represented on the upper-right panel (the slow-roll region, where $\epsilon_1<1$, is identified in orange). Zooms of the potentials around $\phi_0$ are given on the lower-right panel.   \label{fig:potential} }
\end{figure}

To begin with, we discuss and illustrate the shapes of the \Vtree\ and \Vrge\ potentials. For \Vrge, we take the example of a \lletexte\ inflaton for BP1 (\cref{tab:benchmark_point}). The parameter sets ($p_\mathrm{tree}$ and $p_{LLe}^{\textrm{BP1}}$) have been respectively determined for \Vtree\ and \Vrge\ following the methodology described in \cref{sec:methodo}, so that both potentials exhibit a common inflection point at $\phi_0\simeq1.2\times 10^{15}$ \GeV\ and match the inflationary observables (\cref{tab:cosmo_obs}).  
They are shown in blue and red in \cref{fig:potential}. We do not
propagate here the experimental errors on the observables and assume an instantaneous reheating for the purpose of illustration. 

Around this inflection point, the slow-roll approximated first Hubble-flow parameter, as given by
\cref{eq:epsilon:1:pot},
is plotted in the upper-right panel of the same figure. Note that this approximation applies only when $\epsilon_1\ll 1$ and that, in practice, $\epsilon_1$ can never be larger than $3$ — thus values of $\epsilon_1$ of order one or larger in \cref{fig:potential} simply signal a break down of the slow-roll approximation. Slow roll takes place when $\epsilon_1<1$ (orange region). A zoom of the potential is also given in the lower-right panel in the same field range (the red curve is above the Y-axis range of the plot). 

One sees that the inflection points are very close to flat for $\eqVtree\left(p_\mathrm{tree}\right)$ and $\eqVrge\left(p_{LLe}^{\textrm{BP1}}\right)$, with $\epsilon_1$ diving well below one in the slow-roll region (whose narrowness has been quantified in \cref{eq:narrow}). 

Applying the parameters obtained with \Vtree \  
(ie. $p_\mathrm{tree}$) to \Vrge\ leads to a tilted inflection point 
shown in green in \cref{fig:potential}. In this case, $\eqVrge\left(p_\mathrm{tree}\right)$
does not
satisfy the slow-roll conditions: the corresponding values of the parameters are thus ruled out. 

This illustrates that the parameters determined for \Vtree\ cannot be simply applied to a one-loop corrected \Vrge\ potential, suggesting that the one-loop correction of the inflationary potential is a feature that cannot be ignored (contrary to what is often done in the literature). Instead, one needs to re-determine the new set of parameters specifically for \Vrge. Proceeding this way, we recover a shape similar to the \Vtree\ one, with the same order of magnitude for $\epsilon_1$ at the inflection point as shown with $\eqVrge\left(p_{LLe}^{\textrm{BP1}}\right)$ in red.

\subsection{Field phase-space}
\label{sec:phase_space}
As mentioned in \cref{sec:IA1}, slow roll is a dynamical attractor. This implies that, in a given inflationary potential, a successful phase of inflation takes place starting from a large set of initial conditions $(\phi,\dot{\phi})$, all attracted towards the same slow-roll solution~\cite{Remmen:2013eja, Chowdhury:2019otk}. However, it is also known that the size of the basin of attraction depends on the shape of the inflationary potential (for instance, it is larger for plateau and large-field models than for hilltop potentials~\cite{Chowdhury:2019otk}). In order to
determine to which extent eMSSM inflation is robust under changing the field initial conditions, let us thus
study its field phase-space structure.

Both \Vrge\ and \Vtree\ have a slow-roll region whose (narrow) extent is given by \cref{eq:narrow}. We therefore expect the conclusions of this section to be identical when one uses either potential.  Hence, we consider for explicitness \Vtree, where we arbitrarily set $\phi_0=0.395 \Mp$, and apply the methodology defined in \cref{sec:methodo} to determine the potential parameters such that the predicted values of $\ns$ and $\As$ match the measurements given in \cref{tab:cosmo_obs}. This leads to: $\mphi=4.29\times 10^9$ \GeV, $\Asix=5.42\times 10^{10} $\GeV\ and $\lambda_6=2.29\times 10^{-8}$. 

\begin{figure}[htbp]
\includegraphics[width=1\textwidth]{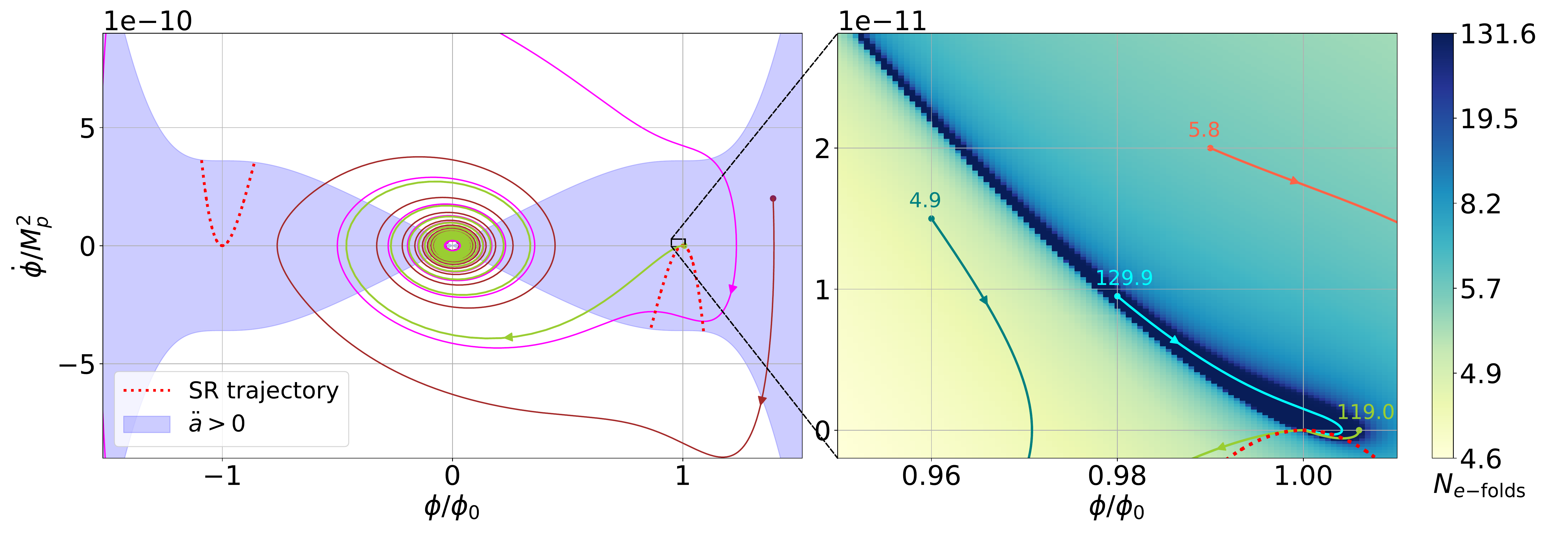}
\caption{ Phase-space diagram of the eMSSM inflationary potential \Vtree, with $\mphi=4.29\times 10^9$ \GeV, $\Asix=5.42\times 10^{10} $ \GeV\ and $\lambda_6=2.29\times 10^{-8}$ (such that the model is compatible with cosmological observations). \textsc{Left panel}: A few trajectories (brown, magenta and light-green curves) are compared to the slow-roll attractor \cref{eq:SR:traj} (red-dotted lines). The blue shaded region indicates where inflation takes place, \ie where the condition of ~\cref{eq:inflate} is satisfied.
\textsc{Right panel}: Zoom-in on the low-velocity inflection-point region (black square on the left). The color encodes the total number of inflationary \textit{e}-folds generated by the trajectory starting at the given point of the phase space.
A few examples of initial conditions (the light-green trajectory is the same on both panels) with their associated \textit{e}-folds are also displayed. 
} \label{fig:attractor}\end{figure}

In the left panel of \cref{fig:attractor}, the phase-space trajectory obtained by numerically integrating \cref{eq:klein-gordon} and \cref{eq:friedmann} is shown 
for three different initial conditions. In particular, the light-green one is starting near the inflection point, in the slow-roll region. 
In this figure, the blue area indicates the region where inflation takes place (\ie where \cref{eq:inflate} is satisfied). 
The red-dotted lines correspond to the slow-roll trajectory (\ref{eq:SR:traj}), where inflation proceeds at small velocity close to the inflection point, \ie around $(\phi=\phi_0,\ \dot{\phi}\simeq 0)$. If initial conditions are set close enough to the slow-roll trajectory, it acts as an attractor, as expected. This is the case for the light-green trajectory.

Otherwise, the inflection point is overshot (this is the case for the magenta and brown trajectories), and the field quickly oscillates around the minimum of its potential at $\phi=0$. Let us note that, during this oscillating phase, the system repeatedly crosses the inflating region, but it does so across very short periods of time and when averaging over several oscillations it does not inflate (but rather behaves as pressure-less matter for a quadratic minimum~\cite{Turner:1983he}). Therefore, the relevant phase of inflation to consider is the one taking place before the oscillations. Unless initial conditions are chosen close to the slow-roll attractor, that phase is very short-lived, as illustrated in the right panel of \cref{fig:attractor}. In this panel, some trajectories and their generated total number of \textit{e}-folds 
$N_{\textit{e}\textrm{-folds}}$ (given by \cref{eq:e-folds}, evaluated between $t=0$ and the time of the end of inflation) are represented. 
One can see that only the ones being attracted by the slow-roll trajectory generate enough \textit{e}-folds (\ie originates in the dark-blue region).
From \cref{eq:implicit_eq}, inflation must at least generate $N_{\textit{e}\textrm{-folds}}\simeq 50$, and one can see that this requires to fine-tune the initial conditions close to the slow-roll attractor. The basin of attraction is therefore very narrow in this model, which constitutes a first level of fine-tuning.

Let us note that, at large-field values, $\phi\gg\phi_0$, the potential is of the ``large-field type'', $\eqVtree\propto \phi^{10}$, hence the slow-roll attractor is very powerful in that region~\cite{Chowdhury:2019otk} (\ie its basin of attraction is very large). If initial conditions are set in that region, it uniquely determines the trajectory along which the inflection point is approached at lower field values. This trajectory is the magenta one in the left panel of \cref{fig:attractor}, which starts at $\phi=100 \Mp$, oscillates a few times around the potential minimum, before entering the range of the plot and overshooting the slow-roll attractor around the inflection point.

Therefore, another mechanism must be invoked to set the initial conditions close to slow roll at the inflection point, possibly involving additional dynamical fields~\cite{Allahverdi:2008bt}. Hereafter, we will assume that such a mechanism takes place and we will restrict the analysis to the slow-roll attractor.

Finally, let us point out that the reason why the slow-roll attractors at large-field values and around the inflection point are disconnected is because, when $\phi_0$ is sub-Planckian, the slow-roll conditions are violated between these two regions.
This implies that the slow-roll attractor is broken, and we found that the problem becomes worse when decreasing $\phi_0$.
Here, for illustrative convenience, we have set $\phi_0$ to a mildly sub-Planckian value, but as will be made explicit below, $\phi_0$ is usually expected to be much lower. This implies that, in practice, the fine-tuning problem of the field initial conditions is even worse than what can be seen in \cref{fig:attractor}. To this respect, eMSSM inflation (and inflection-point models in general) behaves like small-field hilltop models~\cite{Chowdhury:2019otk}.

\subsection{Fine-tuning of the inflationary potential parameters}
\label{sec:fine-tune}

A peculiarity of the eMSSM potential that has been pointed out, in particular in \cite{Allahverdi:2006iq,Allahverdi:2006we,Allahverdi:2010zp,Boehm:2012rh, Martin:2013tda}, 
is that there is a high level of fine-tuning of the parameters when one imposes the condition of a quasi-flat inflection point. We discuss this point in the case of \Vtree\ and \Vrge\ in this section.

\subsubsection{At tree level}
\label{sec:inflection_cond}

To ensure that the potential function remains monotonic with a quasi-flat inflection point for \Vtree,  the parameters have to fulfill
the following very restrictive relation between the bi-trilinear coupling and the combination of soft-SUSY-breaking scalar masses appearing in $m_\phi$ \cite{Allahverdi:2006iq,  Allahverdi:2006we,Allahverdi:2007vy,Allahverdi:2010zp,Boehm:2012rh,Martin:2013tda}:
\begin{equation}
     \label{eq:fine-tuning}
    0< 1-\frac{\Asix^2}{{20} {\mphi}^2 }\ll1\ ,
\end{equation}
where the additional factor $\sqrt{2}$, cf. \cref{eq:potential_form}, is included, as compared to the literature. 

For later use, we define\footnote{Beware the different conventions between \cite{Boehm:2012rh, Allahverdi:2010zp} and \cite{Martin:2013tda}, and the one adopted in this work in the definition of $\alpha$. The relations between the different conventions are as follows: $\alpha_\textrm{\cite{Boehm:2012rh}}=\sqrt{\alpha}/2$ and $\alpha_\textrm{\cite{Martin:2013tda}}=1-\alpha$.}
\begin{equation}
\label{eq:alpha_def}
\alpha\equiv 1-\frac{\Asix^2}{{20} {\mphi}^2 }  .
\end{equation}

We recall here the origin of the  fine-tuned requirement, \cref{eq:fine-tuning}, at tree-level. One can determine
the value of $\phi_0$ that satisfies the inflection-point requirement, the first equation
in \cref{eq:flat_inflection-nu0}. Indeed,  
${\eqVtree}_{, \phi \phi}$ being quadratic in $\phi^4$ irrespective of the magnitude of $\alpha$, there is a two-branch solution for $\phi_0^4$, 
\begin{equation}
\label{eq:phi04pm}
 {\phi_0^4}_{\pm}=   \frac{\mphi\Mp^3}{9 \lambda_6 \sqrt{10}}\left( 5 \sqrt{1-\alpha} \pm \sqrt{16 - 25 \alpha} \right).
\end{equation}
When $\alpha < \frac{16}{25}$, the two branches are {\sl a priori} acceptable
since ${\phi_0^4}_{\pm}$ remain real-valued and positive.
One also finds that ${\eqVtree}_{, \phi}(\phi_{0 +})$ vanishes for $\alpha=0$ and
${\eqVtree}_{, \phi}(\phi_{0 -})$ vanishes asymptotically for $\alpha$ very large and negative, so that
the second equation in \cref{eq:flat_inflection-nu0} can be satisfied with vanishingly
small $\nu$. However the ``-'' branch cannot satisfy \cref{eq:flat_inflection-nu} and will not be further considered. As for the ``+'' branch, $\epsilon_1$, cf. \cref{eq:epsilon:1:pot}, is found to scale as $\lambdasix^\frac12 \Mp^\frac12 \mphi^{-\frac12}$  and vanishes for $\alpha=0$, whence the required high degree of fine-tuning on $\alpha$ around zero in order to keep $\epsilon_1 <1$.

It is thus justified to rely on the lowest order expansion in 
$\alpha$ for the \vevchange\ and the potential function and its first derivative at the inflection 
point, $\phi_0=\phi_{0+}$, that can be derived analytically from \cref{eq:phi04pm}:

\begin{align}
     \label{eq:phi0_def_tree}
    \phi_0^4=\frac{\mphi\Mp^3}{\lambda_6\sqrt{10}} + \mathcal{O}\left(\alpha\right), \\
\label{eq:V}
    \eqVtree(\phi_0)=\frac{4}{15}\mphi^2\phi_0^2 + \mathcal{O}\left(\alpha\right),
\end{align}
and
\begin{equation}
    \label{eq:Vprime}
{\eqVtree}_{, \phi}(\phi_0)=\mphi^2\phi_0\alpha  + \mathcal{O}\left(\alpha^2\right).
\end{equation}
As clear from the above equation, relaxing the monotonicity assumption boils down to allowing for negative $\alpha$ (since $V(\phi)$ increases monotonically  for $0<\phi\ll \phi_0$ and $\phi \gg \phi_0$). Such potentials predict values for $\ns$ that are always incompatible with measurements (see \cite{Martin:2013tda}), and they will not be further investigated here. 

Restricting to $\alpha >0$, \cref{eq:epsilon:1:pot}, \cref{eq:V} and \cref{eq:Vprime} imply a tight relation between $\alpha$ and the first 
slow-roll parameter at $\phi_0$:
\begin{equation}
    \label{eq:ft_eps1}
    \alpha = \frac{4\sqrt{2}}{15}\frac{\phi_0}{\Mp}\sqrt{\epsilon_1(\phi_0)} + \mathcal{O}\left[\frac{\phi_0^2}{\Mp^2}\epsilon_1(\phi_0)\right] \ .
\end{equation}
The required level of fine-tuning of the parameter $\alpha$ is illustrated by this equation 
as slow-roll conditions impose $\epsilon_1 \ll 1$. More precisely, $\epsilon_1<1$ implies that $\alpha < 4\times 10^{-3}$ for $\phi_0=\Mgut$, and this required level of precision increases when $\phi_0$ decreases. It gets even more stringent when one requires that cosmological observables are correctly reproduced (cf.~\cref{tab:cosmo_obs}). 
This is further illustrated in \cref{sec::finetune2}.

In practice, the fine-tuning of $\alpha$ requires a high level of (quadratic) precision in the numerical determination of the parameter space if we solve \cref{eq:implicit_eq} directly for \Vtree\ (see \cite{Martin:2013tda}). However, this would lead to computational accuracy mismatch when interfacing with lower (double-)precision codes such as \texttt{SuSpect3} and \texttt{SFitter}.  To circumvent this problem, and also
set the stage for the generalization to the one-loop effective potential in the next subsection, we give the exact solution of \cref{eq:flat_inflection-nu0}:
\begin{align}
\label{eq:mphi_tree}
 \mphi^2=&  \frac{1}{40}\left( 45 \frac{\nu}{\phi_0} + \Asix^2 + \Asix \sqrt{\Asix^2 -10 \frac{\nu}{\phi_0}}\right), \\
 \label{eq:lambda6_tree}
 \lambdasix=& \frac{\Mp^3}{20 \sqrt{2} \phi_0^4} \left( \Asix + \sqrt{\Asix^2 -10 \frac{\nu}{\phi_0}}\right) . 
 \end{align}
This form is different, though equivalent, to the one discussed above. To obtain these equations we used the fact that both ${\eqVtree}_{, \phi \phi}(\phi)$ and $\frac{1}{\phi}{\eqVtree}_{, \phi}(\phi)$ are quadratic in $\lambdasix \phi^4/\Mp^3$ and retained the ``+'' branch solution as discussed previously.
Using this form of the solution one can reexpress $V_\phi(\phi)$ and $V_{\phi\phi}(\phi)$ by expanding $\lambdasix \phi^4$ around $\lambdasix \phi_0^4$ so that large cancellations are already effected analytically, thus bypassing the need for numerical quadratic precision. Here, we give the corresponding exact expressions:
\begin{align}
    V_{\mathrm{tree},\phi}(\phi)&= (1+ \Delta_4)^{\frac{1}{4}}  
   \left\lbrace\nu + \Delta_4 \lambdasix \frac{\phi_0^5}{\Mp^3}
    \left[ 10 (2 + \Delta_4) \lambdasix \frac{\phi_0^4}{\Mp^3}  -\sqrt{2} \Asix \right]\right\rbrace,\label{eq:treefirst_precision}\\
    V_{\mathrm{tree},\phi\phi}(\phi)&= 5 \Delta_4 \lambdasix \frac{\phi_0^4}{\Mp^3} \left[18 (2 + \Delta_4) \lambdasix \frac{\phi_0^4}{\Mp^3}  -\sqrt{2} \Asix\right],\label{eq:treesecond_precision}
    \end{align}
where $\Delta_4 \equiv \phi^4/\phi_0^4 - 1$ and $\nu$ (\cref{eq:flat_inflection-nu0}) can be taken as input. These expressions are at the core of the methodology described in \cref{sec:methodo}: given a value for $A_6$, one tunes $\phi_0$ and $\nu$ such that the predicted values for $\ns$ and $\As$ match the observations, then one gets $\lambda_6$ and $\mphi$ thanks to \cref{eq:lambda6_tree} and \cref{eq:mphi_tree}.

\subsubsection{At one-loop level}
\label{sec::finetune2}

We now consider the one-loop potential, \Vrge. To estimate the level of fine-tuning in this case, we first note that the successive derivatives of $\eqVrge(\phi)$ will have the same polynomial dependences on $\phi$ as those obtained when differentiating $\eqVtree(\phi)$. This is a direct consequence 
of the form of the RGEs. The first and 
second derivatives of the potential (\ref{eq:potential_RGE}) read 

\begin{eqnarray}
\label{eq:Vprimeloop}
{\eqVrge}_{,\phi}(\phi)&=& \phi \times \left[\mphi^2(\phi) + \frac12 \beta_m(\phi) - \xi_1(\phi)\frac{\lambda_6(\phi)\phi^4}{6\Mp^3}+ \xi_2(\phi)\frac{\lambda_6(\phi)^2\phi^{8}}{6 \Mp^6}\right] 
\\
{\eqVrge}_{,\phi\phi}(\phi)&=& \mphi^2(\phi) + \frac12 {\cal B}_1(\phi) + \xi_3(\phi)\frac{\lambda_6(\phi)\phi^4}{6\Mp^3}- \xi_4(\phi)\frac{\lambda_6(\phi)^2\phi^{8}}{6 \Mp^6}, \label{eq:Vdoubleprimeloop}
\end{eqnarray}
where ${\cal B}_1$ and the $\xi_i$'s depend on the running of $\Asix$, the gauge couplings
and the gauginos masses, as well as the Yukawa and trilinear soft-SUSY-breaking scalar couplings. Their explicit expressions do not require 
the knowledge of the solutions of the RGEs. They are given in 
\cref{app:betafunctions} in the approximation of negligible Yukawa terms. 

At the inflection point introduced in \cref{eq:flat_inflection-nu0} and
taking into account the $\phi$ dependence in 
Eqs.~(\ref{eq:Vprimeloop}) and (\ref{eq:Vdoubleprimeloop}), one finds: 
\begin{align}
\label{eq:cond1}
\mphi^2(\phi_0) &=   \frac{{\cal A}^2(\phi_0,\nu)}{20}, \\
\label{eq:cond2}
\lambda_6(\phi_0) &=  3 \frac{\Mp^3}{\phi_0^4} \frac{{\cal B}_1(\phi_0) \xi_2(\phi_0) - {\cal B}_2(\phi_0) \xi_4(\phi_0) + 2 \mphi^2(\phi_0) \left[\xi_2(\phi_0) + \xi_4(\phi_0)\right]}{\xi_1(\phi_0) \xi_4(\phi_0) - \xi_2(\phi_0) \xi_3(\phi_0)} \ , 
\end{align}
where ${\cal B}_2(\phi)$ is given in \cref{app:betafunctions} for the \lletexte\ case, and ${\cal A}^2(\phi,\nu)$ is given in \cref{app:genA}.

In view of \cref{eq:cond1}, a straightforward generalization at one-loop of the tree-level $\alpha$ parameter, see \cref{eq:alpha_def}, is given by:

\begin{equation}
    \alpha^{(\mathrm{loop})} \equiv   1-\frac{{\cal A}^2(\phi_0, \nu=0)}{20 \mphi^2(\phi_0)}.
    \label{eq:alpha_loop}
\end{equation}

Taking \cref{eq:cond1} into account, one can remark that $\alpha^{(\mathrm{loop})}$ vanishes in the flat-inflection limit, $\nu=0$, and deviates from $0$ for the quasi-flat-inflection, cosmology-consistent configurations. In the limit of negligible one-loop corrections, Eqs. (\ref{eq:cond1}) and (\ref{eq:cond2}) give back the exact tree-level relations  (\cref{eq:mphi_tree} and \cref{eq:lambda6_tree}) and $\alpha^{(\mathrm{loop})}=\alpha$.
 
\begin{figure}[htbp]
\begin{tabular}{cc}
\includegraphics[width=.47\textwidth]{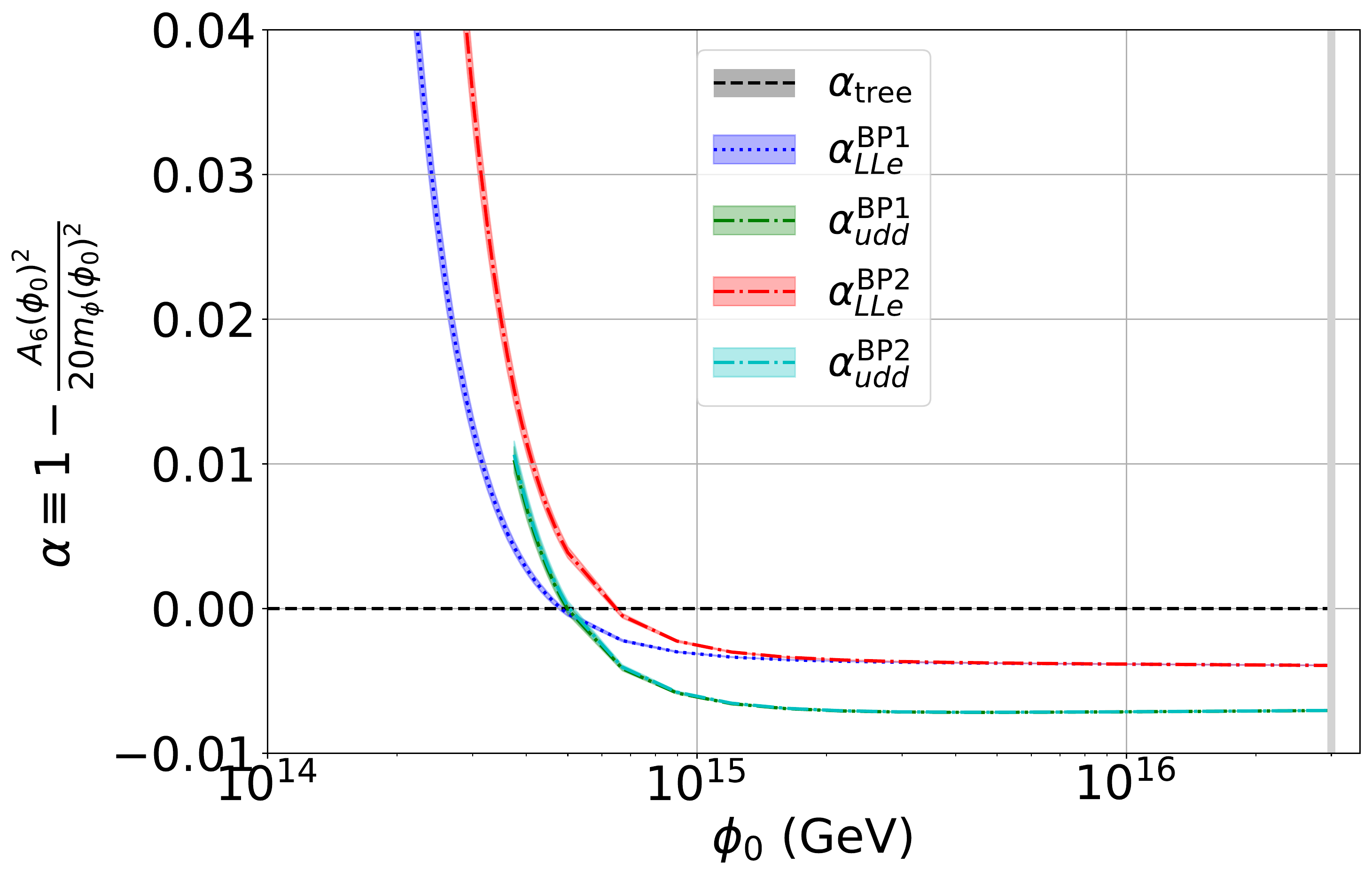}
&
\includegraphics[width=.45\textwidth]{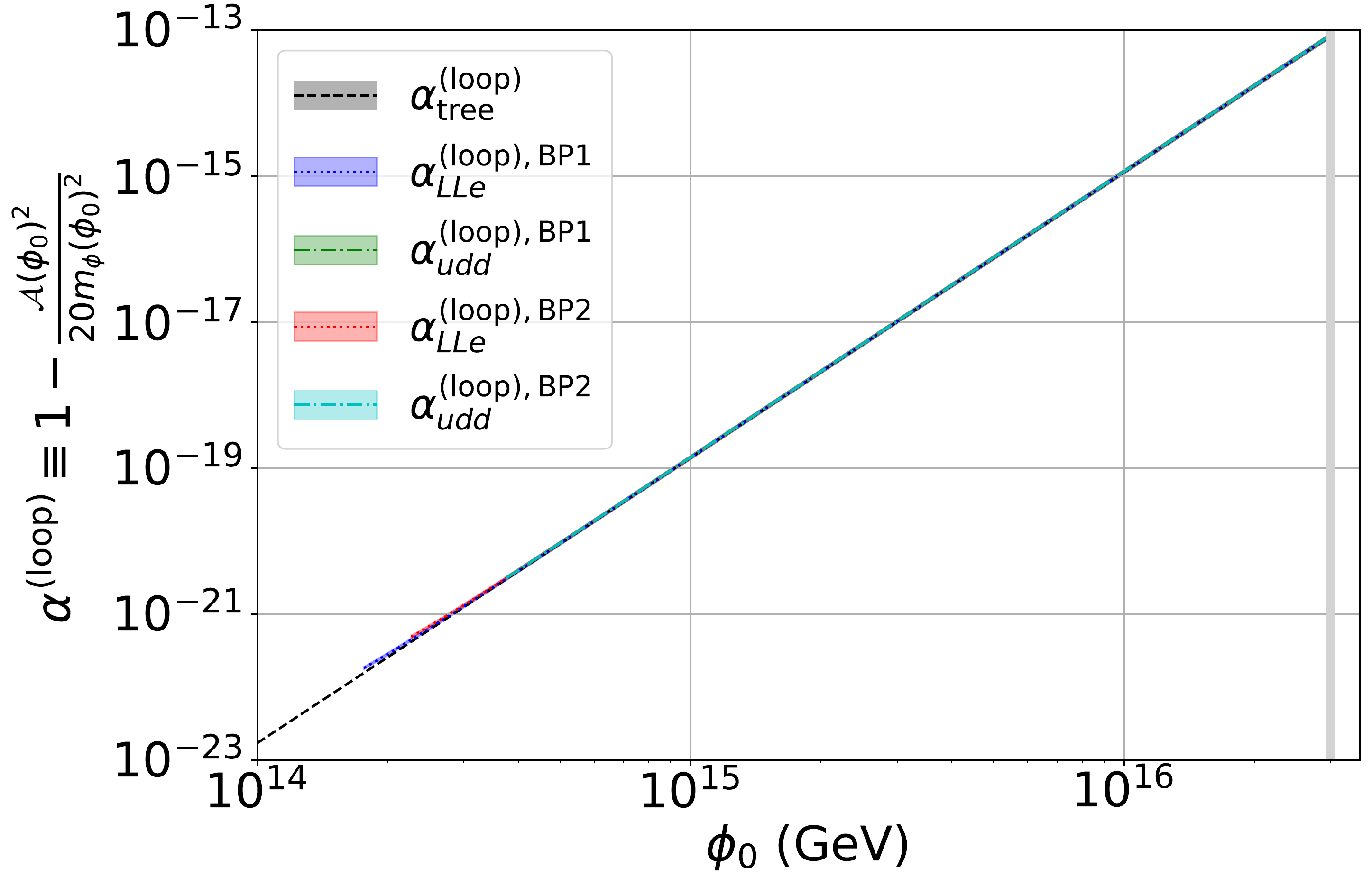}
\end{tabular}
\caption{ $\alpha$ (left panel) and $\alpha^{(\mathrm{loop})}$ (right panel) as a function of $\phi_0$, such that \Vtree\ and \Vrge\ lead to predictions on $\ns$ and $\As$ in agreement with measurements in the case of the two inflaton types (\lletexte\ and \uddtexte) and of the two benchmark points (BP1 or BP2).  On the left panel, the green and cyan \uddtexte\ contours are almost superimposed and the black contour close to zero corresponds to $\alpha_{\mathrm{tree}}$. The vertical gray line indicates the GUT scale. }\label{fig:fine_tuning}
\end{figure}

To illustrate the level of fine-tuning, for \Vtree\ and \Vrge, one can compute and compare $\alpha$  and $\alpha^{(\mathrm{loop})}$ at $\phi_0$ once we have applied the methodology described in \cref{sec:methodo}. This is done in \cref{fig:fine_tuning}, where the tree level case is shown in black, and the \lletexte\ and \uddtexte\ cases for the two benchmark points are illustrated in colors\footnote{In the right panel, the different cases do not overlap exactly at low $\phi_0$: this is linked to the fact that the inflaton mass needs to satisfy \cref{eq:definition_range}. This is further discussed in the next section.}. The right panel of this figure provides evidence  that the level of fine-tuning needed in \Vrge\ is of the same order of magnitude as the one in \Vtree\ (contrary to what is suggested for instance in \cite{Allahverdi:2010zp}).
However, it cannot be estimated by simply replacing the tree-level quantities by their running counterparts in the definition of $\alpha$ (\cref{eq:alpha_def}), as illustrated on the left panel. One needs to compute $\alpha^{(\mathrm{loop})}$ instead.

Similarly to the \Vtree\ case, we can circumvent the issue of the quadratic precision calculations required to solve Eqs.~(\ref{eq:srDeltaN*}) and (\ref{eq:implicit_eq}) for \Vrge\ directly by expanding the first and second derivatives of the potential around $\phi_0$. In contrast to the tree-level case where exact forms can be obtained, \cref{eq:treefirst_precision,eq:treesecond_precision}, here the non-polynomial dependence in $\phi$ is handled analytically through a Taylor expansion around $\phi=\phi_0$, where only the knowledge of the
$\beta$-functions (and not the RGEs solutions) is needed. We have performed this calculation up to the tenth order in $\delta \phi=\phi-\phi_0$. This is absolutely required when one needs to make use of the double-precision calculations of \texttt{SuSpect3} to estimate the radiative corrections. We make use of this trick in  \cref{sec:examples_and_big_table}.

\section{Inflationary constraints on the parameter space}
\label{sec:cosmoconstraints}

The aim of this section is to study the inflationary potential parameter space that satisfies the inflationary constraints. We first address this question using \Vtree\ (cf. \cref{sec:tree_results}). We then compare the results to those obtained with \Vrge\ (\cref{sec:oneloop}). 

\subsection{Results for the  potential at tree level} 
\label{sec:tree_results}
In this section, we consider the case of the tree-level inflationary potential, \Vtree.

\subsubsection{Parameter-space constraints}

\begin{figure}[htbp]
\subfloat{\includegraphics[width=.40\textwidth]{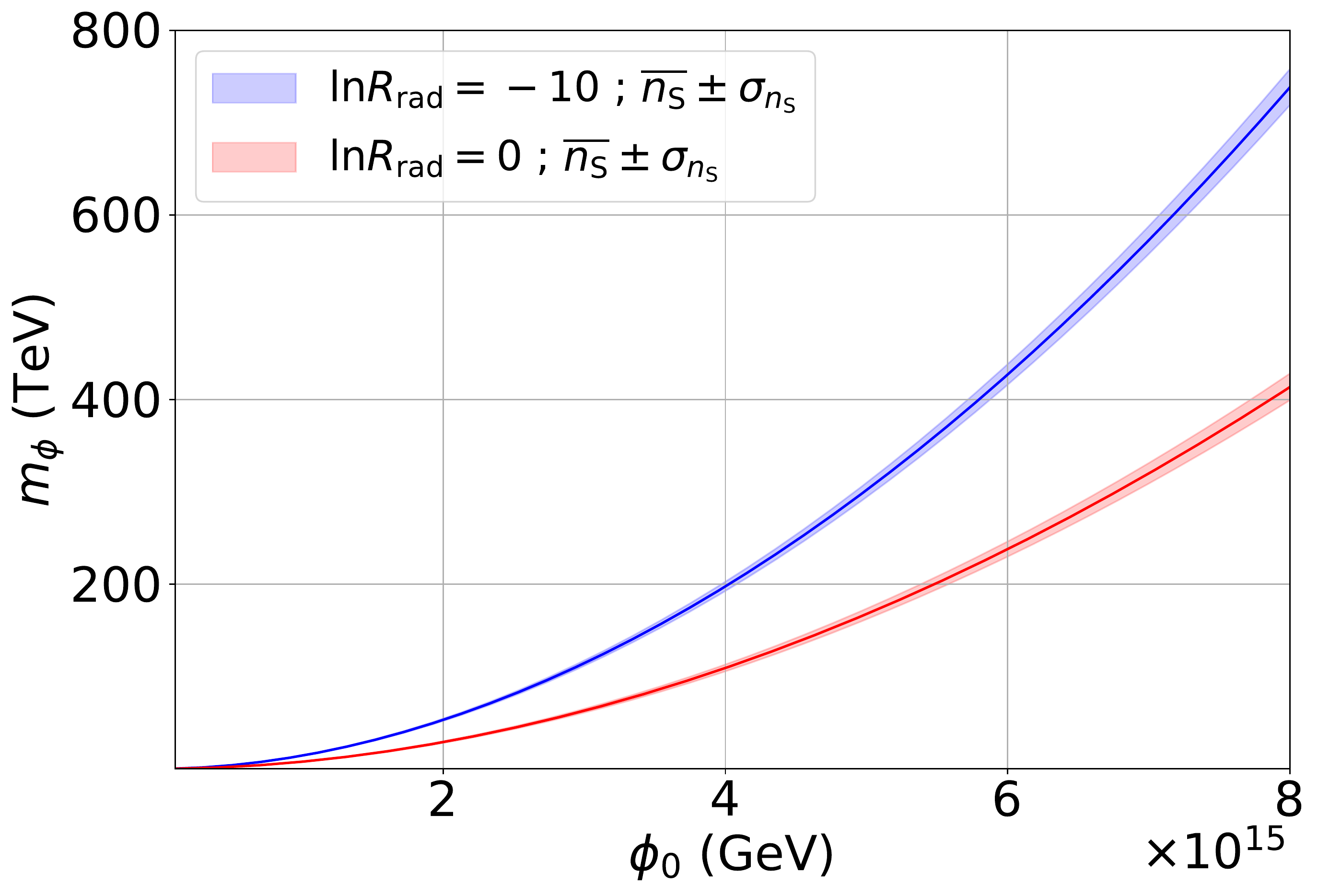}}
\subfloat{\includegraphics[width=.40\textwidth]{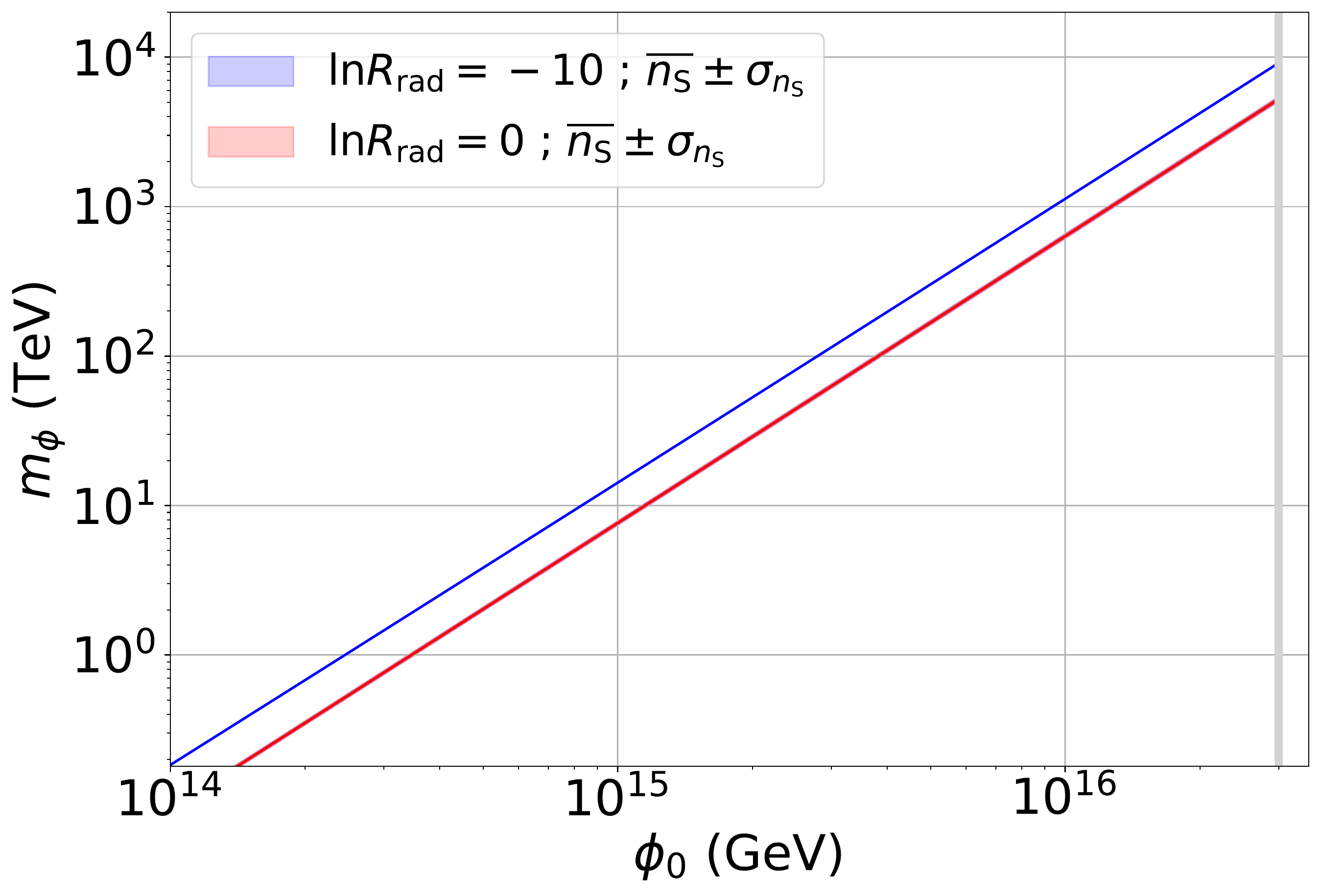}}\\
\subfloat{\includegraphics[width=.40\textwidth]{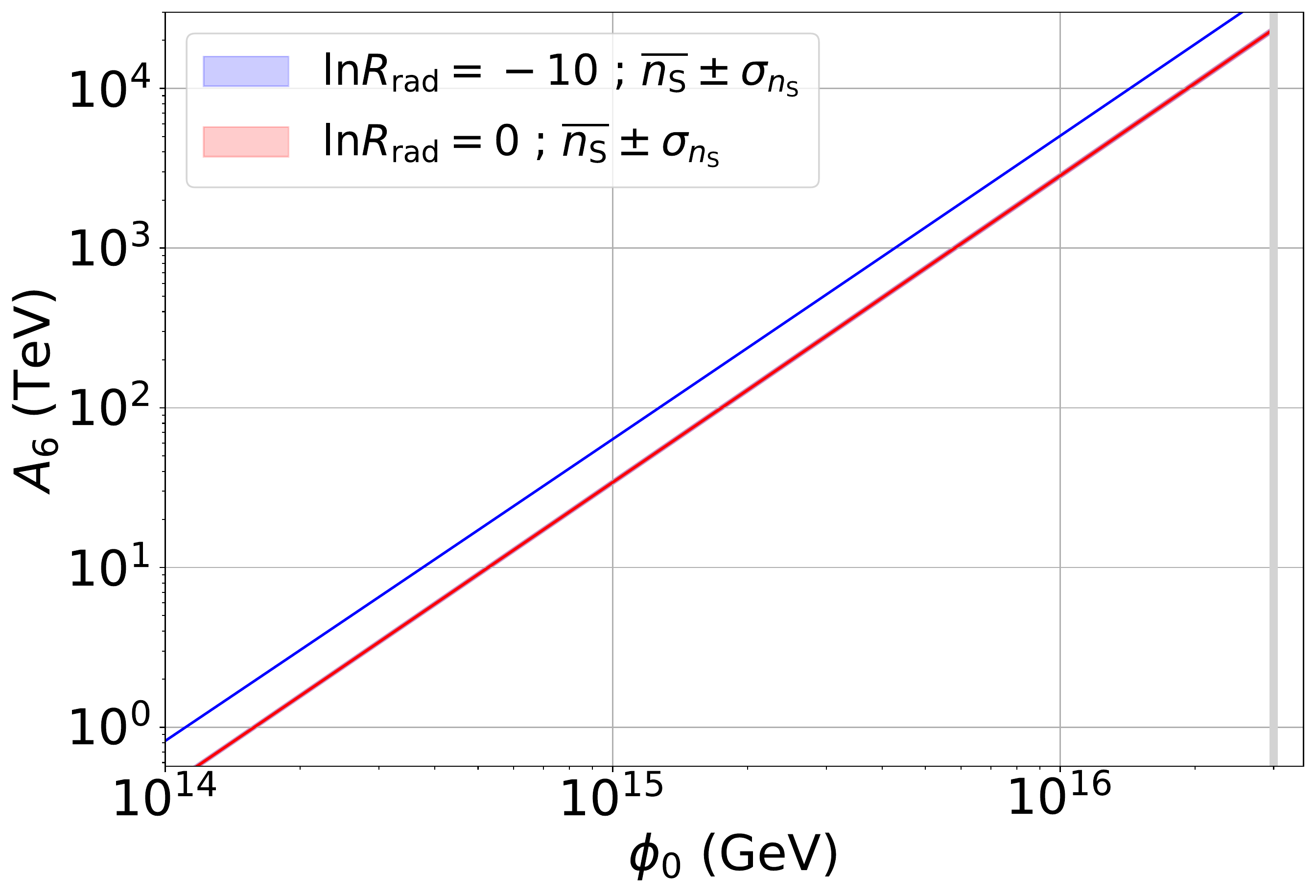}}
\subfloat{\includegraphics[width=.40\textwidth]{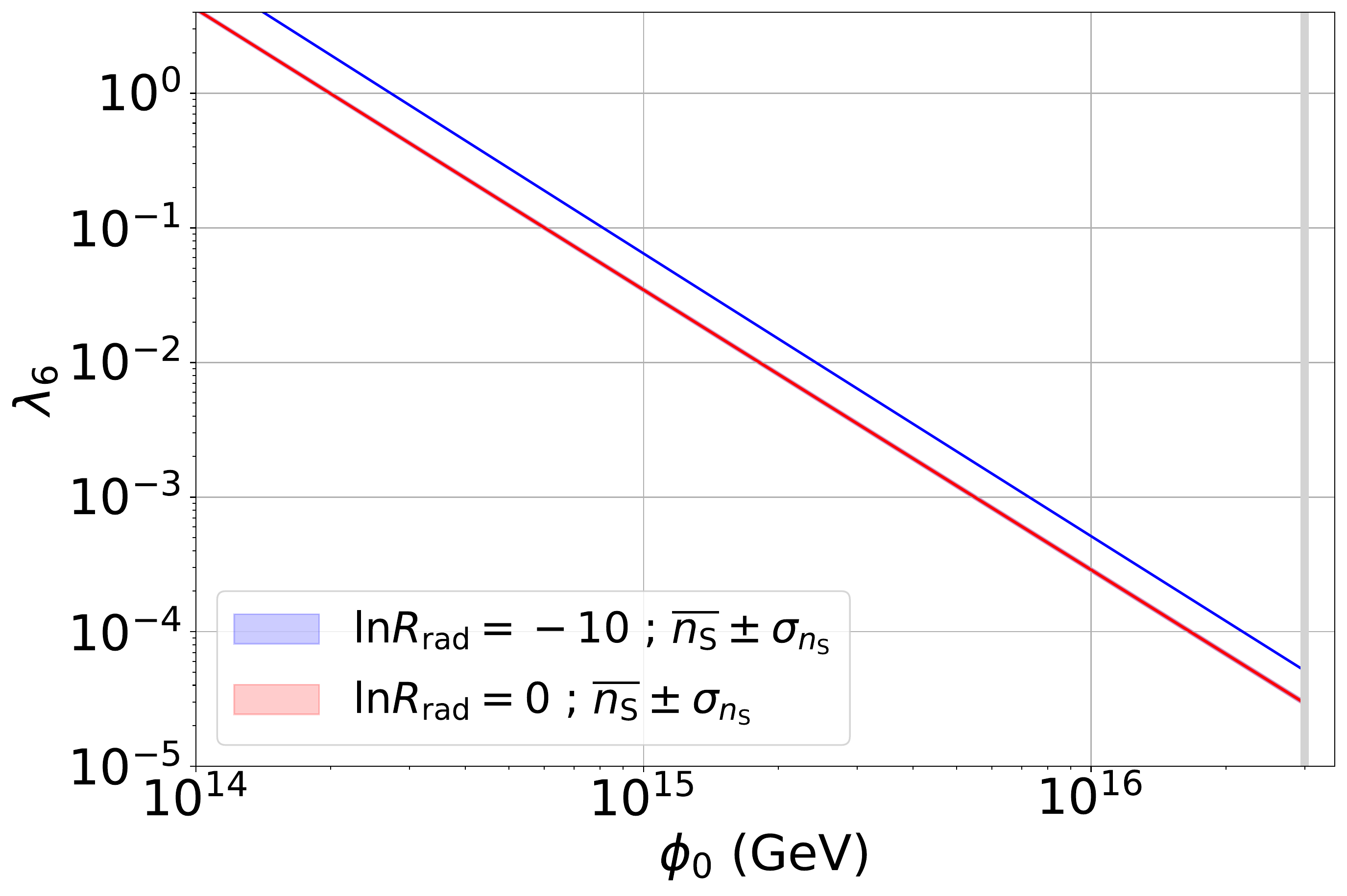}}
\caption{Tree-level inflationary potential parameters compatible with the inflationary observables 
as a function of $\phi_0$:  $\mphi$ on the upper-left panel (low $\phi_0$ in linear scale) and upper-right panel (wider range in log-scale), $A_{6}$ on the lower-left, and $\lambda_6$ the lower-right (log-scale). Two assumptions on $\lnrrad$ are illustrated: -10 in blue and 0 (instantaneous reheating) in red.  
\label{fig:phi0_param} }
\end{figure}

Following the methodology defined in \cref{sec:methodo}, one can determine $\mphi$, $\Asix$ and $\lambdasix$ such that inflation takes place at $\phi_0$, given the current measurements of the inflationary observables. The result depends on the assumptions made on reheating through \cref{eq:implicit_eq}. \Cref{fig:phi0_param} shows the parameter-space projections as a function of $\phi_0$, in log-scale. In red, we illustrate an instantaneous reheating, and in blue, we assume $\lnrrad=-10$ for illustration (see next section).
The figure of the upper-left panel is a zoom on the low-$\phi_0$ region (in linear scale). At tree-level, the factor $\sqrt{2}$ in front of the $\phi^6$ term that was introduced in \cref{eq:potential_RGE} with respect to \cite{Boehm:2012rh}  may appear equivalent to a simple redefinition of the $A_6$ parameters. But this is not the case when one wants to relate inflation and \hep\ constraints and make use of \cref{eq:polonyi} which connects $\Asix$ to $\Atopm$ at the GUT scale (this is further discussed in \cref{sec:examples_and_big_table}). 

The width of the contour corresponds to the one-$\sigma$ error on $\ns$. The relative error due to the $\As$ uncertainty (cf. \cref{tab:hep_obs}) is  roughly of the order of seven per mille. It is therefore not propagated. The $\sigma_{\ns}$ error translates into a $\sigma_{\mphi}/\mphi$ of the order of $3\%$ for a given $\phi_0$. This compares to the approximately $15 \ \%$ previously obtained in \cite{Boehm:2012rh}. This difference is mainly due to the reduction of the error bar on $\ns$ between \WMAP\ and \Planck\ (the improvement factor is approximately $3.5$ between the two measurements).

\subsubsection{Number of \textit{e}-folds and reheating} 
\label{sec:tree_results_reheating}

In order to obtain the contours shown in \cref{fig:phi0_param}, one has to solve \cref{eq:implicit_eq}, which gives the number of \textit{e}-folds $\DeltaNStar$ between the time the pivot scale $k_*$ crosses out the Hubble radius and the end of inflation (at next-to-leading order in slow-roll approximation). The larger $\phi_0$, the larger $\DeltaNStar$: typically at the GUT scale for an instantaneous reheating, $\DeltaNStar \simeq 45$, and for $\phi_0\simeq 1\times 10^{14}$ \GeV, $\DeltaNStar$ reaches 37.5. For slow-rolling potentials with $\phistar\simeq\phi_0$, the $\DeltaNStar$ expression (\ref{eq:implicit_eq}) is often reduced to its slow-roll leading-order approximation\footnote{Beware the difference with \cite{Boehm:2012rh} linked to the $k_*$ values at which the inflationary observables are estimated.},
\begin{align}
\label{eq:BoehmDeltaN}
   \DeltaNStar &\simeq  -\ln\left(\frac{k_*}{a_0\Tilde\rho_{\gamma}^{1/4}}\right)-\frac{1}{4}\ln(9) +\frac{1}{4} \ln \frac{\eqVtree(\phi_0)}{\Mp^4}    \\
   &\simeq  61.2 +\frac{1}{4} \ln \frac{\eqVtree(\phi_0)}{\Mp^4} \ , 
\end{align}
for $k_*=0.05\ \Mpc^{-1}$, $\rho_\gamma$ being given in \cref{sec:observables}.  The difference between both estimates for the number of \textit{e}-folds is of the order of 0.1, showing that \cref{eq:BoehmDeltaN} is a very good approximation for this potential. We nevertheless keep solving Eqs.~(\ref{eq:srDeltaN*}) and (\ref{eq:implicit_eq}) explicitly in the following. 

So far we have not propagated any error on the prediction of $\DeltaNStar$. As already discussed in \Cref{sec:observables,sec:reheating-theo}, such an error can be sourced by various effects: a deviation from an instantaneous reheating (\ie $\lnrrad\neq 0$), a corrective term to account for the MSSM relativistic number of degrees of freedom at reheating, the experimental uncertainty of the $\rho_\gamma$ measurement or the use of the slow-roll approximation. In \cref{fig:phi0_param}, in blue, we illustrate the case where the contributions from these different terms end up shifting $\DeltaNStar$ by $-10$ \textit{e}-folds.
It is a  
very pessimistic example since one expects the various contributing errors to be 
at most of order 1.
Such a value of $\DeltaNStar$ could, for example, arise from an extremely long reheating scenario with $\lnrrad=-10$. 
Such a change is equivalent, for the tree potential, to a shift of the $\ns$ value from 0.9665 to 1.0353 (while keeping the instantaneous reheating assumption). 
Conversely, the current constraint on $\ns$ would propagate into an error on $\lnrrad$ of $0.6$ if all other parameters were to be fixed.

\subsection{Results for the one-loop RGEs potential }
\label{sec:oneloop}
In the two next sections, we compare the parameter space constraints assuming a tree-level potential to the ones obtained when taking into account the one-loop RGEs in the potential. 

\subsubsection{Comparisons for given  $\phi_0$}
\label{sec:oneloop_vs_phi0}

In this section, we determine the parameters such that, for both potentials \Vtree\ and \Vrge, 
inflation takes place at the same value for $\phi_0$ and we first compare their values at this scale. 
We refer to each set of values as:   $p^{\scriptscriptstyle{\mathrm{V_{tree}}}|_{\phi_0}}$ ({\sl resp.} $p^{\scriptscriptstyle{\mathrm{V_{RGE}}}|_{\phi_0}}$)
where $p$ can be $\mphi$, $\Asix$, $\lambdasix$. Note that $\mphi^{\scriptscriptstyle{\mathrm{V_{tree}}}|_{\phi_0}}$ corresponds to $m_\phi$ of the previous section (the same applies for the other parameters). \cref{fig:phi0_param} gives the absolute scale of these parameters at tree level.

\begin{figure}[htb]
    \subfloat{\includegraphics[width=.45\textwidth]{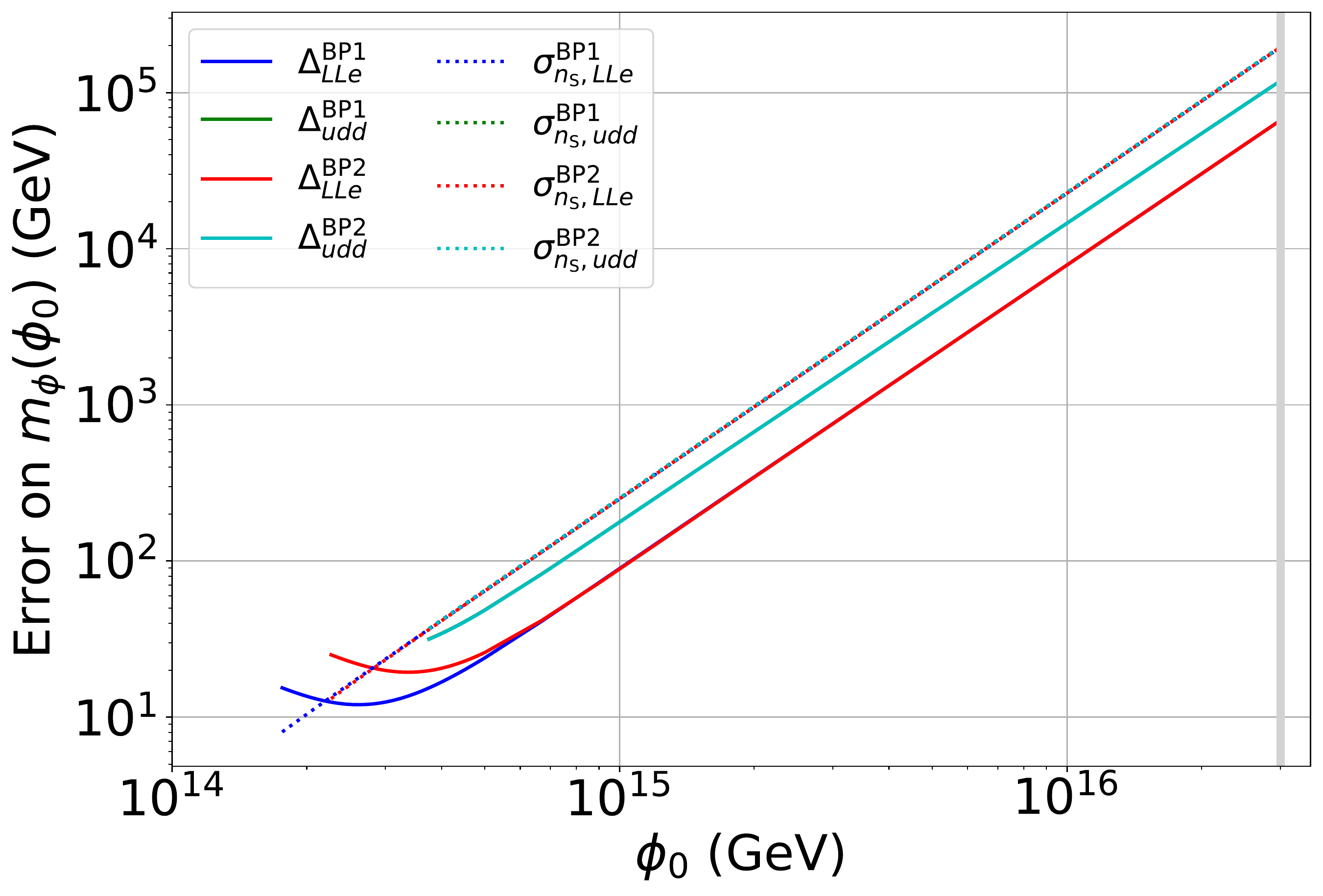}}
    \subfloat{\includegraphics[width=.45\textwidth]{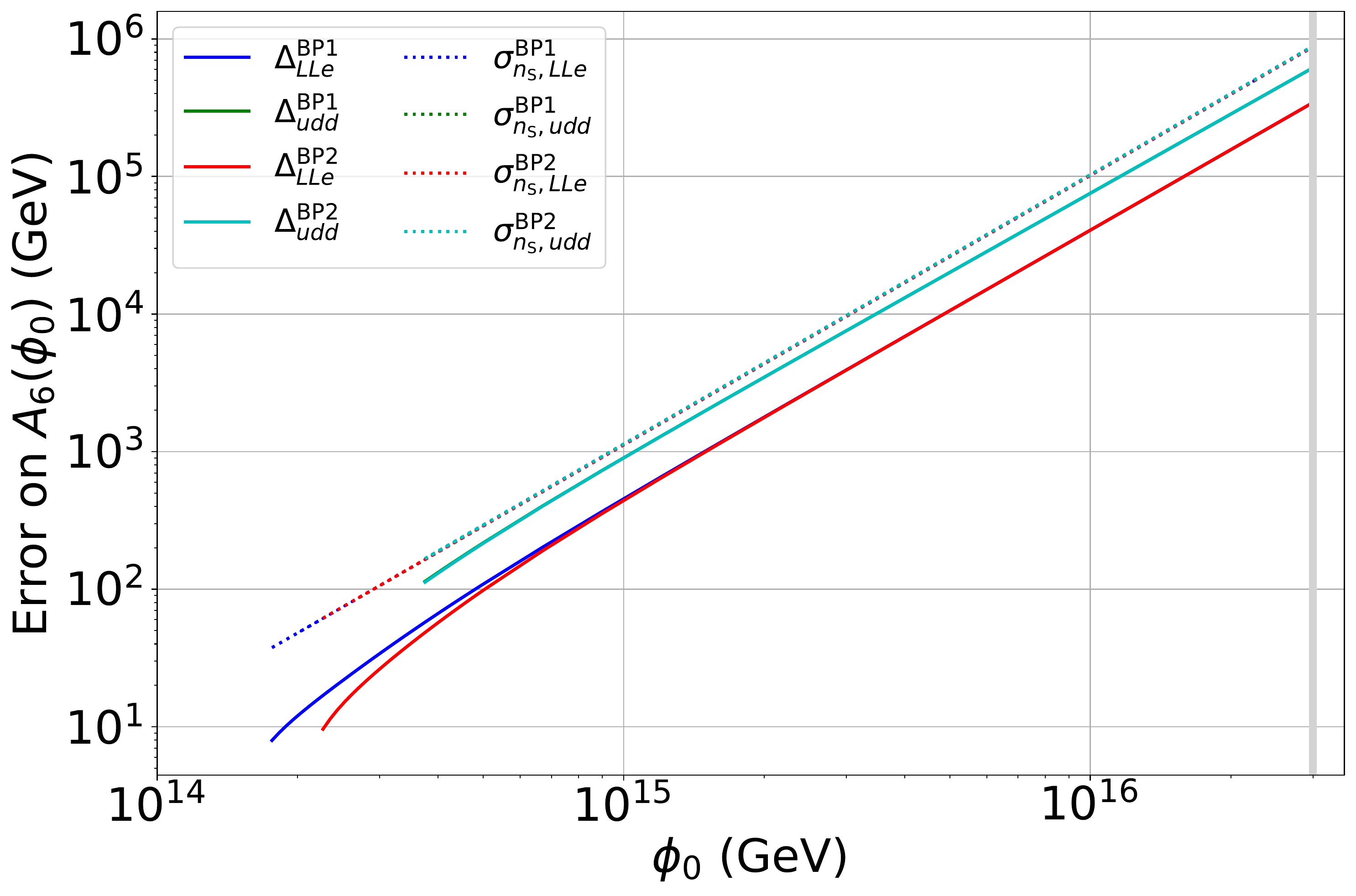}} \\
    \subfloat{\includegraphics[width=.45\textwidth]{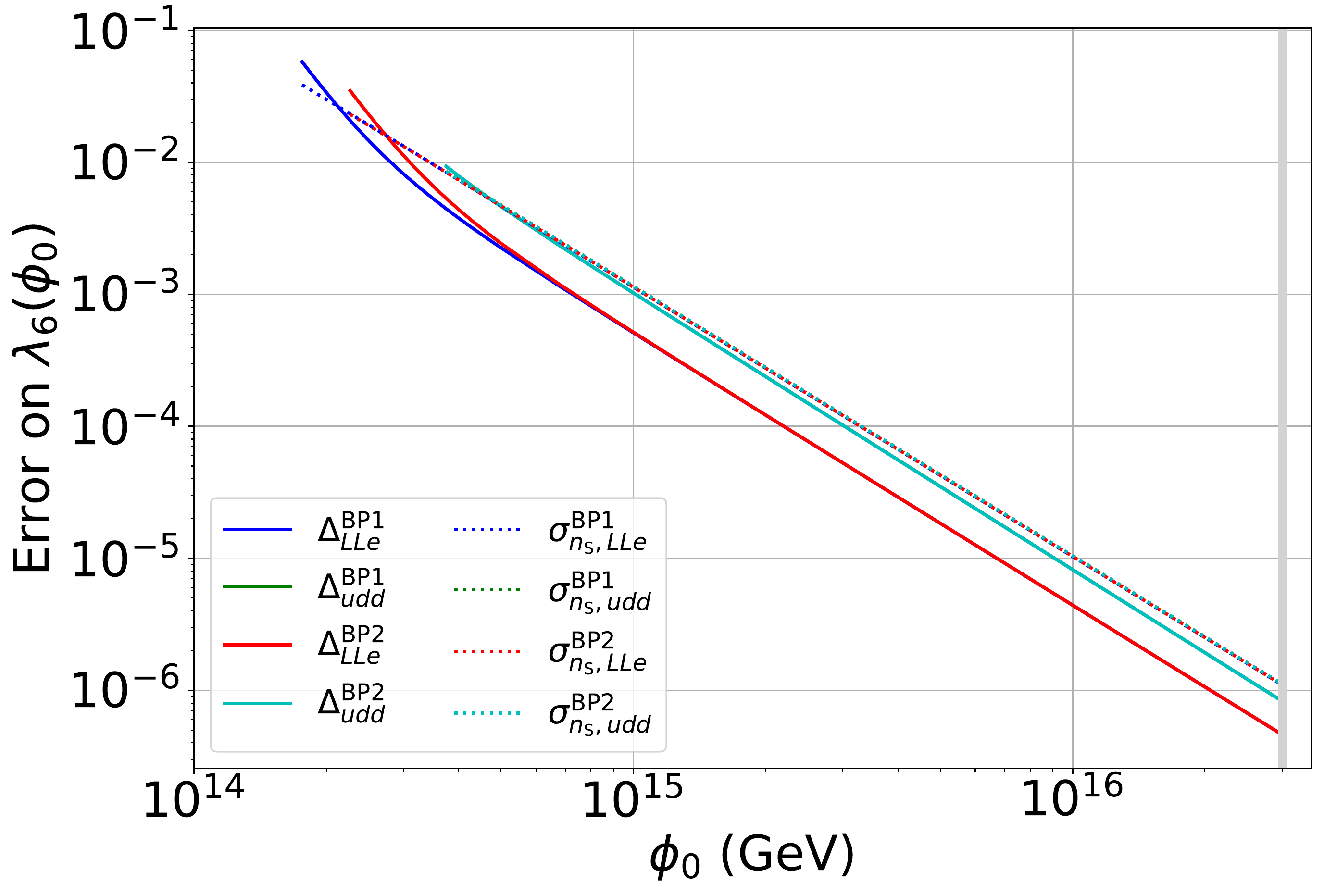}}
    \caption{Systematic biases - $\Delta$'s  as defined by \cref{eq:Deltap} in solid lines - and statistical uncertainties - $\sigma$'s as defined by \cref{eq:sigmap} in dotted lines - on the determination of
    the inflationary potential parameters as a function of $\phi_0$: in the upper-left for $\mphi$, in the upper-right for $\Asix$, and in the lower-left for $\lambdasix$. BP1 ({\sl resp.} BP2) is illustrated in dark blue ({\sl resp.} red) for an \lletexte\ inflaton, and in green ({\sl resp.} light blue) for \uddtexte. 
   \label{fig:srge_deltaparam_v3}}
\end{figure}

To proceed, we introduce, for each parameter $p$, the following notations:
\begin{align}
\label{eq:Deltap}
\Delta_{i}^{\mathrm{BP}j} [p] &= p^{\scriptscriptstyle{\mathrm{V_{tree}}}{|_{\phi_0}}}(\muorq=\phi_0, \ns=\nsmean)- p^{\scriptscriptstyle{\mathrm{V_{RGE}}}{|_{\phi_0}}}(\muorq=\phi_0, \ns=\nsmean), \\
\label{eq:sigmap}
\sigma_{\ns,i}^{\mathrm{BP}j} [p] &= {\frac{1}{2}}\left|p^{\scriptscriptstyle{\mathrm{V_{RGE}}}{|_{\phi_0}}}(\muorq=\phi_0, \ns=\nsmean+\sigma_{\ns})-p^{\scriptscriptstyle{\mathrm{V_{RGE}}}{|_{\phi_0}}}(\muorq=\phi_0, \ns=\nsmean-\sigma_{\ns})\right|,
\end{align}
where $i$ indicates the inflaton type (\uddtexte\ or \lletexte) and the subscript $j$ refers to the benchmark point (as defined in \cref{tab:benchmark_point}). The $\Delta$'s give the biases that are related to the use of \Vtree\ instead of the  one-loop \Vrge\ on the values of the parameter $p$. The $\sigma$'s indicate the statistical errors on the parameter $p$ given the current $1\sigma$ error on $\ns$. In \cref{fig:srge_deltaparam_v3}, these quantities are represented as a function of $\phi_0$
for $\mphi$ (upper-left panel), $\Asix$ (upper-right panel)
and $\lambdasix$ (bottom panel) for all benchmark points. 

The dotted lines in \cref{fig:srge_deltaparam_v3} show that at any inflation scale, for all parameters, and for both inflaton types and benchmark points, we recover the $3\%$ relative statistical error that we have quantified for \Vtree\ (cf \cref{sec:tree_results}). 

In the high-$\phi_0$ region, as illustrated by the comparison of the solid lines, the parameters are systematically shifted towards larger values when using \Vrge\ instead of \Vtree. In these examples where the Yukawa's terms can be neglected, this bias depends mainly on the inflaton type (the red and blue curves are perfectly superimposed). It is almost insensitive to the  values of the gaugino masses and the gauge couplings assumed at the GUT scale.  
It is of the order of 253 \GeV\ at $\phi_0\simeq 1.2\times 10^{15}$ \GeV\ for $\mphi(\phi_0)$ (upper-left panel) and $\simeq 1.3$ TeV\ for $\Asix(\phi_0)$ (upper-right panel) for \uddtexte. This compares to the precisions on the determination of the parameters that are linked to the current uncertainty on $\ns$  which are of the order of $360$ \GeV\ and 1.6 TeV respectively. 
The bias is not negligible: it is roughly $70\%$  of the current statistical error due to $\sigma_{\ns}$. 

The low-$\phi_0$ area exhibits two distinctive behaviors. On the one hand, in the RGEs case, there are unphysical regions where the inflaton masses become tachyonic (\ie \cref{eq:definition_range} is not satisfied). This implies that there exist lower bounds in the parameter space that depend on the benchmark point and on the inflaton type. This is an important difference to the tree-level potential for which the domain of definition of the parameters is unrestricted.  In our examples, $\phi_0$ cannot be lower than $2\times 10^{14}$ GeV to $5\times 10^{14}$ GeV depending on the cases. 
On the other hand, one can see a change of slope in the $\Delta$'s curves at low-$\phi_0$. It is the same feature one could already see in \cref{fig:fine_tuning}. In this area, the systematic bias gets even larger than the statistical error and dominates the error budget. These two combined effects show how important it is to properly take into account the runnings of the parameters in the potential.

To assess the effect of these biases on the particle-physics phenomenology, the RGEs (\cref{eq:rgemphi_lle} to \cref{eq:rgelambda6_udd}) are used to calculate the values of these parameters at the EWSB scale taken to be 2 TeV as an illustration. For \Vtree, by definition, the parameters do not depend on $\phi$ in the potential, the RGEs are therefore taken into account only at this stage, as done in \cite{Boehm:2012rh}.

The induced difference between the parameters using \Vtree\ and \Vrge\ is almost the same at 2 TeV as at $\phi_0$. To give orders of magnitude, for the \lletexte\ BP1, while fixing $\phi_0=1.2\times 10^{15}$ \GeV, if one assumes \Vtree\ and runs the parameters from $\phi_0$ to 2 TeV, one obtains:
\begin{align}
m_\phi^{\scriptscriptstyle{\mathrm{V_{tree}}}|_{\phi_0}}(\muorq=2\mathrm{ TeV}) &= 10892\ _{-349}^{358} \ \GeV, \\
A_6^{\scriptscriptstyle{\mathrm{V_{tree}}}|_{\phi_0}}(\muorq=2\mathrm{ TeV}) &=  47681\ _{-1565}^{1605}\ \GeV, \\ 
\lambda_6^{\scriptscriptstyle{\mathrm{V_{tree}}}|_{\phi_0}}(\muorq=2\mathrm{ TeV}) &= 0.045 \pm 0.001 \ .
\end{align}
In addition, while using \Vrge\ and running the parameters to 2 TeV, one gets:
\begin{align}
m_\phi^{\scriptscriptstyle{\mathrm{V_{RGE}}}|_{\phi_0}}(\muorq=2\mathrm{ TeV}) &= 11020\ _{-354}^{363} \ \GeV, \\
A_6^{\scriptscriptstyle{\mathrm{V_{RGE}}}|_{\phi_0}}(\muorq=2\mathrm{ TeV}) &=  48336\ _{-1586}^{1627}\ \GeV, \\ 
\lambda_6^{\scriptscriptstyle{\mathrm{V_{RGE}}}|_{\phi_0}}(\muorq=2\mathrm{ TeV}) &= 0.045 \pm 0.001. 
\end{align}

In such an example, the predicted value of the inflaton running mass at the EWSB scale assuming \Vtree\ instead of \Vrge, is under-estimated by 128 \GeV\ (more than one third of the statistical error bar linked to the propagation of the current value of $\sigma_{\ns}$). For $A_6$, this systematic shift is of the order of 655\ \GeV, about $40\%$ of the statistical error, hence it is not negligible.   

The precise values of the induced systematic effect does depend, eventually, on the inflaton type, the gauginos masses and the gauge couplings. Still, as illustrated in the present example, using the simplified tree version of the inflationary potential may induce non-negligible bias in the end-results when one wants to combine constraints from cosmological and HEP observables: a shift of more than 100 GeV on sparticle masses could lead to erroneous conclusions in the determination of the favored/disfavored area of the eMSSM parameter space if one wants to sample it extensively and couple it to particle-physics observables. 

\subsubsection{Comparisons for given {$A_6(\Mgut)$}}
\label{sec:oneloop-a6}

To illustrate the favored areas of the parameter space, one can also choose to compare their values assuming the same value of $A_6$ at the GUT scale, a particularly interesting quantity since it
relates to \Atop\ through the Polonyi relation, and, thus, to the particle-physics phenomenology (see next section). 
This implies that the comparison of the values of the \Vtree\ and \Vrge\ parameters is performed at different values of $\phi_0$. 

\begin{figure}[htb]
    \subfloat{\includegraphics[width=.45\textwidth]{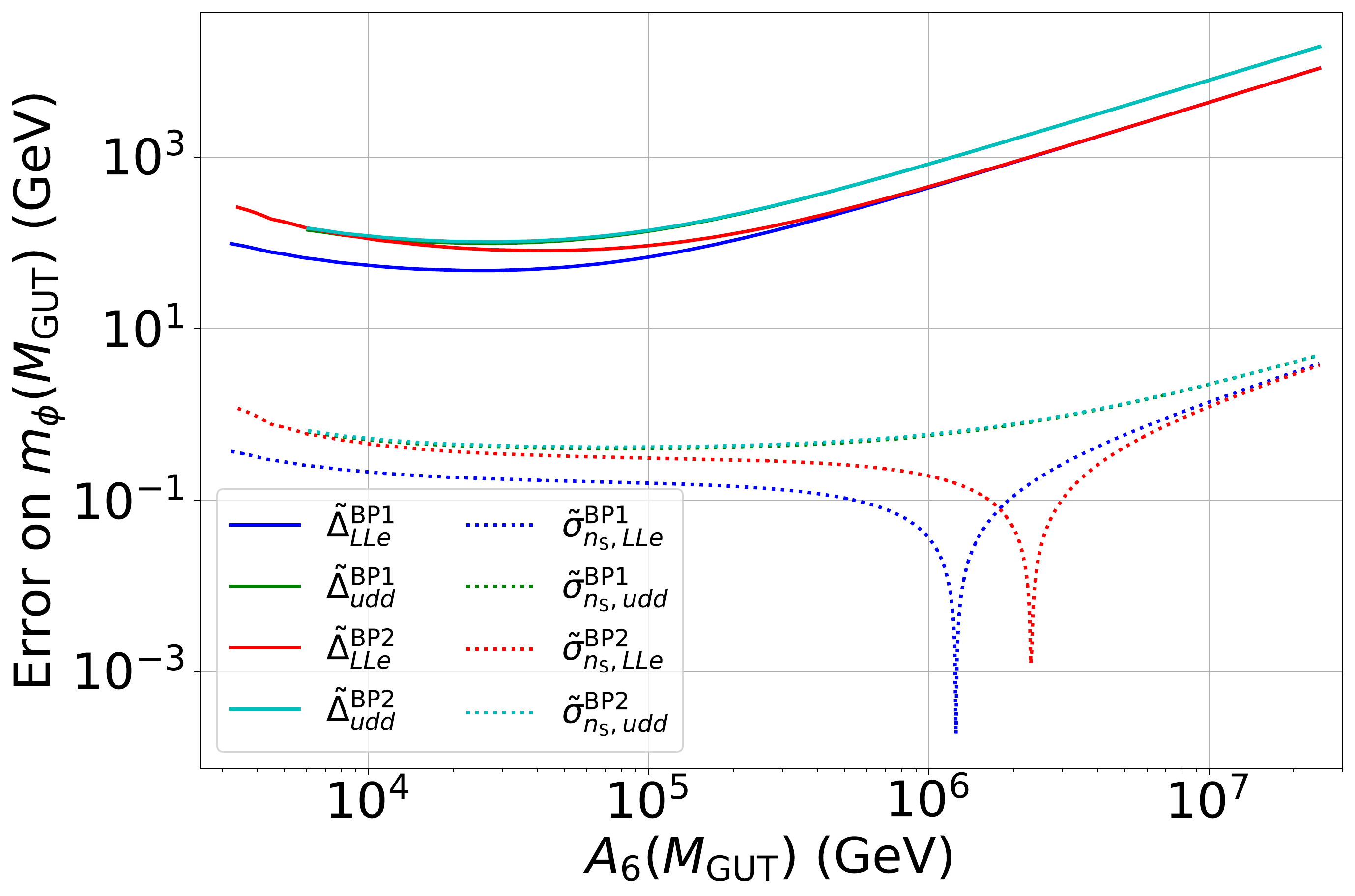}} 
    \subfloat{\includegraphics[width=.45\textwidth]{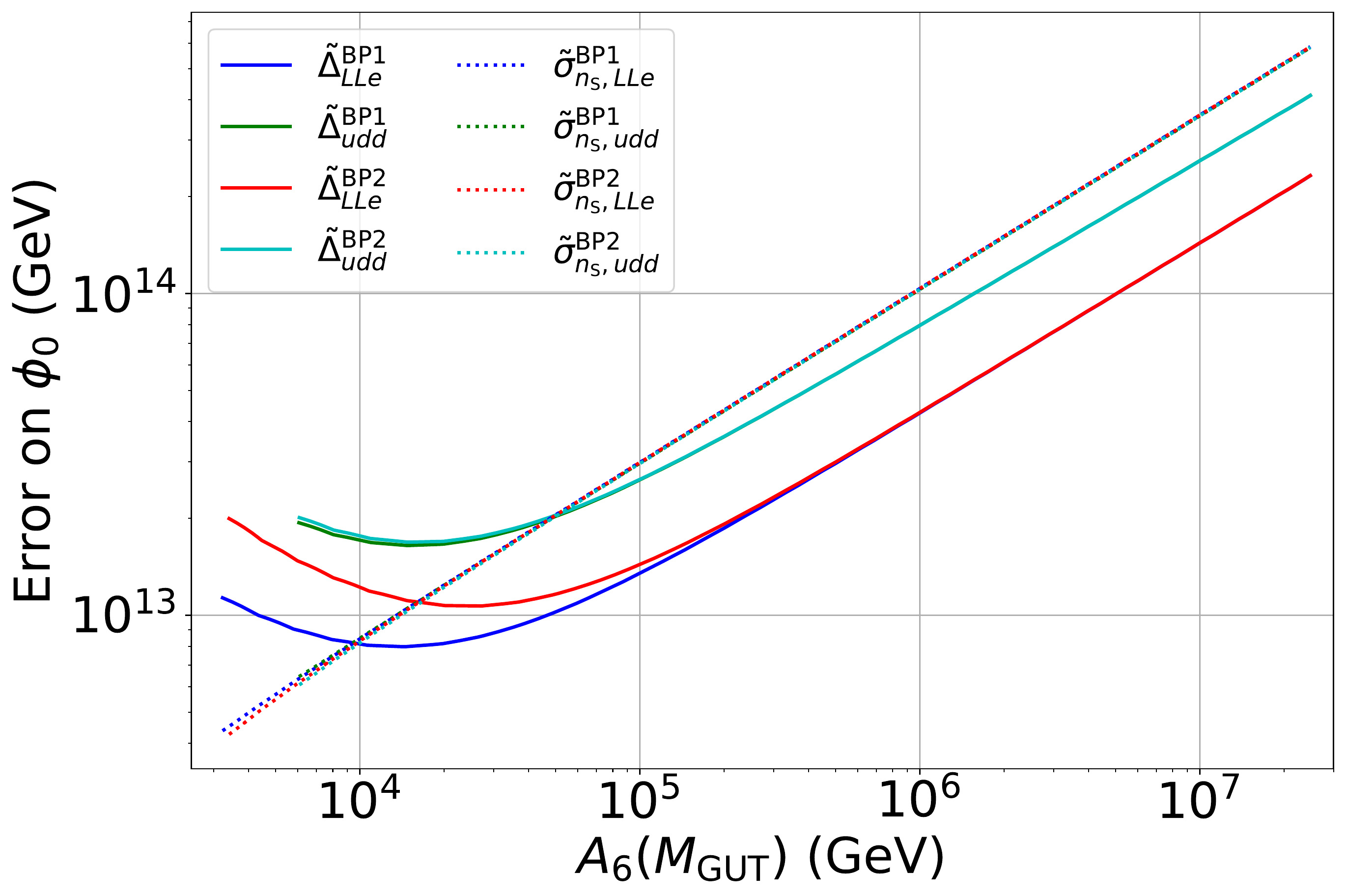}}\\
    \subfloat{\includegraphics[width=.45\textwidth]{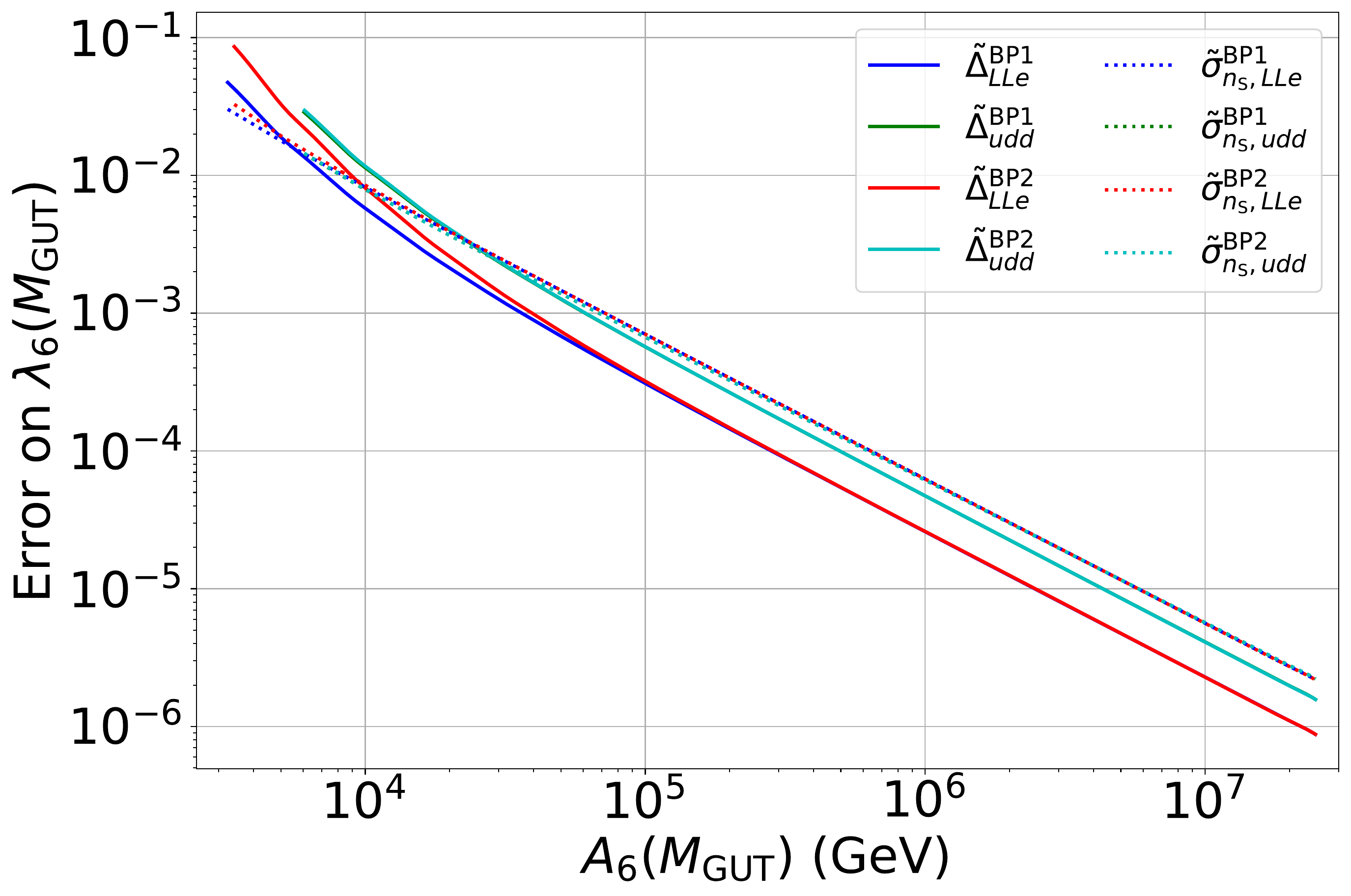}}
    \caption{
    Systematic biases - $\tilde\Delta$'s  as defined by \cref{eq:DeltatildeA} in solid lines - and statistical uncertainties - $\tilde\sigma$'s as defined by \cref{eq:sigmatildepA} in dotted lines - on the determination of
    the inflationary potential parameters as a function of $A_6(\Mgut)$: on the upper-left panel for $m_\phi(\Mgut)$, on the upper-right panel for $\phi_0$, and on the lower panel for $\lambda_6(\Mgut)$. BP1 ({\sl resp.} BP2) is illustrated in dark blue ({\sl resp.} red) for an \lletexte\ inflaton, and in green ({\sl resp.} light blue) for \uddtexte. 
   \label{fig:newfigA6}}
\end{figure}

Similarly to what was done in \cref{sec:oneloop}, one defines:

\begin{align}
\label{eq:DeltatildeA}
\tilde{\Delta}_{i}^{\mathrm{BP}j} [p] &= 
p^{\scriptscriptstyle{\mathrm{V_{tree}}}{|_{A_6(\Mgut)}}}(\muorq=\Mgut, \ns=\nsmean)- p^{\scriptscriptstyle{\mathrm{V_{RGE}}}{|_{A_6(\Mgut)}}}(\muorq=\Mgut, \ns=\nsmean), \\
\label{eq:sigmatildepA}
\tilde{\sigma}_{\ns,i}^{\mathrm{BP}j} [p] &= {\frac{1}{2}}\left|p^{\scriptscriptstyle{\mathrm{V_{RGE}}}|_{A_6(\Mgut)}}(\muorq=\Mgut, \ns=\nsmean+\sigma_{\ns})-p^{\scriptscriptstyle{\mathrm{V_{RGE}}}|_{A_6(\Mgut)}}(\muorq=\Mgut, \ns=\nsmean-\sigma_{\ns})\right|.
\end{align}

These errors are shown in \cref{fig:newfigA6} for 
$\mphi$, $\lambdasix$ and $\phi_0$. When proceeding this way, because of the fine-tuning relation the propagation of the error on $\ns$ on the $m_\phi$ axis is almost negligible (of the order of a few $0.1$ \GeV). The systematic bias when one compares the results using \Vrge\ versus \Vtree\ for $\mphi$ at the GUT scale is 250 times larger than this statistical error. This implies that tree-level and one-loop predicted inflaton masses are not compatible with each other.
Such a high precision in the prediction of the inflaton mass comes with a worse determination of the $\phi_0$
 values. 

For example, for $A_6(\Mgut)=49562$ \GeV, one gets for \Vrge:

\begin{align}
m_\phi^{\scriptscriptstyle{\mathrm{V_{RGE}}}|_{A_6(\Mgut)}}(\muorq=\Mgut, \ns=\nsmean) &= 10999.4 \pm 0.2 \ \GeV, \\
\lambda_6^{\scriptscriptstyle{\mathrm{V_{RGE}}}|_{A_6(\Mgut)}}(\muorq=\Mgut, \ns=\nsmean) &= 0.022 \pm 0.001, \\
\phi_0^{\scriptscriptstyle{\mathrm{V_{RGE}}}|_{A_6(\Mgut)}}(\muorq=\Mgut, \ns=\nsmean) &= (1.20  \pm 0.02 ) \times 10^{15} \ \GeV \ , 
\end{align}
which compare, for \Vtree, with:
\begin{align}
m_\phi^{\scriptscriptstyle{\mathrm{V_{tree}}}|_{A_6(\Mgut)}}(\muorq=\Mgut, \ns=\nsmean)&= 11049.03 \pm 0.01 \ \GeV, \\
\lambda_6^{\scriptscriptstyle{\mathrm{V_{tree}}}|_{A_6(\Mgut)}}(\muorq=\Mgut, \ns=\nsmean) &= 0.021 \pm 0.001, \\
\phi_0^{\scriptscriptstyle{\mathrm{V_{tree}}}|_{A_6(\Mgut)}}(\muorq=\Mgut, \ns=\nsmean) &= (1.21  \pm 0.02 ) \times 10^{15} \ \GeV \ . 
\end{align}

\subsection{Error-budget summary}
\label{sec:error-budget}
\begin{figure}[htb]
  \subfloat{\includegraphics[width=.45\textwidth]{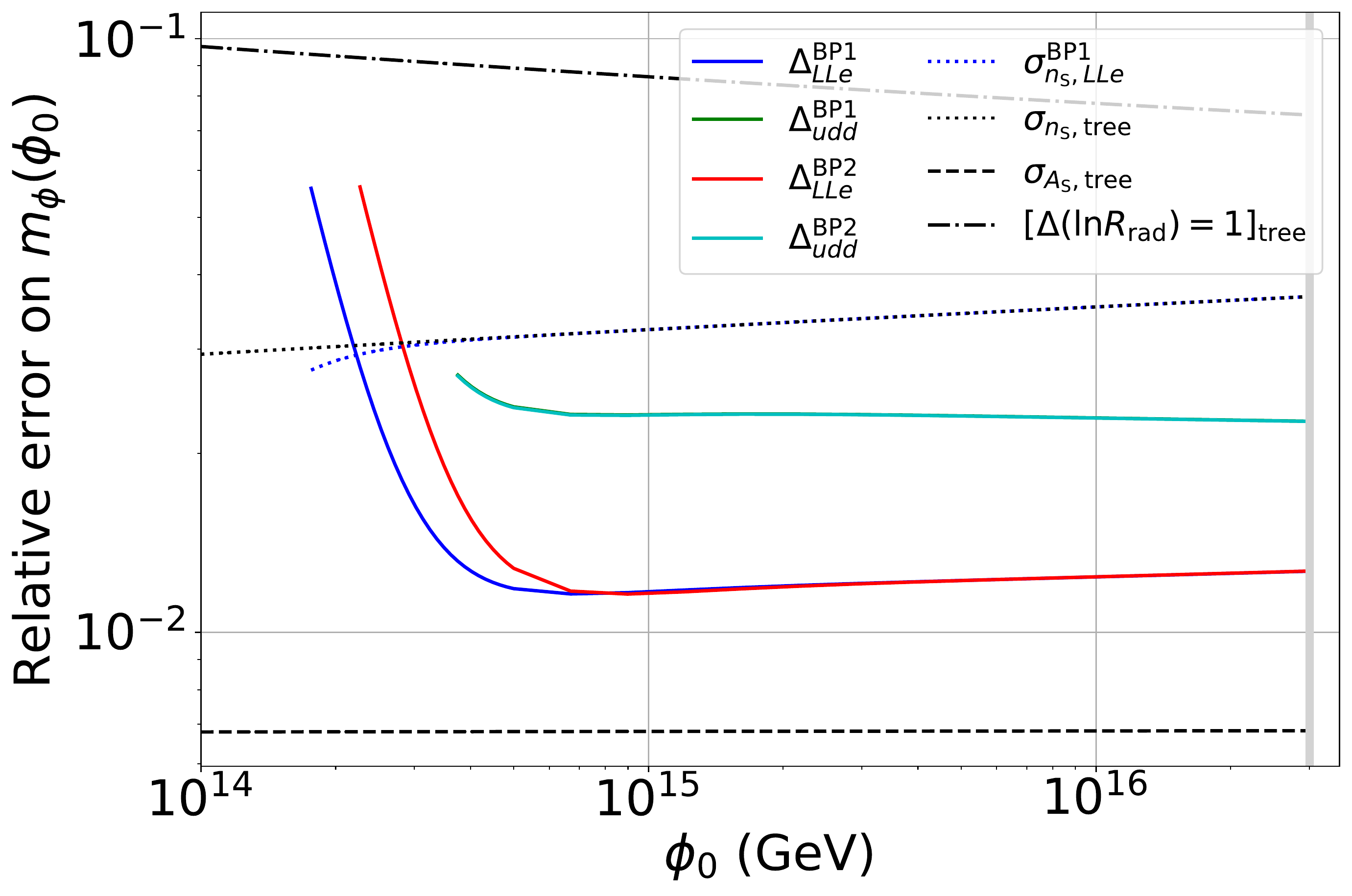}}\subfloat{\includegraphics[width=.45\textwidth]{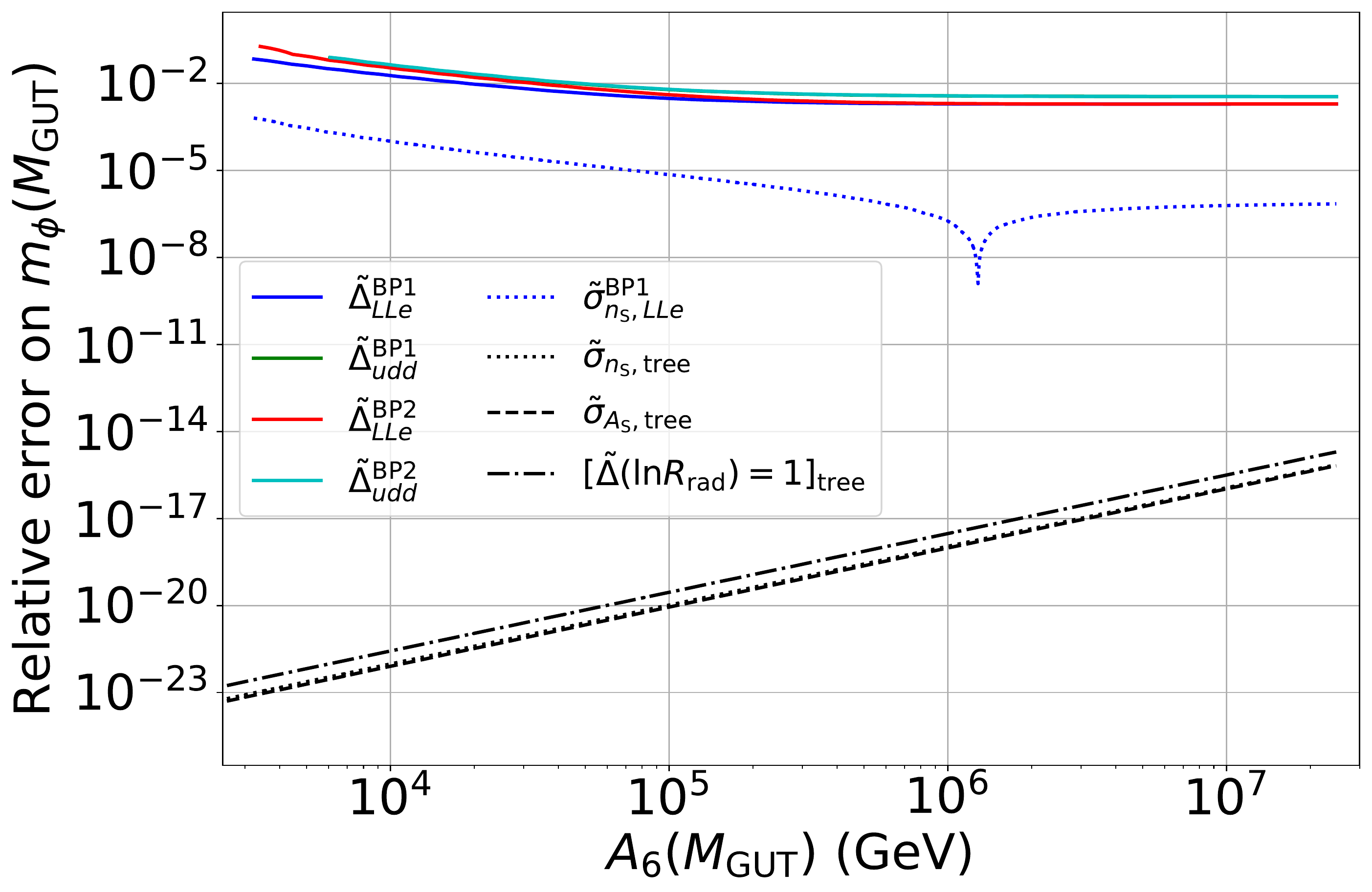}}
\caption{Summary of the different contributions to the relative errors on the predicted inflaton mass for different benchmark points and inflaton types. \textsc{Left panel:} for $\muorq=\phi_0$ as a function of $\phi_0$. \textsc{Right panel:} for $\muorq=\Mgut$ as a function of $A_6(\muorq=\Mgut)$}. \label{fig:errors}
\end{figure}

A summary plot of the different contributions to the error budget on the determination of the inflaton mass is given in  \cref{fig:errors}, which shows the relative errors on the determination of $\mphi$ for \Vtree\ and \Vrge: at $\muorq=\phi_0$ assuming inflation occurs at $\phi_0$ (left panel) and at $\muorq=\Mgut$ assuming given values of $A_6(\muorq=\Mgut)$ (right panel). The solid lines show the systematic bias induced when adopting a tree-level approximation instead of taking into account the one-loop corrections in the potential for the four configurations already discussed in the previous sections. The dotted lines represent the propagation of the error on $\ns$ in the \lletexte\ BP1 case (in blue) and in the tree case (in black). Finally, the black-dash-dotted line represents the propagation of the $\As$ error, and the black-dotted line corresponds to the propagation of a theoretical error of one unit on $\lnrrad$. 

This shows that the one-loop corrections to the inflationary potentials cannot be neglected. The tree-level fine-tuning that implies a tight relation between $\mphi$ 
and $\Asix$ at $\phi_0$ is broken by the radiative corrections. It leads to a systematic underestimation of the parameters when one uses \Vtree\ instead of \Vrge\ at high $\phi_0$. In addition, because of the RGEs' runnings of the parameters, their definition domain is reduced, implying a minimal energy scale below which the parameters are not physical. Such a behavior is completely absent from the tree-level treatment of the potential. Moreover, for the same $A_6(\muorq=\Mgut)$
the predicted inflaton masses using \Vtree\ or \Vrge\ are not compatible with each other anymore, the systematic error dominating the statistical error  by several orders of magnitude. Finally, we also show the contribution of a theoretical error of $\lnrrad$ of one unit, which dominates the error budget if one wants to fix $\phi_0$ in the analysis.

We have demonstrated that the measurements of the inflationary observables are, to date, accurate enough for this analysis to be sensitive to radiative corrections on the parameters of the inflationary potential. 
RGEs cannot be ignored anymore. 

In addition, the theoretical error on the reheating phase is, today, a limiting factor in the constraints one can put on the eMSSM parameter space.

\section{Constrained MSSM spectra by \hep\ and cosmological data}  

\label{sec:examples_and_big_table}

In this section, we combine cosmological and \hep\ measurements using \Vrge\ and discuss some MSSM points in the context of the current data. We assume an instantaneous reheating (see \cref{sec:reheating-theo}), and we do not propagate any error on the prediction of $\DeltaNStar$. The goal of this section is to pin-point some specific examples of a combined fit but not to perform an exhaustive scan of the parameter space. 

\subsection{Combined fit}
\subsubsection{Method} 

We base our study on the best-fit MSSM points that have been identified in \cite{Henrot-Versille:2013yma}, which were found to match particle-physics observations and $\omegacdm$ measurements.
We have updated the experimental constraints (cf. \cref{tab:hep_obs} and \cref{tab:cosmo_obs}), and refitted the MSSM parameters in order for the corresponding predictions of the observables to be in agreement with the current measurements, using \texttt{SFitter}.

For each of these points, we proceed as follows. We set \Atop\ at the GUT scale such that the Higgs sector is almost unchanged (while keeping a low $\chi^2$). This allows us to fix $A_6(\Mgut)$ through \cref{eq:polonyi}. We are therefore in the case discussed in  \cref{sec:oneloop-a6}. 
We then determine the corresponding values of $\mphi$, $\lambdasix$ and $\phi_0$ imposed by $\As$ and $\ns$. 
To do this, we bypass the quad-precision computation as explained at the end of \cref{sec::finetune2}, in order to compute the one-loop corrections with the double-precision code \texttt{Suspect3} and feed the \texttt{ASPIC}-like part of the code.  
Finally, we tune the eMSSM parameters in order for the soft-SUSY-breaking masses to match the inflaton mass (according to \cref{eq:mphi_udd} or \cref{eq:mphi_lle}): either the mass of the second generation squarks (for a \uddtexte\ inflaton), or the mass of the sleptons (for \lletexte), given the relations defined by \cref{eq:mphi_udd} and \cref{eq:mphi_lle}. To keep the Yukawa's terms low enough to be neglected, we only focus on the first two generations for the inflaton candidates: $u_1d_1d_2$ and $L_1L_2e_1$. Finally, we ensure the full consistency of the procedure in a global $\chi^2$ minimisation using \texttt{SFitter}. In this way, the obtained corresponding MSSM spectra are compatible with all current observational constraints described in \cref{sec:all_observ}. 

\begin{table}[t]
\begin{center}
  \begin{tabular}{l|c|c|c|c|c|c|c}

ID      &	$H_1$ & $H_2$& $h_1$ & $h_2$ & $A_1$ & $A_2$ & $A_3$  \\ 
\hhline{=|=|=|=|=|=|=|=}
DM channel	      &	\multicolumn{2}{c|}{Higgsino}           & \multicolumn{2}{c|}{h-funnel}	   &	\multicolumn{3}{c}{A-funnel}  \\
inflaton	      & $u_1d_1d_2$ &	$L_1L_2e_1$ & $u_1d_1d_2$ & $L_1L_2e_1$ & $u_1d_1d_2$  &	$L_1L_2e_1$ & $L_1L_2e_1$ \\

\hhline{=|=|=|=|=|=|=|=}

 EWSB scale	      &   2556.9 & 2556.8	   &	2713.1 & 2713.0	   &	4008.9 & 4008.9	   & 4009.7  \\
 $\tan\beta$  &                 \multicolumn{2}{c|}{29} &	\multicolumn{2}{c|}{26.6}	   & 	\multicolumn{2}{c|}{18.3}	   &    24.7 \\
 $\sgn(\mu)$        &                  \multicolumn{2}{c|}{+} &	\multicolumn{2}{c|}{+}	   &	\multicolumn{3}{c}{-}  \\
 $M_{\mathrm{GUT}}$        &          \multicolumn{2}{c|}{$1.237\times 10^{16}$}	   &	\multicolumn{2}{c|}{$3\times 10^{16}$} 	   & 	\multicolumn{3}{c}{$1.295\times 10^{16}$}   \\ 
 $M_1$        &          \multicolumn{2}{c|}{1571}	   &	\multicolumn{2}{c|}{61} 	   & 	\multicolumn{2}{c|}{400}   	   &     362    \\
 $M_2$        &          \multicolumn{2}{c|}{2917} 	   &	\multicolumn{2}{c|}{967} 	   &   \multicolumn{2}{c|}{1515}	   &        1662    \\
 $M_3$        &          \multicolumn{2}{c|}{1931}	   &	\multicolumn{2}{c|}{1934} 	   &   \multicolumn{2}{c|}{1898}	   &        1098    \\
 $M_{\tilde{\mu}_L}$     &          2864 &	12108   &	4210 & 9892	   &   2564 & 10567	   & 4773    \\
 $M_{\tilde{\mu}_R}$   	 & 2785 & 12108	&	4221 & 9892  &                    3976 & 10400	   &    4723    \\
 $M_{\tilde{\tau}_L}$	 &          \multicolumn{2}{c|}{3342}	   &	\multicolumn{2}{c|}{4068}	   &   \multicolumn{3}{c}{2039}	  \\
 $M_{\tilde{\tau}_R}$    &          \multicolumn{2}{c|}{2064}	   &	\multicolumn{2}{c|}{3931}	   &   \multicolumn{3}{c}{2801}	  \\
 $M_{\tilde{q}_{2L}}$    &  11959 & 9090  &	9860 & 7370 &  10452    &   7299	  & 2984    \\
 $M_{\tilde{q}_{3L}}$    &           \multicolumn{2}{c|}{2176}	   &	\multicolumn{2}{c|}{3155}	   &   \multicolumn{3}{c}{3950}	   \\
 $M_{\tilde{t}_R}$       &          \multicolumn{2}{c|}{3003}	   &	\multicolumn{2}{c|}{2330}	   &   \multicolumn{3}{c}{4066}	  \\
 $M_{\tilde{b}_R}$    	 &          \multicolumn{2}{c|}{3474}	   &	\multicolumn{2}{c|}{1952}	   &   \multicolumn{3}{c}{2422}	  \\
 $A_{\tau}$          	 &          \multicolumn{2}{c|}{-3499}	   &	\multicolumn{2}{c|}{-2564}      &   \multicolumn{3}{c}{-3010}	 \\
 $\Atopm$          	 &   	 3180 & 3223  &	2985	   & 3020 &   3450	   & 3493 &   1190    \\
 $A_{b}$       		 &          \multicolumn{2}{c|}{148}	   	   &	\multicolumn{2}{c|}{143}	   &   \multicolumn{3}{c}{187}	\\

\hhline{=|=|=|=|=|=|=|=}

$m(\tilde{\chi}_1^0)$	& 	  1108 & 1108	&  60.0 & 60.1	   & 397 & 398	   &    357    \\
$m(\tilde{\chi}_2^0)$       & 	 -1113 & -1112  & 497 & 496	   &	766 & 766	   &  760   \\
$m(\tilde{\chi}_3^0)$       & 1578 & 1587 	& -504 & -503	   &	-769 & - 770	   &   -763  \\
$m(\tilde{\chi}_4^0)$	 & 2992 & 3005 	& 1041 & 1041  &	1589 & 1592	   &     1710    \\
$m(\tilde{\chi}_1^+)$	 & 	1111 & 1111  	& 496 & 495	   &	765 & 766	  & 760 \\
$m(\tilde{\chi}_2^+)$	 &  2992 & 3005	& 1041 & 1041	   &	1589 & 1592   & 1711    \\
$m(\tilde{g})$		 & 	2371 & 2333 	& 2351 & 2310	   &	2382 & 2332	  & 1389  \\
$m(\tilde{q}_L)$           & 	 12072 & 9173	& 9956 & 7442   &	10557 & 7376  & 3036   \\
$m(\tilde{q}_R)$           & 	 12056 & 9162	& 9939 & 7429   &	10540	& 7364 & 3024 \\
$m(\tilde{b}_1)$ 		& 	 2243 & 2244  	& 2031 & 2031	   &	2499 & 2499   & 2453   \\
$m(\tilde{b}_2)$ 		& 	 3512 & 	 3512 	&3177 & 3177   &	3974 & 3974	 & 3971    \\	   
$m(\tilde{t}_1)$ 		& 	2244 & 2244 & 2347 & 2347  &	3946 & 3945	   &    3972   \\
$m(\tilde{t}_2)$ 		& 	 2992 & 2992 & 3188 & 3188	   &	4074 & 4075   & 4059    \\
$m(\tilde{\tau}_1^-)$   	& 2066 & 2066	& 3931 & 3931   &	2055 & 2055   & 2055   \\

$m_h$         & 125.3 & 125.2  	& 125.3 & 125.2 & 125.4	& 125.3   & 122.2 \\
$m_A$         & 	 784.9 & 783.9 		& 3625.9 & 3625.8   & 782.2 & 784.4	   & 757.2   \\
$m_{3/2}$ & 12596 & 12557 & 10352 & 10383 & 11005 & 11010 & 4886\\
\hline

$\phi_0$             & 	 $1.25\times 10^{15}$ 	& $1.26\times 10^{15}$ & $1.13\times 10^{15}$	& $1.14\times  10^{15}$ &  $1.16\times 10^{15}$	& $1.17\times 10^{15}$&  $7.68\times 10^{14}$  \\
$m_{\phi}(\phi_0)$    &	 11982 		    & 11973 & 9847		        & 9892 & 10459	& 10487 &  
4661
\\
$\Asix(\phi_0)$               & 	 53757 		    & 53593 & 44181		& 44312 &  46970	& 46990 &  
20855 
\\
$\lambda_6(\phi_0)$         & 	 0.0224 		& 0.0218 & 0.0278 & 0.0269& 0.0260        & 0.0252  & 0.0609\\

\hhline{=|=|=|=|=|=|=|=}

$\chi^2_\mathrm{HEP}$ (d.o.f. = 78)     & 	 50.9 	 	& 51.7 & 46.1	& 46.5 &  47.5		& 47.0 &  49.2 \\
LHC searches &  &  & \xmark  &\xmark &  &   & \xmark\\

\end{tabular}
\end{center}
\caption{Fitted benchmark MSSM spectra and their status regarding the observables. The parameters are given at $M_{\textrm{EWSB}}$. All the masses and energy scales are given in GeV. The \xmark\ symbol highlights the points excluded by direct LHC searches (see \cref{sec:lhc}). \label{tab:all_spectra}}
\end{table}

\subsubsection{Results for different dark-matter annihilation channels}

We have considered the MSSM spectra that correspond to three dark-matter annihilation channels: a Higgsino\footnote{The channel is called as such because the neutralino is mainly Higgsino in this example.} channel, for example $\tilde\chi_1^0\tilde\chi_1^0\rightarrow W^+W^-$ via a t-channel (hereafter points $H_1$ and $H_2$), a h-funnel channel $\tilde\chi_1^0\tilde\chi_1^0\rightarrow h$  (points $h_1$ and $h_2$), and a A-funnel channel $\tilde\chi_1^0\tilde\chi_1^0\rightarrow A$ (points $A_1$, $A_2$ and $A_3$).

The corresponding spectra are summarized in \cref{tab:all_spectra} for different inflaton hypotheses (\uddtexte\ and \lletexte). For each point, the first block provides the MSSM fundamental parameters at the EWSB scale. The second one gives the determined mass spectrum (the slepton physical masses are of the same order as the soft masses), together with the inflationary parameters, $\mphi$, $\Asix$ and $\phi_0$ determined with \Vrge, whose values are given for $\muorq=\phi_0$. A $\chi^2$ value is also given, it refers to the difference between predictions and all \hep\ observables described in \cref{sec:hep-obser} as well as \omegacdm. The contributions to the $\chi^2$ are evenly distributed among the different observational constraints taken into account.

This table shows that we are able to combine all current observations to study, within the coherent framework of the eMSSM, the underlying parameter space and relate inflation to LHC physics at the level of the one-loop RGEs corrections on the inflationary potential. For the different DM annihilation channels we have considered, we have found eMSSM points that are compatible with the Higgs mass and the cold-dark-matter energy density but also with $\ns$ and $\As$ (together with all the \hep\ observables detailed in \cref{sec:all_observ}). 
Apart from $A_3$, all the points have a similar value of \Atop, for this reason the inflaton masses are of the same order of magnitude ($\simeq 1\times 10^{4}$ \GeV), so are the values for $\phi_0$ ($\simeq 1\times 10^{15}$ \GeV). 

Using \Vtree, we would have obtained an inflaton mass (hence squark masses) roughly 30 \GeV\ below the one obtained with \Vrge\ for the $A_1$ case. This difference is significantly larger than the statistical error linked to $\sigma_{\ns}$, which is lower than one \GeV. Even though these $30$\GeV\ will highly depend on the inflaton type and on the values of the gauge couplings and gaugino masses, it is far from being negligible when considering \hep\ data. This further reinforces the fact that the measurements of inflationary observables are, to date, sufficiently accurate for inflationary potential analyses to be sensitive to radiative corrections on the parameters.

\subsection{Additional cross-checks}

\subsubsection{LHC direct searches}
\label{sec:lhc}

As the gluino and the neutralino masses are, for some points, close to the limit of the current LHC searches,
we have performed additional \textit{a posteriori} cross-checks of these points, adding the constraints from direct searches for supersymmetric particles at the LHC.

For this, we use \texttt{SmodelS} \cite{Kraml:2013mwa,Ambrogi:2017neo,Ambrogi:2018ujg,Alguero:2020grj,Alguero:2021dig,Buckley:2013jua} with cross-sections calculated using \texttt{Pythia6} \cite{Sjostrand:2006za} and \texttt{NLL-fast} \cite{Beenakker:1996ch,Beenakker:1997ut,Kulesza:2008jb,Kulesza:2009kq,Beenakker:2009ha,Beenakker:2010nq,Beenakker:2011fu} comparing to a database of Run~2 analyses \cite{Dutta:2018ioj}. It should be noted that the phenomenology of the points illustrated in \cref{tab:all_spectra} is far from the ``simplified models'' topologies used to express exclusions in LHC experiments. In particular, gluinos masses are close to the excluded regions for simplified models but in these points they rather tend to decay via sbottom or in long cascade decays. For most of the points, a definite conclusion would require a complete event simulation and analysis. Nevertheless, the h-funnel points ($h_1$ and $h_2$) with their low-mass charginos and neutralinos are excluded with a small margin by searches for di-leptons and missing transverse momentum (MET).
The A-funnel $A_3$ point with a low-mass gluino has a large enough cross section to be excluded despite the complex decay chain by multiple analyses for topologies like jets and MET, and leptons jets and MET. 

These additional cross-checks on the $h_1$, $h_2$ and $A_3$ points are interesting examples of how particle-physics measurements can disfavor values of the inflationary potential parameters. It also shows how important it is to perform dedicated and detailed analysis of the particle-physics inputs from observations if one wants to perform a full analysis of the eMSSM (as was done for instance in \cite{ATLAS:2015wrn} for the extensive study of the pMSSM).

\subsubsection{Cosmology and LSP}

We proceed here to two other cross-checks: one on the parameters related to cosmology, and the other on the nature of the LSP.

As discussed in \cref{sec:observables}, the predicted value of the tensor-to-scalar ratio $r$ is too low to be meaningful, since it is expected to be lower than the threshold of secondary gravitational waves induced by scalar fluctuations through gravitational non-linearities. This remains true for \Vrge, implying that $r$ cannot be used in the global fit. We have also checked the predicted values of $\alphas$. For the eMSSM points discussed in this section, we obtain $\alphas\simeq -4.4\times10^{-3}$. This is perfectly consistent with current measurements (quoted in \cref{tab:cosmo_obs}) and reinforces the fact that the eMSSM points are robust to the current observational constraint. As a side comment and to give orders of magnitude, the energy density at the pivot scale can also be derived thanks to the following relation: $\rho_* = 3\Mp^2H^2_* \srlo 24\pi^2\Mp^4 \As \epsilon_{1*}$. We obtain $\rho_*^{1/4}\simeq 2\times10^9$ GeV (corresponding to $H_*\simeq 1$ GeV) for the points of \cref{tab:all_spectra}.

The second a posteriori cross-check one can perform is linked to the nature of the LSP. We have made the assumption for our analysis that the LSP is the lightest neutralino, which is ensured by the global fit for all sparticles 
except for the gravitino, which
is not part of the MSSM. Given the mSUGRA relation (\cref{eq:msugra}), one can deduce  $m_{3/2}$  from $A_6$: the corresponding values are given in \cref{tab:all_spectra}, where we show  that the gravitino mass is always larger than $m(\tilde{\chi}^0_1)$, ensuring the consistency of the analysis.  
Sufficiently fast gravitino decays will thus evade a potential problem of thermal gravitino overabundance (a detailed study taking into account the reheating and decay temperatures is however beyond the scope of the present paper, see \eg \cite{Kawasaki:2004qu,Kawasaki:2017bqm}).

\subsubsection{Consistency of the theory}
\label{sec:consistency_theory}
Finally we have checked two theoretical assumptions: the bounds on the parameters and the choice of the UV scale.

In order to avoid tachyonic masses for the inflaton, we have checked that \cref{eq:definition_range} is satisfied at all scales. One also needs  to check that $\widetilde{\lambda_6}$ remains of order 1 between $\Mp$ and $\phi_0$. Since the RGEs enhance the values of $\lambda_6$ at low scales, one only needs to check that $\widetilde{\lambda_6}(\muorq=\phi_0)=18\sqrt{2}\lambda_6(\muorq=\phi_0)\lesssim1$. As shown by the values of $\lambda_6(\muorq=\phi_0)$ of \cref{tab:all_spectra}, this is verified for our points.

The last check concerns the choice of the UV completion scale in the definition of $\lambda_6$, cf. \cref{eq:superpotential1,eq:def-tildelambda6,eq:norma3}.
It has to be noted that the $\Mp$ dependence in the potential only appears through $\lambda_6/\Mp^3$. The choice of a different UV scale (say $\Mgut$), as mentioned in \cref{sec::flatMSSM}, would be equivalent to a rescaling of $\lambda_6$ by a factor $(\Mgut/\Mp)^3$. In addition, such a rescaling is valid at any energy scale, due to the fact that \cref{eq:rgelambda6_lle,eq:rgelambda6_udd} are invariant under a re-scaling of $\lambda_6$. Assuming $M_{\textrm{UV}}=\Mgut$ instead of $\Mp$ preserves the consistency of the theory, since $\lambda_6$ (required to be $\lesssim \mathcal{O}{\left(1/(18 \sqrt{2})\right)}$) now takes an additional $10^{-6}$ factor.

Since new physics is expected both at $\Mp$ and at $\Mgut$, one could instead consider natural to have contributions from these two scales to the superpotential $W_6$ (\cref{eq:superpotential1,eq:def-tildelambda6}). In that case, one would need to apply the following transformation: 
\begin{equation}
    \frac{\widetilde{\lambda_6}}{\Mp^3}\rightarrow\frac{{\widetilde{\lambda_6}}_{,\Mp}}{\Mp^3}+
    \frac{{\widetilde{\lambda_6}}_{,\Mgut}}{\Mgut^3}\ ,
\end{equation}
where ${\widetilde{\lambda_6}}_{,\Mp}$ and ${\widetilde{\lambda_6}}_{,\Mgut}$ are expected to be of order 1. Thus, unless  ${\widetilde{\lambda_6}}_{,\Mgut}$  is unnaturally suppressed, the dominant contribution is the one from $\Mgut$. One can then read off the resulting inflationary configuration from our study by replacing therein formally $\lambda_6(\phi)$ by  $\left(\Mp^3/\Mgut^3\right) {\lambda_6}_{,\Mgut}(\phi)$. Again, unless  ${{\lambda_6}}_{,\Mgut}(\phi)$  is unnaturally suppressed, these significantly increased values of $\lambda_6(\phi)$ would come with significantly smaller values of $\phi_0$, $\Asix$ and $\mphi$, leading to too light sfermions and Higgs mass that are excluded by \hep\ constraints. 
Our conclusion is strengthened when $M_{\textrm{UV}}$, the scale at which new physics arise, is smaller than $\Mgut$, but still large enough to justify the use of effective operators. Hence, for the \uddtexte\ and \lletexte\ inflatons studied in this work, our results strongly suggest that potential new physics effects at the GUT (or any other sufficiently heavy UV) scale have to be (surprisingly) suppressed with respect to those originating from the Planck scale.

\section{Conclusions and Outlooks}

In this work, we have shown how a consistent analysis of cosmological and particle-physics constraints can be performed in the context of the eMSSM model.

We have studied the field phase-space structure and showed that eMSSM inflation behaves as a small-field  model, with a narrow basin of attraction. 

We have identified the region of parameter-space that can support inflation when the one-loop RGEs corrections are included, and compared it with the one obtained at tree level. We have derived a new way to estimate the level of fine-tuning including one-loop RGEs corrections. We have demonstrated that this level remains the same as the tree-level one. We have proposed a solution to overcome the resulting accuracy requirement. Furthermore, we have shown that the parameters of the one-loop inflationary potential are bounded at low scale in order to avoid tachyonic inflaton masses.

We then detailed the area of the parameter space compatible with the $\As$ and $\ns$ measurements, when neglecting or not the one-loop corrections in the expression of the inflationary potential. We have shown that the small changes in the potential due the RGEs induce a significant modification in the prediction for the model. While this shift depends on the masses of the gauginos and on the gauge couplings at the GUT scale, we have given examples for which the induced bias can be, for example for the inflaton mass, almost comparable to the statistical error linked to the $\ns$ measurements for a fixed value of $\phi_0$. 
We also demonstrated that this effect is even more important when one fixes $A_6$ at the GUT scale. This shows that, with the current constraints on $\ns$, one cannot neglect the one-loop corrections in the inflationary potential. 

We have also compared these shifts to the ones induced by the uncertainty on the reheating duration, and have found that the value predicted for $\ns$ is very sensitive to the reheating details. More precisely, changing the reheating temperature by one order of magnitude is enough to shift the spectral index by more than its measurement error. On the one hand, this means that CMB measurements weakly constrain the MSSM parameters, due to this large degeneracy with the reheating sector. On the other hand, this also implies that if those parameters were measured in particle-physics experiments, the CMB data would already be accurate enough to deliver a precise measurement of the reheating temperature. This contrasts with other single-field models of inflation where the reheating is still poorly constrained, even when the inflationary potential does not contain additional parameters \cite{Martin:2016oyk, Martin:2014nya}.

Finally, we have found points in the MSSM that are compatible with current measurements (in particular the Higgs mass, the cold-dark-matter energy density, and the inflationary observables). We have shown how conclusions about their compatibility with inflationary constraints can be affected by the way we take into account RGEs at inflation scale. In particular, beyond the Higgs funnel example, we have highlighted an A-funnel point compatible with most \hep\ and cosmological observations, yet excluded by the LHC beyond Standard Model searches, which opens a door to constrain inflation using \hep\ measurements. We have also given other examples of Higgsinos and A-funnel points that are at the limit of the current LHC constraints, for which the SUSY phenomenology shows many cascade decays. For all these reasons, new insights into the eMSSM inflationary potential are expected in the coming years. But to fully exploit the LHC data, a detailed implementation of the full exclusions for all analyses would be required.

We have used the eMSSM as a test case to combine all \hep\ and cosmological observations.  In order to proceed, we had to make choices to specify the theoretical framework.
For example, we relied on the simple correlation between $\Asix$ and $\Atopm$ obtained in mSUGRA assuming a Polonyi hidden sector and 
setting the SUSY breaking energy scale to the GUT scale, although we carried out the analysis in the the more general phenomenological MSSM. Finally, we did not consider additional terms in the inflationary potentials when we derived \Vrge\ from \Vtree,
such as anomalous dimension running effects in the inflaton field, or possible induced runnings if considering an additive constant to the potential, or non-RGEs loop-induced operators in the effective potential. 
These assumptions would require further investigations in future work.

Needless to say that a detection of the tensor-to-scalar ratio (for instance by LiteBIRD \cite{LiteBIRD:2022cnt}), or of primordial non-Gaussianities, 
would question this model. However, we have demonstrated that this work is very timely given the fact that, to date, the measurements of $\ns$ and $\As$ are already sufficiently accurate for inflationary potential analyses to be sensitive to radiative corrections on the parameters (and their accuracy will be significantly reduced in a close future \cite{Errard:2015cxa}). This analysis opens the door towards full-scan studies of the parameter space combining all observational constraints, but it also paves the way for future studies of theories that are able to describe the physical processes at the low-energy scale of the LHC up to the inflation scale in a well-defined theoretical framework. 

\begin{acknowledgements}
The authors thank Anja Butter and Dirk Zerwas for helpful discussions in a preliminary phase of this analysis, for their help with the use of \texttt{SFitter} and \texttt{SuSpect3},
and for the organization of a dedicated workshop.

\JuneCorrections{Gilbert Moultaka has received partial support from the European Union’s Horizon 2020 research and innovation programme under the Marie Skłodowska-Curie grant agreement No 860881-HIDDeN.}

The work of Richard von Eckardstein was partially supported by the Deutsche Forschungsgemeinschaft (DFG) through the Research Training Group, GRK 2149: Strong and Weak Interactions – from Hadrons to Dark Matter.
\end{acknowledgements}

\appendix

\section{Runnings} \label{app:analytical_Yuk0}
For completeness, we give hereafter the explicit dependence of the Yukawa contributions to 
\cref{eq:rgemphi_lle,eq:rgeA6_lle,eq:rgelambda6_lle,eq:rgemphi_udd,eq:rgeA6_udd,eq:rgelambda6_udd} for
four flat directions involving the third generation of leptons or quarks.

$$L_e L_\mu \tau:$$
\begin{align}
    \textrm{Y}_{m_\phi}^{L_1L_2e_3}=&   \frac12 ({\bf Y}_{\!\!E_{\,3 3}})^2 \left[({\bf A}_{\!E_{\,3 3}})^2 + m_{H_d}^2 + m_{\tilde \tau_R}^2 + m_{\tilde \tau_L}^2 \right]\nonumber \\
    &+ \frac14 \sum_{i=1}^2 ({\bf Y}_{\!\!E_{\,i i}})^2 \left[({\bf A}_{\!E_{\,i i}} )^2 + m_{H_d}^2 + m_{\tilde e_R^{i}}^2 + m_{\tilde l_L^{i}}^2\right],\\
    \textrm{Y}_{\Asix}^{L_1L_2e_3} =& \frac12 \sum_{i=1}^2 {\bf A}_{\!E_{\,i i}} ({\bf Y}_{\!\!E_{\,i i}})^2 
+ {\bf A}_{\!E_{\,3 3}} ({\bf Y}_{\!\!E_{\,3 3}})^2,\\
    \textrm{Y}_{\lambda_6}^{L_1L_2e_3}=& - \frac12 \sum_{i=1}^2({\bf Y}_{\!\!E_{\,i i}})^2 -
 ({\bf Y}_{\!\!E_{\,3 3}})^2,
\end{align}

%
%
$$L_\tau L_\mu \tau:$$
\begin{align}
    \textrm{Y}_{m_\phi}^{L_3L_2e_3}=&  \frac34 ({\bf Y}_{\!\!E_{\,3 3}})^2 \left[({\bf A}_{\!E_{\,3 3}})^2 + m_{H_d}^2 + m_{\tilde \tau_R}^2 + m_{\tilde \tau_L}^2 \right] \nonumber\\
    &+ \frac14  ({\bf Y}_{\!\!E_{\,2 2}})^2 \left[({\bf A}_{\!E_{\,2 2}} )^2 + m_{H_d}^2 + m_{\tilde e_R^{2}}^2 + m_{\tilde l_L^{2}}^2\right],\\
    \textrm{Y}_{\Asix}^{L_3L_2e_3} =& \frac12  {\bf A}_{\!E_{\,2 2}} ({\bf Y}_{\!\!E_{\,2 2}})^2 
+ \frac74{\bf A}_{\!E_{\,3 3}} ({\bf Y}_{\!\!E_{\,3 3}})^2,\\
    \textrm{Y}_{\lambda_6}^{L_3L_2e_3}=& - \frac12 ({\bf Y}_{\!\!E_{\,2 2}})^2 - 2
 ({\bf Y}_{\!\!E_{\,3 3}})^2,
\end{align}

%
$$tds:$$
\begin{align}
\textrm{Y}_{m_\phi}^{u_3d_1d_2}=&  \frac12 ({\bf Y}_{\!\!U_{\,3 3}})^2 \left[({\bf A}_{\!U_{\,3 3}})^2 + m_{H_u}^2 + m_{\tilde t_R}^2 + m_{\tilde q_L^{3}}^2 \right]\nonumber \\
&+ \frac12 \sum_{i=1}^2 ({\bf Y}_{\!\!D_{\,i i}})^2 \left[({\bf A}_{\!D_{\,i i}} )^2 + m_{H_d}^2 + m_{\tilde d_R^{i}}^2 + m_{\tilde q_L^{i}}^2\right],\\
\textrm{Y}_{\Asix}^{u_3d_1d_2} =& \sum_{i=1}^2 {\bf A}_{\!D_{\,i i}} ({\bf Y}_{\!\!D_{\,i i}})^2 
+ {\bf A}_{\!U_{\,3 3}} ({\bf Y}_{\!U_{\,3 3}})^2,\\
\textrm{Y}_{\lambdasix}^{u_3d_1d_2} =& -\sum_{i=1}^2({\bf Y}_{\!\!D_{\,i i}})^2 -
({\bf Y}_{\!U_{\,3 3}})^2,
\end{align}
%
%
$$tsb:$$

\begin{align}
\textrm{Y}_{m_\phi}^{u_3d_2d_3}=&  \frac12 ({\bf Y}_{\!\!U_{\,3 3}})^2 \left[({\bf A}_{\!U_{\,3 3}})^2 + m_{H_u}^2 + m_{\tilde t_R}^2 + m_{\tilde q_L^{3}}^2 \right]\nonumber \\
&+ \frac12  ({\bf Y}_{\!\!D_{\,3 3}})^2 \left[({\bf A}_{\!D_{\,3 3}} )^2 + m_{H_d}^2 + m_{\tilde b_R}^2 + m_{\tilde q_L^{3}}^2\right]\nonumber\\
&+ \frac12  ({\bf Y}_{\!\!D_{\,2 2}})^2 \left[({\bf A}_{\!D_{\,2 2}} )^2 + m_{H_d}^2 + m_{\tilde d_R^{2}}^2 + m_{\tilde q_L^{2}}^2\right],\\
\textrm{Y}_{\Asix}^{u_3d_2d_3} =& \sum_{i=2}^3 {\bf A}_{\!D_{\,i i}} ({\bf Y}_{\!\!D_{\,i i}})^2 
+ {\bf A}_{\!U_{\,3 3}} ({\bf Y}_{\!U_{\,3 3}})^2,\\
\textrm{Y}_{\lambdasix}^{u_3d_2d_3} =& -\sum_{i=2}^3({\bf Y}_{\!\!D_{\,i i}})^2 -
({\bf Y}_{\!U_{\,3 3}})^2.
\end{align}
%
Following the notations of \cite{Allanach:2003eb}, the bold-faced $\bf Y$ and $\bf A$ are matrices in the lepton or quark flavor space of the R-parity-conserving MSSM Yukawa and trilinear soft-SUSY-breaking couplings. We take these matrices to be real-valued and assume for simplicity vanishing off-diagonal components.

From \cref{eq:couplings}, one can deduce in the general case that
\begin{align}
    g_i(\muorq)&=\frac{g_i(\muorq_0)}{\sqrt{1-\frac{b_i}{8\pi^2}g_i(\muorq_0)^2\ln\left(\frac{\muorq}{\muorq_0}\right)}},\\
    M_i(\muorq)&=M_i(\muorq_0)\left[\frac{g_i(\muorq)}{g_i(\muorq_0)}\right]^2.
\end{align}
Neglecting the Yukawa terms it gives for $LLe$:
\begin{align}
    \mphi ^2(\muorq)&=\mphi ^2(\muorq_0)+\left[\Mtwo^2(\muorq_0)-\Mtwo^2(\muorq)\right]+\frac{1}{11}\left[\Mone^2(\muorq_0)-\Mone^2(\muorq)\right],\\
    \Asix(\muorq)&=\Asix(\muorq_0)-6\left[\Mtwo(\muorq_0)-\Mtwo(\muorq)\right]-\frac{6}{11}\left[\Mone(\muorq_0)-\Mone(\muorq)\right],\\
    \lambda_6(\muorq)&=\lambda_6(\muorq_0)\left[\frac{g_2(\muorq_0)}{g_2(\muorq)}\right]^6\left[\frac{g_1(\muorq_0)}{g_1(\muorq)}\right]^{6/11},
\end{align}
while for $udd$: 
\begin{align}
    \mphi ^2(\muorq)&=\mphi ^2(\muorq_0)-\frac{8}{9}\left[\Mthree^2(\muorq_0)-\Mthree^2(\muorq)\right]+\frac{4}{99}\left[\Mone^2(\muorq_0)-\Mone^2(\muorq)\right],\\
    \Asix(\muorq)&=\Asix(\muorq_0)+\frac{16}{3}\left[\Mthree(\muorq_0)-\Mthree(\muorq)\right]-\frac{8}{33}\left[\Mone(\muorq_0)-\Mone(\muorq)\right],\\
    \lambda_6(\muorq)&=\lambda_6(\muorq_0)\left[\frac{g_3(\muorq_0)}{g_3(\muorq)}\right]^{-16/3}\left[\frac{g_1(\muorq_0)}{g_1(\muorq)}\right]^{8/33}.
\end{align}

\section{Useful functions}
\label{app:betafunctions}
The functions appearing in \cref{sec::finetune2} are given below. 

\begin{align}
\xi_1(\phi) &= \sqrt{2}\left\{\beta_A(\phi)  + \Asix(\phi) \left[6 + \bar\beta_\lambda(\phi)\right]\right\}, \\
\xi_2(\phi) &= 12 \left[ 5 + \bar\beta_\lambda(\phi) \right],\\
\xi_3(\phi) &= -\frac{3}{2\sqrt{2}\pi^2}\sum_{i=1,k} b_ie_iM_i(\phi)g_i(\phi)^4-\sqrt{2}\beta_A(\phi)\left[11+2\bar\beta_\lambda(\phi)\right]+\nonumber\\
&\qquad \sqrt{2}\Asix(\phi)\left\{\frac{3}{2}\sum_{i=1,k} e_i\beta_{g_i^2}(\phi)-\left[5+\beta_\lambda(\phi)\right]\left[6+\bar\beta_\lambda(\phi)\right]\right\},\\
\xi_4(\phi) &= 12 \left\{ \frac{3}{2}\sum_{i=1,k} e_i \beta_{g_i^2}(\phi) - \left[ 5 + \bar\beta_\lambda(\phi) \right] \left[ 9 + 2 \bar\beta_\lambda(\phi) \right]\right\},\\
{\cal B}_1(\phi) &= 3 \beta_m(\phi) - \frac{3}{8\pi^2} \sum_{i=1,k} b_i e_i M_i^2(\phi) g_i^4(\phi), \\
{\cal B}_2(\phi) &= \frac{2 \nu}{\phi} - \beta_m(\phi).
\end{align} 

These functions depend themselves on the $\beta$-functions given by:

\begin{align}
    \beta_{g_i^2}(\phi) &= \frac{b_i}{8\pi^2}g_i^4(\phi),   \\ \beta_{M_i}(\phi) &= \frac{b_i}{8\pi^2}g_i^2(\phi)M_i(\phi),  \\ \bar\beta_{\lambda}(\phi) &= \frac{\beta_{\lambda}(\phi)}{\lambda(\phi)} = -\sum_{i=1,k}\frac{3}{2}e_ig_i^2(\phi), \\
    \beta_m(\phi) &= -\sum_{i=1,k}e_ig_i^2(\phi)M_i^2(\phi), \\
    \beta_A(\phi) &= \sum_{i=1,k}3e_ig_i^2(\phi)M_i^2(\phi). 
\end{align}
To obtain the \lletexte\ ({\sl resp.} \uddtexte\ ) case, take $k=2$ ({\sl resp.} $k=3$) in the equations above, 
with
\begin{align}
    e_{1,2} &= \frac{3}{20\pi^2},\frac{1}{4\pi^2} \ \ \text{for \ \lletexte,} \\
    e_{1,3} &= \frac{1}{15\pi^2},\frac{2}{3\pi^2} \ \ \text{for \ \uddtexte.}
\end{align}

%
%

%

\section{New fine-tuning parameter}

\label{app:genA}

The general definition of ${\cal A}^2$ is:
\begin{equation}
 {\cal A}^2(\phi,\nu)  = \frac{1}{2}\left(\sqrt{r_2^2 - 4 r_1} - r_2\right),
\end{equation}

with
\begin{align}
r_1 &= \frac{c_1(\phi)}{c_3(\phi)}, \  \   r_2=  \frac{c_2(\phi)}{c_3(\phi)}, \\
c_1(\phi) &= 3 \left[{\cal B}_1(\phi) \xi_2(\phi) -{\cal B}_2(\phi) \xi_4(\phi) \right]^2 - \left[ {\cal B}_1(\phi) \xi_1(\phi) - {\cal B}_2(\phi) \xi_3(\phi)\right] \left[  \xi_1(\phi) \xi_4(\phi) - \xi_2(\phi) \xi_3(\phi)\right], \\
c_2(\phi) &=\frac{1}{10} \left\{ 6\left[\xi_2(\phi)+\xi_4(\phi)\right]\left[{\cal B}_1(\phi) \xi_2(\phi) -{\cal B}_2(\phi) \xi_4(\phi) \right] - \left[\xi_1(\phi) + \xi_3(\phi)\right]\left[  \xi_1(\phi) \xi_4(\phi) - \xi_2(\phi) \xi_3(\phi)\right]\right\}, \\
c_3(\phi) &=\frac{3}{100} \left[\xi_2(\phi)+\xi_4(\phi)\right],
\end{align}

where the $\xi_i$'s and $\cal B_i$'are given in \cref{app:betafunctions}.

\newpage
\bibliography{MSSM-Inflation}

\end{document}

%% file: macros.tex
\def\WMAP{{WMAP}}

\newcommand{\mksym}[1]{\ifmmode {\rm #1}\else #1\fi}

\newcommand{\vevchange}{{{field value}}\xspace}
\newcommand{\Vevchange}{{{Field value}}\xspace}

\providecommand{\Planck}{\textit{Planck}}

\providecommand{\text}[1]{\rm{#1}}

\newcommand{\Mpc}{\text{Mpc}}

\providecommand{\muK}{\mu\rm{K}}

\newcommand{\GeV}{\,\text{GeV}}

\newcommand{\begm}{\begin{pmatrix}}
\newcommand{\enm}{\end{pmatrix}}

\newcommand\ba{\begin{eqnarray}}
\newcommand\ea{\end{eqnarray}}
\newcommand\bea{\begin{eqnarray}}
\newcommand\eea{\end{eqnarray}}

\newcommand\be{\begin{equation}}
\newcommand\ee{\end{equation}}

\def\pmb#1{\setbox0=\hbox{#1}%
    \kern-.025em\copy0\kern-\wd0
    \kern.05em\copy0\kern-\wd0
    \kern-.025em\raise.0433em\box0}

\def\p2Y{\;_2Y}
\def\m2Y{\;_{-2}Y}
\def\beglet{
  \addtocounter{equation}{1}%
  \setcounter{parentequation}{\value{equation}}%
  \setcounter{equation}{0}%
  \def\theequation{\arabic{parentequation}\alph{equation}}%
  \ignorespaces
}
\def\endlet{
  \setcounter{equation}{\value{parentequation}}%
  \def\theequation{\arabic{equation}}%
}
\providecommand{\beglet}{\begin{subequations}}
\providecommand{\endlet}{\end{subequations}}

\def\setsymbol#1#2{\expandafter\def\csname #1\endcsname{#2}}
\def\getsymbol#1{\csname #1\endcsname}

\def\Planck{\textit{Planck}}

\newbox\tablebox    \newdimen\tablewidth
\def\leaderfil{\leaders\hbox to 5pt{\hss.\hss}\hfil}
\def\tablenote#1 #2\par{\begingroup \parindent=0.8em
    \abovedisplayshortskip=0pt\belowdisplayshortskip=0pt
    \noindent
    $$\hss\vbox{\hsize\tablewidth \hangindent=\parindent \hangafter=1 \noindent
    \hbox to \parindent{$^#1$\hss}\strut#2\strut\par}\hss$$
    \endgroup}

\def\L2{\ifmmode L_2\else $L_2$\fi}

\def\DeltaT{\ifmmode \Delta T\else $\Delta T$\fi}
\def\deltat{\ifmmode \Delta t\else $\Delta t$\fi}
\def\fknee{\ifmmode f_{\rm knee}\else $f_{\rm knee}$\fi}
\def\Fmax{\ifmmode F_{\rm max}\else $F_{\rm max}$\fi}
\def\solar{\ifmmode{\rm M}_{\mathord\odot}\else${\rm M}_{\mathord\odot}$\fi}
\def\Msolar{\ifmmode{\rm M}_{\mathord\odot}\else${\rm M}_{\mathord\odot}$\fi}
\def\Lsolar{\ifmmode{\rm L}_{\mathord\odot}\else${\rm L}_{\mathord\odot}$\fi}
\def\inv{\ifmmode^{-1}\else$^{-1}$\fi}
\def\mo{\ifmmode^{-1}\else$^{-1}$\fi}
\def\sup#1{\ifmmode ^{\rm #1}\else $^{\rm #1}$\fi}
\def\expo#1{\ifmmode \times 10^{#1}\else $\times 10^{#1}$\fi}
\def\,{\thinspace}
\def\lsim{\mathrel{\raise .4ex\hbox{\rlap{$<$}\lower 1.2ex\hbox{$\sim$}}}}
\def\gsim{\mathrel{\raise .4ex\hbox{\rlap{$>$}\lower 1.2ex\hbox{$\sim$}}}}

\def\simprop{\mathrel{\raise .4ex\hbox{\rlap{$\propto$}\lower 1.2ex\hbox{$\sim$}}}}
\def\deg{\ifmmode^\circ\else$^\circ$\fi}
\def\pdeg{\ifmmode $\setbox0=\hbox{$^{\circ}$}\rlap{\hskip.11\wd0 .}$^{\circ}
          \else \setbox0=\hbox{$^{\circ}$}\rlap{\hskip.11\wd0 .}$^{\circ}$\fi}
\def\arcs{\ifmmode {^{\scriptstyle\prime\prime}}
          \else $^{\scriptstyle\prime\prime}$\fi}
\def\arcm{\ifmmode {^{\scriptstyle\prime}}
          \else $^{\scriptstyle\prime}$\fi}
\newdimen\sa  \newdimen\sb
\def\parcs{\sa=.07em \sb=.03em
     \ifmmode \hbox{\rlap{.}}^{\scriptstyle\prime\kern -\sb\prime}\hbox{\kern -\sa}
     \else \rlap{.}$^{\scriptstyle\prime\kern -\sb\prime}$\kern -\sa\fi}
\def\parcm{\sa=.08em \sb=.03em
     \ifmmode \hbox{\rlap{.}\kern\sa}^{\scriptstyle\prime}\hbox{\kern-\sb}
     \else \rlap{.}\kern\sa$^{\scriptstyle\prime}$\kern-\sb\fi}
\def\ra[#1 #2 #3.#4]{#1\sup{h}#2\sup{m}#3\sup{s}\llap.#4}
\def\dec[#1 #2 #3.#4]{#1\deg#2\arcm#3\arcs\llap.#4}
\def\deco[#1 #2 #3]{#1\deg#2\arcm#3\arcs}
\def\rra[#1 #2]{#1\sup{h}#2\sup{m}}

\def\dots{\relax\ifmmode \ldots\else $\ldots$\fi}
\def\WHzsr{\ifmmode $W\,Hz\mo\,sr\mo$\else W\,Hz\mo\,sr\mo\fi}
\def\mHz{\ifmmode $\,mHz$\else \,mHz\fi}
\def\GHz{\ifmmode $\,GHz$\else \,GHz\fi}
\def\mKs{\ifmmode $\,mK\,s$^{1/2}\else \,mK\,s$^{1/2}$\fi}
\def\muKs{\ifmmode \,\mu$K\,s$^{1/2}\else \,$\mu$K\,s$^{1/2}$\fi}
\def\muKRJs{\ifmmode \,\mu$K$_{\rm RJ}$\,s$^{1/2}\else \,$\mu$K$_{\rm RJ}$\,s$^{1/2}$\fi}
\def\muKHz{\ifmmode \,\mu$K\,Hz$^{-1/2}\else \,$\mu$K\,Hz$^{-1/2}$\fi}
\def\MJysr{\ifmmode \,$MJy\,sr\mo$\else \,MJy\,sr\mo\fi}
\def\MJysrmK{\ifmmode \,$MJy\,sr\mo$\,mK$_{\rm CMB}\mo\else \,MJy\,sr\mo\,mK$_{\rm CMB}\mo$\fi}
\def\microns{\ifmmode \,\mu$m$\else \,$\mu$m\fi}

\def\muK{\ifmmode \,\mu$K$\else \,$\mu$\hbox{K}\fi}
\def\microK{\ifmmode \,\mu$K$\else \,$\mu$\hbox{K}\fi}
\def\muW{\ifmmode \,\mu$W$\else \,$\mu$\hbox{W}\fi}
\def\kms{\ifmmode $\,km\,s$^{-1}\else \,km\,s$^{-1}$\fi}
\def\kmsMpc{\ifmmode $\,\kms\,Mpc\mo$\else \,\kms\,Mpc\mo\fi}
\newcommand\aap{\ref@jnl{A\&A}}%

\providecommand{\sorthelp}[1]{}

\def\lle{LLe}
\def\udd{udd}
\def\lletexte{$\lle$}
\def\uddtexte{$\udd$}
\def\lleind{L_iL_je_k}
\def\uddind{u_id_jd_k}
\def\Atopm{A_{\mathrm{t}}}
\def\Atop{$\Atopm$}

\def\epsilon{\varepsilon}

\def\wbarreh{{\bar{{w}}}_{\mathrm{reh}}}
\def\Pscalar{\mathcal{P}_\zeta}
\def\Ptensor{\mathcal{P}_h}

\def\Mgut{M_{\rm{GUT}}}
\def\Msusy{M_{\rm{SUSY}}}
\def\mphi{m_{\phi}}
\def\Asix{A_{6}}
\def\lambdasix{\lambda_{6}}